%% file: main.tex
\documentclass[sigconf]{acmart}

\newif\ifpreview
\previewtrue

\AtBeginDocument{%
  \providecommand\BibTeX{{%
    \normalfont B\kern-0.5em{\scshape i\kern-0.25em b}\kern-0.8em\TeX}}}

\usepackage{acmart-taps}

\usepackage{subcaption}
\usepackage{graphicx}
\usepackage{svg}
\usepackage{cleveref}
\usepackage{enumitem}
\usepackage{listings}

\setlist[description]{leftmargin=\parindent,labelindent=\parindent}

\newcommand{\TextViewNoFormat}{transcript view}
\newcommand{\TextView}{\textit{\TextViewNoFormat}}
\newcommand{\demographics}[2]{(gender: #1, age: #2 years)}
\newcommand{\forceSpace}[1]{#1}

\copyrightyear{2024}
\acmYear{2024}
\setcopyright{rightsretained}

\acmConference[UIST '24]{The 37th Annual ACM Symposium on User Interface Software and Technology}{October 13--16, 2024}{Pittsburgh, PA, USA}
\acmBooktitle{The 37th Annual ACM Symposium on User Interface Software and Technology (UIST '24), October 13--16, 2024, Pittsburgh, PA, USA}
\acmDOI{10.1145/3654777.3676334}
\acmISBN{979-8-4007-0628-8/24/10}

\begin{document}

\title[DrawTalking: Building Interactive Worlds by Sketching and
Speaking]{DrawTalking: \\ Building Interactive Worlds by Sketching and
Speaking}

\author{Karl Toby Rosenberg}
\orcid{0000-0001-7744-5188}
\affiliation{%
  \institution{New York University}
  \country{New York, New York, USA}
}
\email{ktr254@nyu.edu}

\author{Rubaiat Habib Kazi}
\orcid{0009-0007-9781-0032}
\affiliation{%
  \institution{Adobe Research}
  \country{Seattle, Washington, USA}
}
\email{rhabib@adobe.com}

\author{Li-Yi Wei}
\orcid{0000-0002-1076-6339}
\affiliation{%
  \institution{Adobe Research}
  \country{San Jose, California, USA}
}
\email{liyiwei@acm.org}

\author{Haijun Xia}
\orcid{0000-0002-9425-0881}
\affiliation{%
  \institution{University of California, San Diego}
  \country{La Jolla, California, USA}
}
\email{haijunxia@ucsd.edu}

\author{Ken Perlin}
\orcid{0000-0003-0701-4379}
\affiliation{%
  \institution{New York University}
    \country{New York, New York, USA}
}
\email{perlin@nyu.edu}

\renewcommand{\shortauthors}{Rosenberg, et al.}

\begin{abstract}
    We introduce DrawTalking, an approach to building and controlling interactive worlds by sketching and speaking while telling stories.
    It emphasizes user control and flexibility, and gives programming-like capability without requiring code.
      
    An early open-ended study with our prototype shows that the mechanics resonate and are applicable to many creative-exploratory use cases, with the potential to inspire and inform research in future natural interfaces for creative exploration and authoring.
\end{abstract}



\begin{teaserfigure}
\centering
  \includegraphics[width=\textwidth]{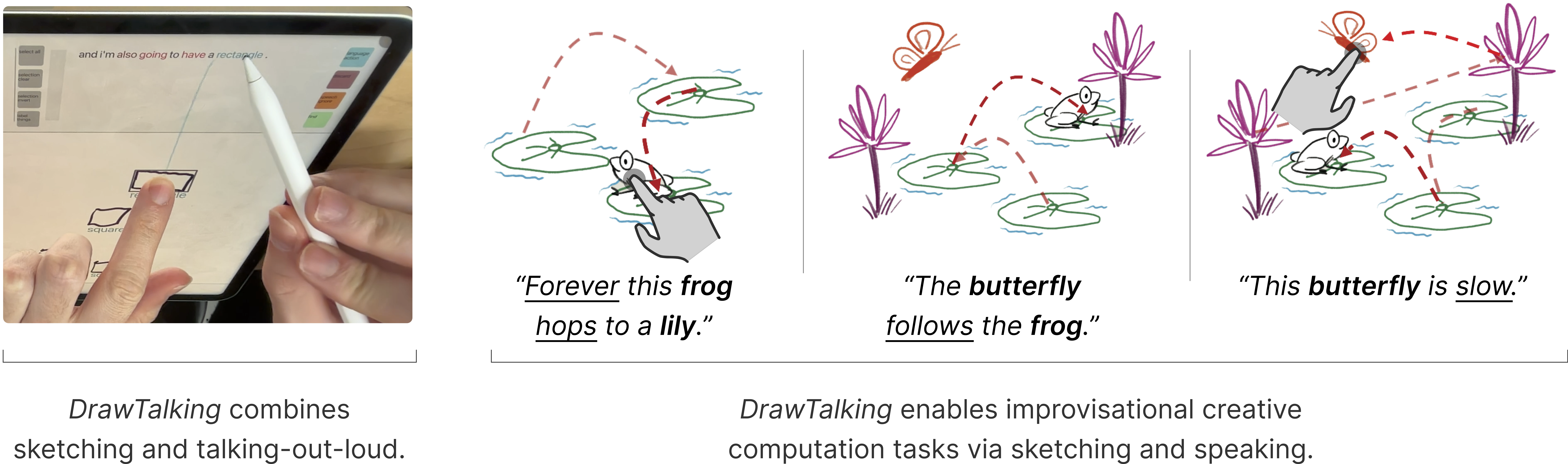}
  \caption[interfaceTeaser]{\textbf{Our approach \textit{DrawTalking}} mediates sketching and talking-out-loud through direct manipulation, enabling many use cases across improvisational creative tasks.}
  \label{fig:teaser}
  \Description[This figure showcases the DrawTalking sketching + speaking controls.]{This figure showcases the DrawTalking sketching + speaking controls. (1) On the left is a real-world photo of the user labeling a sketch on the digital canvas. The canvas contains a few user-drawn rectangle objects. The user narrates "and I'm also going to have a rectangle," which shows-up in a text transcript. Simultaneously, the user holds one of the rectangles with their finger and tapping the word "rectangle" in the aforementioned transcript. This labels the user's sketch with the word "rectangle." This is important because now the system knows what the name of the user's object is, meaning it can now be controlled with more speech commands. (2) On the right is a mock-up of an example scenario; 3 panels. Panel 1: 3 lily pads have been drawn, as well as a frog. A caption reads a speech command: "Forever this frog hops to a lily." The user's finger points to the frog to indicate which frog is being commanded. Dotted arrows indicate that the frog will now hop between the lilies once the command stops. Panel 2: Now a butterfly and some flowers have been drawn. The new command caption reads, "The butterfly follows the frog." Panel 3: The command caption reads "This butterfly is slow" and the user's finger is touching the butterfly, meaning the butterfly's speed will now be decreased. Dotted arrows are now indicating that the butterfly has been moving around, following the frog as previously commanded. Slowing the butterfly does not require redoing the previous command.}
\end{teaserfigure}

\maketitle

\section{Introduction}
Sketching while speaking aids innovation, thinking, and communication --- with applications in animation, game design, education, engineering, rapid prototyping, storytelling, and many other creative and spontaneous activities \cite{fan_drawing_2023, tversky_visualizing_2011}. The combination enables us to think about and share anything through make-believe -- including things that do not or cannot exist. We achieve this by assigning \textit{representations} (sketches) to \textit{semantic concepts} (objects, behaviors, properties) \cite{turner_make-believe_2016}.
For example we might suspend our disbelief so a square represents anything such as a house, a playing card, a map, a dog, or a person.

Furthermore, we often engage with an audience or conversational partner while drawing and talking, whether it's through storytelling to one's child or explaining concepts during a whiteboard lecture. These types of drawing and talking interactions have a spontaneous, playful, exploratory, and improvisational feel with fluid interactions. Storytelling and drawing while interacting with a standard GUI is arguably challenging if we wish to achieve this feel and fluidity. The goal, then, is to explore drawing and talking for these scenarios as an alternative, additional input scheme. 

\textbf{Specifically, we explore how we can take advantage of the presence of narrative to support richer human-computer interactions in creative world-building,} where speech can both be used as verbal storytelling addressed to people \textit{and} as a simultaneous input to the machine. \textbf{We are motivated to design interactions built around this presence of narrative to enable richer human-computer interactions in creative world-building.}

This work is an attempt to realize a style of spontaneous interaction that seamlessly integrates sketching and talking-out-loud to build interactive, explorable worlds, thus greatly increasing our range of computational expression \cite{victor_humane_2014}. In the design of these interactions, we explore the use of speech and direct input in the context of storytelling. 
Prior works in interactive sketching \cite{sutherland_sketchpad_1963, kazi_kitty_2014, suzuki_realitysketch_2020, chalktalk_sketch_commands_2018, saquib_constructing_2021}, language/AI-mediated interfaces \cite{winograd_understanding_1972, bolt_put-that-there_1980, liu_visual_2023_fix, xia_crosstalk_2023, huang_scones_2020, xia_crosspower_2020, subramonyam_taketoons_2018, laput_pixeltone_2013}, and visual programming-adjacent interfaces or games \cite{sutherland_-line_1966, resnick_scratch_2009, landay_silk_1996, perlin_improv_1996, Max:Cycling74, game:LittleBigPlanet2, game:Dreams} have laid valuable groundwork. However, they often require content to be pre-built or assume that users have predetermined goals, limiting the user's potential to be spontaneous. Tools focus on generating a specific output rather than facilitating an ongoing creative process \cite{compton_casual_2015}. They might enforce given representations of objects (e.g. realistic or specific sketch recognition). They might feature complex UI, and in the case of programming-oriented tools, require explicit programming knowledge (whether via text, nodes, or blocks).

In our prototype system, \textit{DrawTalking}, users speak while freehand-sketching to create, control, and iterate on interactive visual mechanisms, simulations, and animations, with the freedom to tell stories using the same speech input as part of the narrative\footnote{Informally, we distinguish between storytelling and narration: storytelling is a process achieved by mixing possibly many modalities (narration, speech, text, visuals). Narration refers to the use of speech or text — possibly, but not always for storytelling.}. Via speech and direct manipulation, the user names their sketches to provide semantic information to the system, and narrates desired behaviors and rules to influence the interactive simulation, as if by explaining to others or telling a story. This is inspired by the way in which people might speak and/or write text to describe objects as they are being drawn. 

\textbf{The core aspect of the user-machine interaction design is the simultaneous use of speech as narrative, and as input to the system, to enable fluid storytelling and building of interactive worlds.} The user narrates, points, and draws as if telling the story of their drawings to an audience. The user's narration communicates intent to the machine by labeling (naming) or commanding (directing) objects. As a result, the user gains fluid programming-like capability \textit{through} their narration, as opposed to needing to narrate while multitasking with a complex GUI and the potential audience. This allows for a natural extension to the GUI in the context of storytelling.

The second key insight is that narration and pointing as in storytelling can provide the machine with names and properties belonging to sketches, so in this storytelling context, the machine does not need to guess the user's intended meaning behind their drawings. As a result, users can build and control interactive worlds by simple combinations of sketching and speaking while telling a story.

\textit{DrawTalking}, in addition, requires no preparation or staging step, and supports the user in making changes to content and behavior anytime. By design, we balance AI-automation with user-direction such that the user is the one who chooses how to represent content, when to do operations, and what logic to define.

\textbf{In sum, we contribute}:
\begin{itemize}
    \item\textit{DrawTalking}: a novel sketching+speaking interaction with a design supporting fluid worldbuilding within the context of storytelling. It aims to balance user-agency with machine automation.
    \item\textit{A demonstration of \textit{DrawTalking} as a multi-touch prototype application for the iPad.}
    \item \textit{An early qualitative study of \textit{DrawTalking} that reveals its use for emergent creativity and playful ideation.}
\end{itemize}

\section{Related Work}

Our research is based on a confluence of advances in multimodal, sketching, and programming interfaces.

\subsection{Natural Language-Adjacent Interfaces}

Systems such as SHRDLU \cite{winograd_understanding_1972} and Put That There \cite{bolt_put-that-there_1980} pioneered the vision of employing natural language to communicate with computers. Due to recent advances in speech recognition and natural language understanding, the popularity of this interaction modality has exploded, and has been used in a wide range of domains. For example, VoiceCut \cite{kim_vocal_2019} and PixelTone \cite{laput_pixeltone_2013} allow users to speak short phrases or sentences to perform desired operations in image editing applications, but these applications are heavily domain-specific.
Tools like WordsEye~\cite{coyne_wordseye_2001}, Scones~\cite{huang_scones_2020}, and CrossPower~\cite{xia_crosspower_2020} enable scene generation or content editing via language, and interfaces such as Visual Captions~\cite{liu_visual_2023_fix}, RealityTalk~\cite{liao_realitytalk_2022}, and CrossTalk~\cite{xia_crosstalk_2023} use language structure to make content appear during talks or conversations.
However, many of these interfaces tend to assume that the user knows what they want to create in advance — i.e. an end-product with an initial goal. They require an initial phase in which the user must define content and behavior up-front. This could limit open-ended exploration during the creative process, when the user does not necessarily have an end-goal in mind. Further, the majority of such interfaces use language input to generate or spawn content without the user in the loop. An alternative is to enable greater interactive control and definition of the behavior of the content.
Our approach emphasizes flexibility and user-control during the creative process. The user can define and iterate on content at any point. Our prototype specifically supports definition of behaviors within an interactive simulation. Within this prototype, we explore language input, combined with direct manipulation, as a way of empowering the user to program and control scenes interactively.
Notably, although speech+pointing commands aren't new (e.g. "Put That There" \cite{bolt_put-that-there_1980}), we focus on the use of speech input as leveraging a narrative that is required \textit{even when} using standard input.

\subsection{Dynamic Sketching Interfaces}

HCI researchers have extensively explored sketching interfaces for dynamic and interactive visualizations ever since the first graphical user interface (GUI) SketchPad \cite{sutherland_sketchpad_1963}
and William Sutherland's thesis, the forerunner of the visual programming language \cite{sutherland_-line_1966}. Many works use direct manipulation and sketching techniques to help users craft interactive behaviors and toolsets for illustrated animation, UI, and visual-oriented programs. For example, works by Kazi et al.~\shortcite{kazi_draco_2014, kazi_kitty_2014}, Landay et al.~\shortcite{landay_silk_1996}, Saquib et al.~\shortcite{saquib_constructing_2021}, and Jacobs et al.~\cite{jacobs_dynamic_2018} focused on mixing illustration, programming, and prototyping. Programming by-demonstration is featured in works such as K-Sketch \cite{davis_k-sketch_2008} and Rapido \cite{leiva_rapido_2021}.
Scratch \cite{resnick_scratch_2009} is a well-known visual programming environment mixing game-like interactions with user-provided content and images for a playful experience. texSketch \cite{subramonyam_texsketch_2020} supports the user in forming connections between texts and concepts to learn via active diagramming. Our interface is framed in a complementary way around correspondences between sketches and language elements, but we use sketch-language mapping as a control mechanism, enabling the user to create interactive simulations and behaviors for open-ended exploration.

Prior work also explored supporting development of\\pre-programmed simulations and domain specific behaviors to craft interactive diagrams. In Chalktalk, for example, the system uses sketch recognition to map a user's hand drawn sketches into corresponding dynamic, pre-programmed behaviors/visualizations \cite{chalktalk_sketch_commands_2018}.
More domain-specific tools like MathPad2 \cite{laviola_mathpad2_2004}, Eddie \cite{sarracino_user-guided_2017}, PhysInk \cite{scott_physink_2013} and SketchStory \cite{lee_sketchstory_2013} use hand-drawn sketches and direct manipulation interactions to create interactive simulations in physics, math, and data visualization.\

\subsection{Programming-Like Interfaces}

The customizability and flexibility of such a more general interface implies a need for programmability. Examples of these include real-time world simulation systems and programmable environments such as the SmallTalk programming language\cite{goldberg_smalltalk-80_1983}, Scratch \cite{resnick_scratch_2009, maloney_scratch_2010} , Improv \cite{perlin_improv_1996}, ChalkTalk\cite{chalktalk_sketch_commands_2018}, and creative world-building games like the Little Big Planet series \cite{game:LittleBigPlanet, game:LittleBigPlanet2, ross_playing_2012} and Dreams \cite{game:Dreams}.
These encourage interactive building of scenes, games and stories. They combine 
 elements of interactive visual programming and drawing/sculpting with many types of content (2D, 2.5D, 3D, images). However, all use explicit interfaces for programming or programming-like functionality (nodes, wires, text). For example, voice-enabled AI interfaces like StoryCoder \cite{dietz_storycoder_2021} integrate a collaborative agent with which to communicate to create story sequences, but still use the block-based programming GUI as in e.g. Scratch. PUMICE \cite{pumice2019} is an AI agent specializing in defining new tasks on existing mobile GUIs via a conversation with the user. To learn unknown functionality, the agent repeatedly asks the user to demonstrate by direct manipulation with the app or to explain it by speech. Although comparable to our approach in terms of using speech+direct manipulation as a form of programming, PUMICE requires interaction with a chat-bot-like interface by design, and requires existing target applications (a different, specialized use case as opposed to ours). In contrast, we wanted to support simultaneous narration and (potentially) interaction with an audience while building a scene, with no pre-existing application or goal in-place. This leads to different interaction design criteria: we choose not to use an AI assistant because this could divert attention from an audience. Rather, the user speaks in narrative form as in storytelling and mixes direct manipulation (touch, drawing). The same input functions as both storytelling to an audience and commands to the machine, so the user does not need to address an assistant.

Overall, we depart from explicit UI for sketching+programming-like capability, and largely replace much programming-like functionality with the use of language. Our direction explores the use of verbal, descriptive story narration together with other input modalities (i.e., touch and pen input) to create animated and interactive graphics through sketching while potentially communicating to an audience as one might during teaching or storytelling.

\section{Formative Steps}

\label{sec:formativeSteps}
We derive a set of design goals by examining existing practices. This would also inform the development of interaction techniques and a prototype interface.  

\subsection{Methodology}
We used a mixed-method approach for our study consisting of an analysis of online instructional videos and conducting a set of sketching design sessions.

To get a sense of the relationships between speech, text, and drawings, and to find examples of content that people created, we conducted an informal (non-exhaustive) search for example content online. We looked for examples that involved creating (or showing) rough sketches while speaking, e.g. from popular video channels and educational course recordings.
(Refer to appendix \autoref{fig:appendix:formative-content} for additional sample materials.)

To gain further insights about workflow and technical requirements, we conducted informal exercises with 6 participants P1$_{init}$-P6$_{init}$ to observe casual sketching and speaking process with little to no preparation. All participants had some experience with sketching-out ideas (e.g. concepts, storyboarding, project/presentation sketching for game design). Each participant was asked to think of (at least) 1 personal topic they would feel comfortable narrating while drawing with freehand sketches. 
Participants used their choice of tools during the session (eg, MS Paint, Notability, basic tools in Photoshop), but we only allowed basic color selection and transformations to keep tool usage roughly the same. (Refer to appendix \autoref{fig:formative-style-p1} for P1's result.)

\subsection{Results and Observations}
\label{sec:formativeResults}

\subsubsection{Association between spoken language, text, and drawing:} Participants used a variety of visual styles (abstract or cartoon), symbols, and diagrammatic elements to express their ideas. They would sometimes label their objects with names or use in-line text to describe elements in the screen for them to reference later \cite{agrawala_design_2011}. Pointing or proximity to sketches while using deictics also acts as a way to refer to objects (as observed in content recorded with cursors or speakers' hands.)  In short, semantics are associated with sketches in many ways, including through narration, or explicit text labels. \emph{The natural association between spoken language, text, and sketches inspired our our main interaction technique.}

\subsubsection{Temporal synchronization between speech and drawing:}
\label{formativeMinTemporalSync}
In the content search and exercises, people do not tend to synchronize their speech exactly when drawing specific objects (or if they do, the delay between actions is unreliable). However the order in which people mention sketches verbally usually will correspond to pointing and sketching actions\cite{oviatt_ten_1999}. \emph{This means we can use this ordering property with respect to speech to let the user map semantics to objects in sequence, but temporal synchronization is unreliable.}

\subsubsection{Drawn versus non-drawn spoken content:}
\label{formativeDrawnVsNot}
Although narration was used to explain the content on-screen, not all content (such as additional animations in the case of produced-content or additional details) and not all narration mapped to each other. In other words, there is content that the user doesn't necessarily describe via language and some content might be considered unnecessary to represent. \emph{Hence, the modalities complement each other to represent the complete story.}

\subsubsection{Speed and flow:}
\label{formativeSpeedAndFlow}
People operate at many different speeds when drawing and talking, and  unpredictably move back and forth between sketches as they're iterated on. This means that \emph{we can't force people to use a specific speed or cadence.}

Additionally, participants would sometimes pause to locate nested UI components (e.g. opening a color-picking menu) or switch between pen and eraser modes. \textit{To complement the GUI, speech could allow users access functionality by referring to it without needing to find it. This could help users keep the flow of their narration.}

\subsubsection{Content modification:}
\label{formativeModContent}
For the exercises, a participant would usually multitask and refine their sketches over time after initially mentioning the entity. 
For example, P1$_{init}$ added colors to their gray bird and pond sketches as they described their experience at the scene. 
Additionally, the participant might make verbal corrections (e.g. P1$_{init}$ initially called a "pond" a "lake," but self-corrected verbally). This means that speech errors are natural in real-life and we ought \emph{to provide ways to correct it} \cite{suhm_multimodal_2001}. \emph{As users' intentions change, users should be able to update their content.}

\subsubsection{Hierarchical object model:} When referring to objects on-screen, narrators would name objects, refer to existing objects to draw attention to them, and define hierarchies and relationships between objects. In short, sketched presentations and content implicitly encoded a \textit{hierarchical object model, describing entities and the relationships between them, as well as ways to refer back to them}.

Based on our formative observations, we needed to prioritize user control over a variety of possible creative workflows. Rather than a fully-automatic solution, we needed a synthesis of user-directed input and system-feedback in support of spontaneous creative processes.

\subsection{Design Goals}
\label{sec:designGoals}
We envision an interface for creative exploration with interactive capability that (a) is controlled via drawing and talking in the context of narration, (b) does not impose many assumptions about the user's intent or content, (c) focuses on the process, not just on the artifact, and (d) does not require programming knowledge. Above all, the user should have control. 

To that end, we wanted an interface that:

\textbf{D1 \textit{makes minimal system assumptions}}
\\where the user controls the creative process, and the representation and behavior of sketches.

\textbf{D2 \textit{is flexible, mutable, fluid, and playful}}
\\in that it supports improvisation, quick changes, and rapid iteration of ideas with a spontaneous feel, where operations are easily accessible and doable in any order.

\textbf{D3 \textit{is transparent and error tolerant}} 
\\by telegraphing what the system's understanding is and providing multiple opportunities for users to make changes or recover from system or user error.

\textbf{D4 \textit{supports programming-like capability}} 
\\without the need for a coding interface.

\section{DrawTalking}
\label{sec:system}

\begin{figure*}[h]
\centering
    \includegraphics[width=0.901\linewidth]{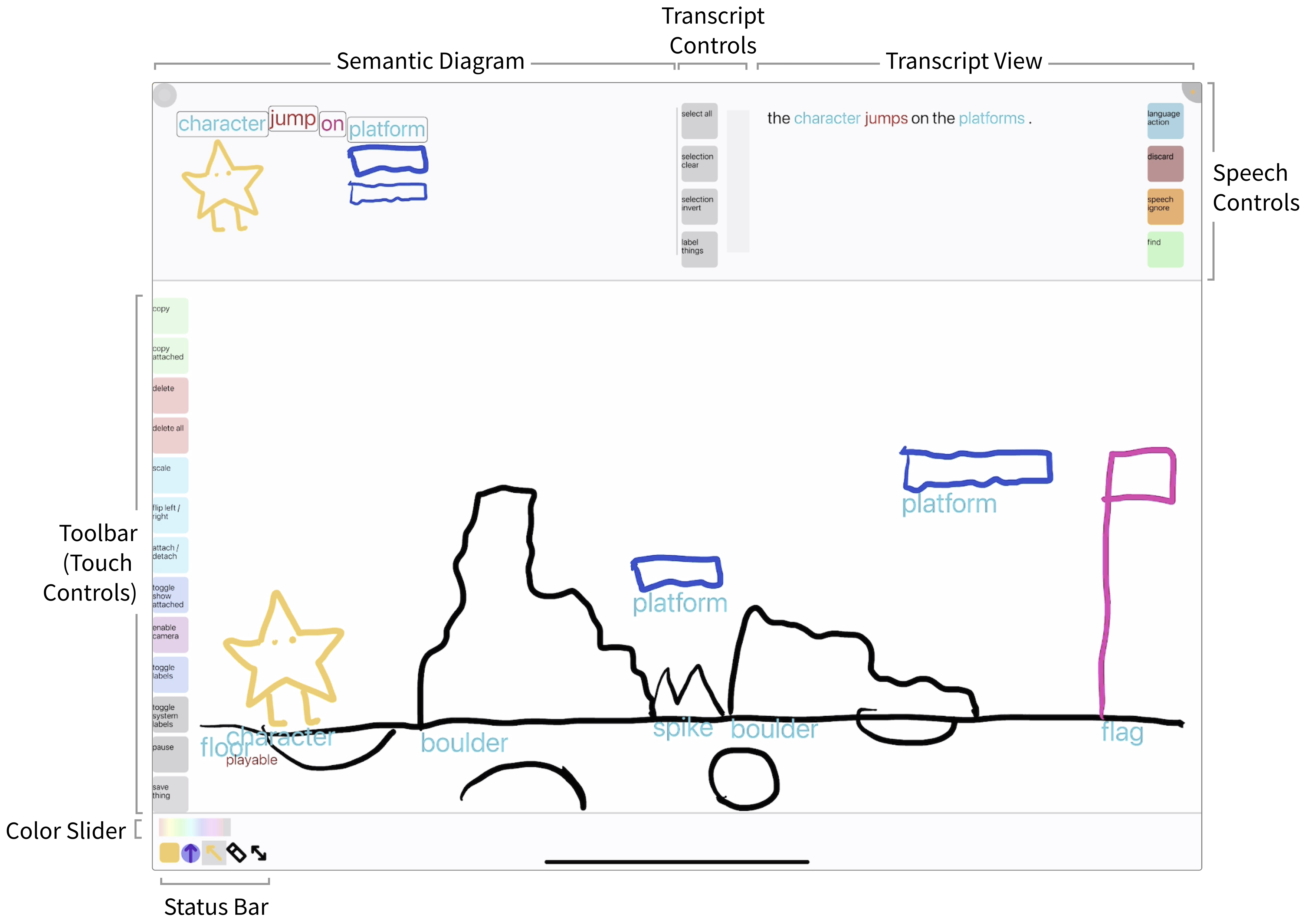}
    \caption{\textbf{Interface Overview}: An interface screenshot (from P4 in \autoref{sec:open_ended_user_study}). The toolbar (left) enables edit operations, e.g. copy, delete, attach/detach, save sketch. The transcript view displays the user's speech input in an interactive panel, and the semantic diagram displays the machine's understanding of the input. "Speech controls" stage/confirm an action, discard input, and toggle speech recognition. "Transcript controls" offer quick transcript text selection. The status bar displays the current color, a compass pointing "up," and has the pen/eraser state-change buttons and indicators. The scene shown is just before the user confirms a command for the utterance \textit{"The character jumps on the platforms"}; it selects the sketch labeled "character" and all sketches labeled "platform." Tapping the "language action" button (top-right) will stage the command and display the semantic diagram (top-left) representing the machine's understanding of the input with selected objects displayed under their respective words; the user confirms the command by tapping again. This causes the character to jump on all of the platforms. For details on the workflow and commands, see \autoref{sec:user_interface_elements} and \autoref{sec:language_commands_and_functionality}.}  
    \label{fig:interface-overview}
    \Description[A screen-shot of the DrawTalking interface, showing full canvas interface elements and semantics diagram mechanics.]{A screen-shot of the DrawTalking interface, showing full canvas interface elements and semantics diagram mechanics. On bottom is the canvas containing a scene of hand-drawn sketches: left-to-right, a cartoon star character labeled "character" and "playable", the surface of the moon "floor," below the entire scene, a rocky cliff labeled "boulder", spikes labeled "spikes", another rocky cliff called "boulder," two moving platforms labeled "platform," one above the spikes, the other after the second boulder, a flag labeled "flag." Around the frame of the interface, bottom-left to top-right: bottom-left: a status bar containing a horizontal color slider, and indicators for the pen control states; left: a vertical toolbar of touchable buttons, to-to-bottom (the relevant ones are): "copy," "copy attached," "delete," "delete all," "scale," "flip left/right," "toggle show attached," "toggle labels," "pause," "save thing;" top-left: a rectangular region containing the semantics diagram, which looks like the simple building blocks for a sentence with objects from the scene corresponding to the nouns as small proxy objects underneath the noun blocks. The diagram reads "character, jump, on, platform" with the character underneath the character block, and both platforms underneath "platform" because the user has specified "The character jumps on the platforms." in their speech command, shown in the transcript on the right. Between the diagram and the transcript are three control buttons for manipulating the text in the transcript: "select all," "selection clear," "selection invert," "label things" (which will label unlabeled objects connected to the semantics diagram with the corresponding nouns in the diagram). On the far top-right are controls for speech commands: (the relevant ones are) "language action," "discard," "speech ignore/enable." Top-right corner: a circular indicator colored orange when the user is speaking, grey otherwise.}
\end{figure*}

\begin{figure*}[!htbp]
    \centering
    \includegraphics[width=\linewidth]{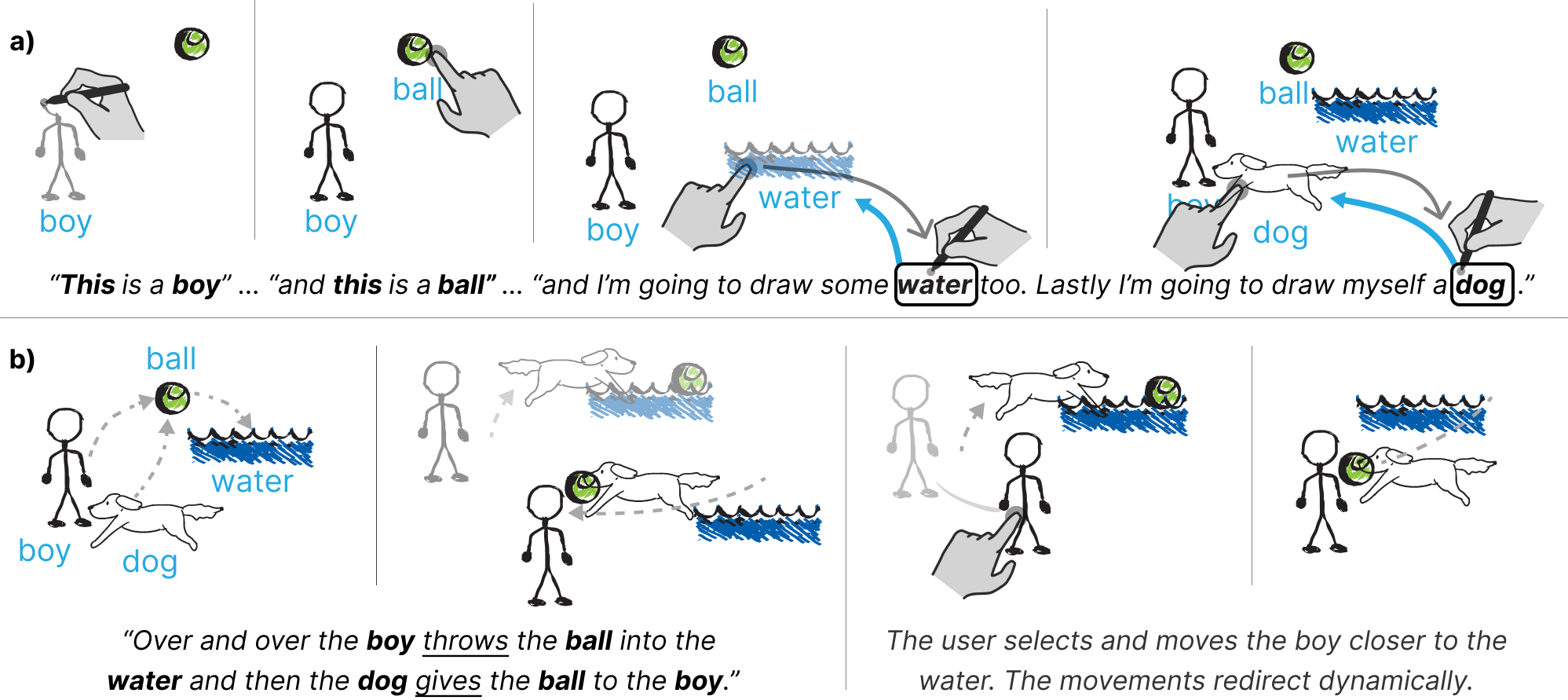}
    \caption{\textbf{Overall workflow: Dog and boy's infinite game of fetch.} \\
    \textit{a)} From left to right, the user draws and labels them using multiple approaches at different stages of drawing.
    a.1) The user is midway through drawing the boy, but can label it using "This is a boy" as the pen is interacting with the object.
    a.2) The ball is already drawn, but unlabeled; during the same sentence the user can quickly tap the ball to label it.
    a.3) The user draws some water and uses free-form speech to say "water" without deixis (such as "this"/"that"). The user can select the object with touch and simultaneously tap the word "water" in the transcript label the object with the word in the transcript. 
    a.4) Touch+pen on a word will remove the label. Adjectives work in the same way.\\
    \textit{b)} \textit{Left:} Labeled sketches are commanded. \textit{Right:} Interactive user-participation. The user can move objects as the system is simulating their movements, which will dynamically adjust as the user plays-around spontaneously.
    }
    \label{fig:dog_fetch}
    \Description[An example of creating objects, naming them, commanding them, and interacting with them by moving the characters around, for a simulation of a game of fetch between a boy and his dog.]{An example of creating objects, naming them, commanding them, and interacting with them by moving the characters around, for a simulation of a game of fetch between a boy and his dog. Two sub-panels a) and b). a): the process of creating and naming the necessary objects. Panel 1: a ball has been drawn and the user is in the process of drawing a boy with the pen, the boy has the name "boy" underneath. Caption: "This is a boy"; Panel 2: The boy is fully drawn and has the name "boy" below. The ball is shown tapped by the user's finger and now has the name "ball" underneath it; caption "and this is a ball; Panel 3: a drawing of a pool of water is added, the user's hand is holding the water, caption: "and I'm going to draw some water too, the user's pen is holding the word water, creating an arrow between the finger-held water and the word "water." A blue link points from "water" to the water sketch, establishing the link, and now naming the water sketch with the text "water" below the sketch; Panel 4: now the touch+pen is done between a new "dog" sketch and the word "dog" in the remainder of the user input caption: "Lastly I'm going to draw myself a dog." b): Panels 1-2 show the boy repeatedly throwing a ball into a pool of water, with the dog giving the ball to the boy afterwards, in an infinite cycle. Below the command reads, "Over and over the boy throws the ball into the water and then the dog gives the ball to the boy." Panels 3-5 show the user moving the boy sketch around as the command continues, forcing the dog to change its trajectory as the targets change during the simulation. The full example is a demonstration of user-in-the-loop interaction.}
\end{figure*}

We designed and developed DrawTalking, an approach to creating interactive worlds using freehand-sketching and speech. It is based on design goals derived from our formative study and implemented as a pen+multi-touch application.

\subsection{Concept}

In DrawTalking, speaking serves a dual-purpose, enabling simultaneous narration and interaction with the interface for storytelling and world-building: the user can explain concepts and tell stories, and at the same time \forceSpace{draw, } refer to\forceSpace{, } and label the objects in their sketched world with semantics (e.g. nouns, adjectives, adverbs) determining objects' names and behaviors.

The \textit{user-specified} labeling tells the system what the objects are, irrespective of their visual representation, making them controllable via the narration. 

The user can furthermore create the rules to automate behavior between objects in the simulation. Touch controls also allow the user to interact directly with the simulated world. This results in a free-form sandbox for animation and programming-like behavior that mixes direct user-control with machine automation. An overall workflow of DrawTalking in shown in \autoref{fig:dog_fetch}.

\subsection{User Interface Components}
\label{sec:user_interface_elements}
\begin{figure*}[h!]
    \centering
    \includegraphics[width=\textwidth]{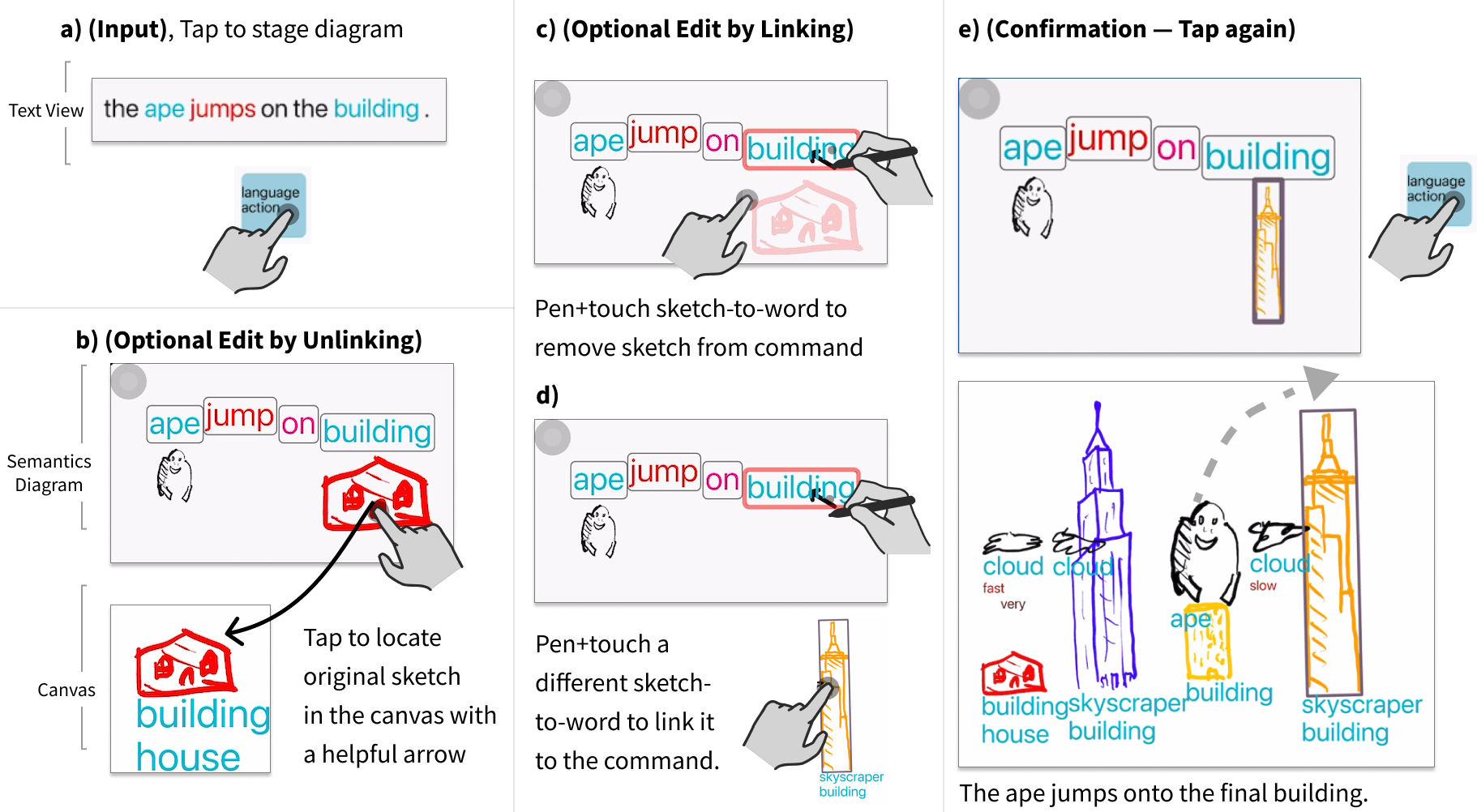}
    \caption{\textit{\textbf{Semantics Diagram}}: An example workflow of error-correction (changing the target "building" from red house to yellow skyscraper). (a) spoken user input in the transcript panel. (b) The semantics diagram  generated from the user's input to visualize system's interpretation. (c-d) Pen+touch interaction between diagram nouns and objects in the scene (including unlabeled ones) allows for re-linking to modify or correct the machine's selections. (e) Confirm language command. If a verb is unknown for a command, the user can pick from an auto-presented list of similar verbs, or otherwise cancel (\autoref{fig:verb-substitution}). In our implementation, if a sentence produces too long a diagram to fit in-view, the user can zoom-out independently from the canvas.
    }
    \label{fig:semantics-diagram}
    \Description[The workflow is shown by example for editing the linking between nouns in a command and sketch objects in the scene.]{The workflow is shown by example for editing the linking between nouns in a command and sketch objects in the scene. a) Subtitled "(Input), Tap to stage diagram." Shows the spoken user input in the transcript view: "the ape jumps on the building." The user taps the "language action" button with their finger to stage the command. b) is subtitled "Optional Edit by unlinking." It shows the semantics diagram representation of the input in a) as understood by the machine. The diagram shows the blocks ape, jump, on, building. Beneath "ape" is a miniature proxy for an ape sketch. Beneath "building" is a red house proxy. A finger is shown tapping the house, generating an arrow pointing from the proxy on the semantics diagram into the sketch canvas where the actual, original house drawing labeled "building," "house" is located. A description reads "Tap to locate original sketch in the canvas with a helpful arrow." c) Subtitled "Optional Edit by Linking." A finger is shown tapping the red building on the diagram and simultaneously tapping the word "building" with the pen to unlink the building from the diagram. A description reads "Pen+touch sketch-to-word to remove sketch from command." d) Pen+touch between another yellow sketch in the canvas labeled "skyscraper," "building" and the word "building" is shown, indicating that now a different sketch is being linked into the command to replace the previous red building. A description reads "Pen+touch a different sketch-to-word to link it to the command. e) Subtitled "(Confirmation — Tap again)" The diagram is shown with the updated yellow skyscraper sketch under the building, indicating that the re-linking was successful. The user taps the "language action" button again to confirm. Below, the full scene in the canvas is shown with various building and skyscraper sketches and some clouds. A dashed-arrow between the ape and the final-selected skyscraper is shown to indicate that the command is about to make the ape jump on the skyscraper. A description reads, "The ape jumps onto the final building."}
\end{figure*}

The interface (\autoref{fig:interface-overview}) exposes a \TextView{} (\autoref{fig:interface-overview}) and semantics diagram (\autoref{fig:semantics-diagram}) to make system understanding of input transparent, quickly editable and accessible, and error-robust, as per our interface design goals (\autoref{sec:designGoals}).

\begin{figure}
    \centering
    \includegraphics[width=\linewidth]{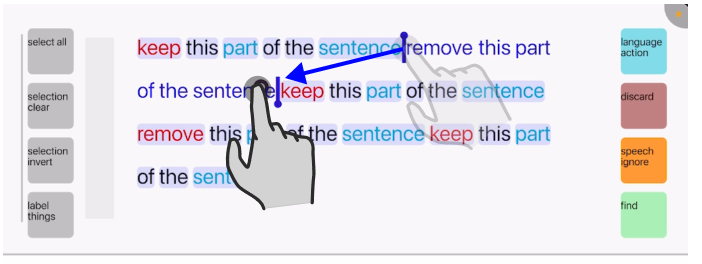}
    \caption{\textit{Transcript View}: Selection / Deselection of Words: Toggle off/on words via touch-dragging to modify input for the next command.
    Small direct edits could be a desirable alternative to repeating the command verbally. For a fallback, the user can type with a keyboard to replace the text. (Left: quick operation buttons for selecting pieces of the text)}
    \label{fig:text_selection_in_view}
    \Description[A screen-shot focused on the rectangular transcript view, its selection/de-selection mechanism, and associated buttons]{A screen-shot focused on the rectangular transcript view, its selection/de-selection mechanism, and associated buttons. The user can toggle selection of ranges of words in the view with touch. The user's finger is shown dragging from a word to another word to define a sub-range of words to remove from the command. The selected words are used to generate a command from the transcript view, so the de-selection of words will remove them. For the input sentence (that will be described now), <B> indicates the beginning position and <E> indicates the end position of the user's finger, and | indicates new-line splits within the transcript view's input: "keep this part of the sentence <B> remove this part | of the sentence <E> keep this part of the sentence | remove this part of the sentence keep this part of the sentence"}
\end{figure}

Speech recognition is continuous for interactive use, so the \TextView{} lets the user visualize and optionally edit speech input (see \autoref{fig:text_selection_in_view}), assign names and attributes to sketches using the pen linking modality, and stage/confirm commands.

When ready, the user taps "language action" to stage a command. The diagram appears and displays a visual for the machine's understanding of the input. It provides a way to reassign objects within the command (if the user wants). The user confirms with the same button to execute the command, or cancels with the discard button.

\begin{figure*}
\centering
\begin{subfigure}[s]{0.45\textwidth}
 \frame{\includegraphics[width=\textwidth]{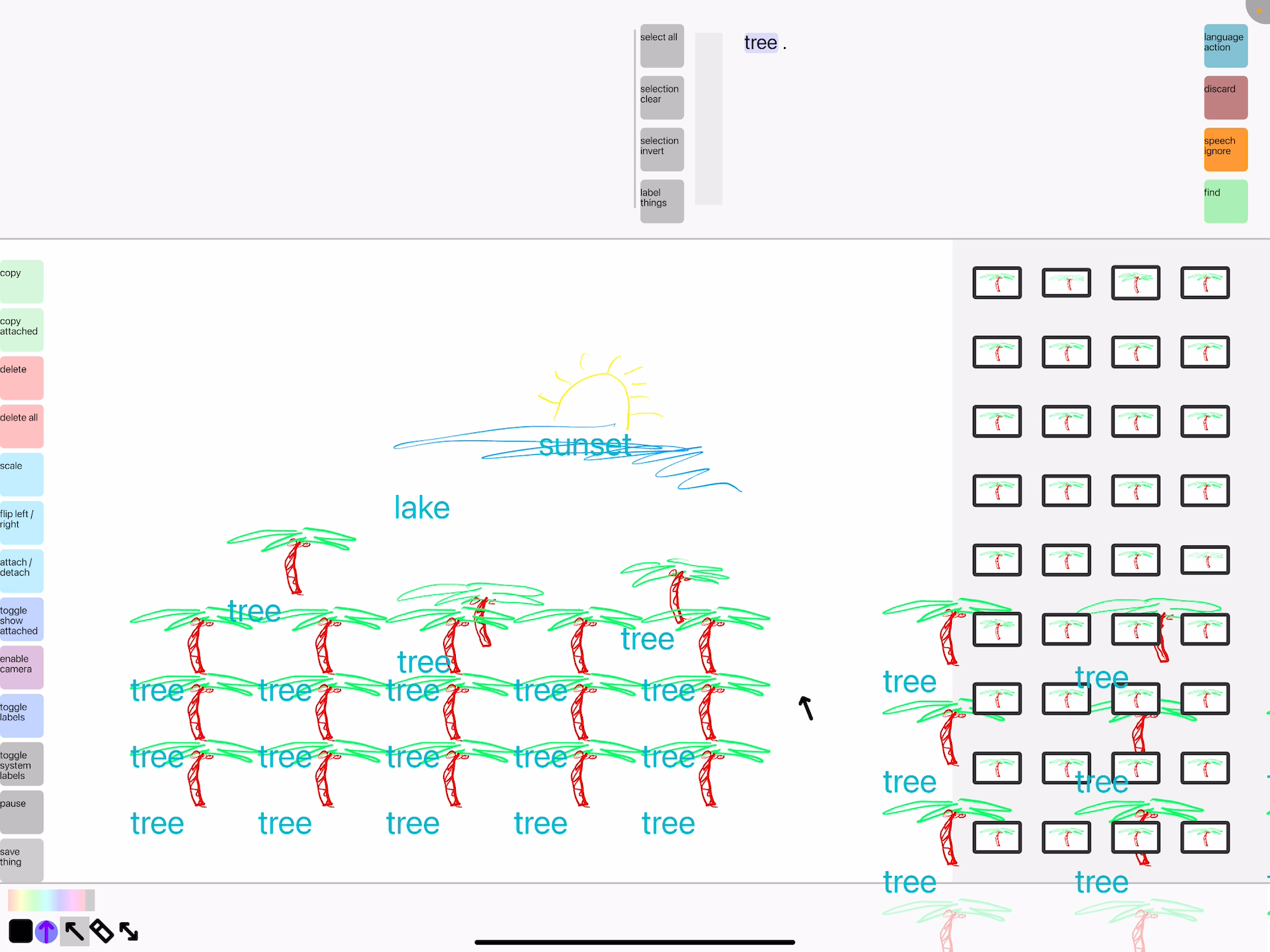}}
\caption{Finding Objects by Labels for Teleporting, Copying, Deleting}
\Description[Find by name functionality using the "find panel"]{Find by name functionality using the "find panel": a scene in the DrawTalking interface with many palm tree ("tree") sketches, a lake sketch, and a sunset sketch. In the transcript view, the word "tree" is selected, indicating a search for all objects named "tree." On the right, the find panel is shown containing miniature selectable entries corresponding to all the tree objects.}
\end{subfigure}
\hfill
\begin{subfigure}[s]{0.45\textwidth}
 \frame{\includegraphics[width=\textwidth]{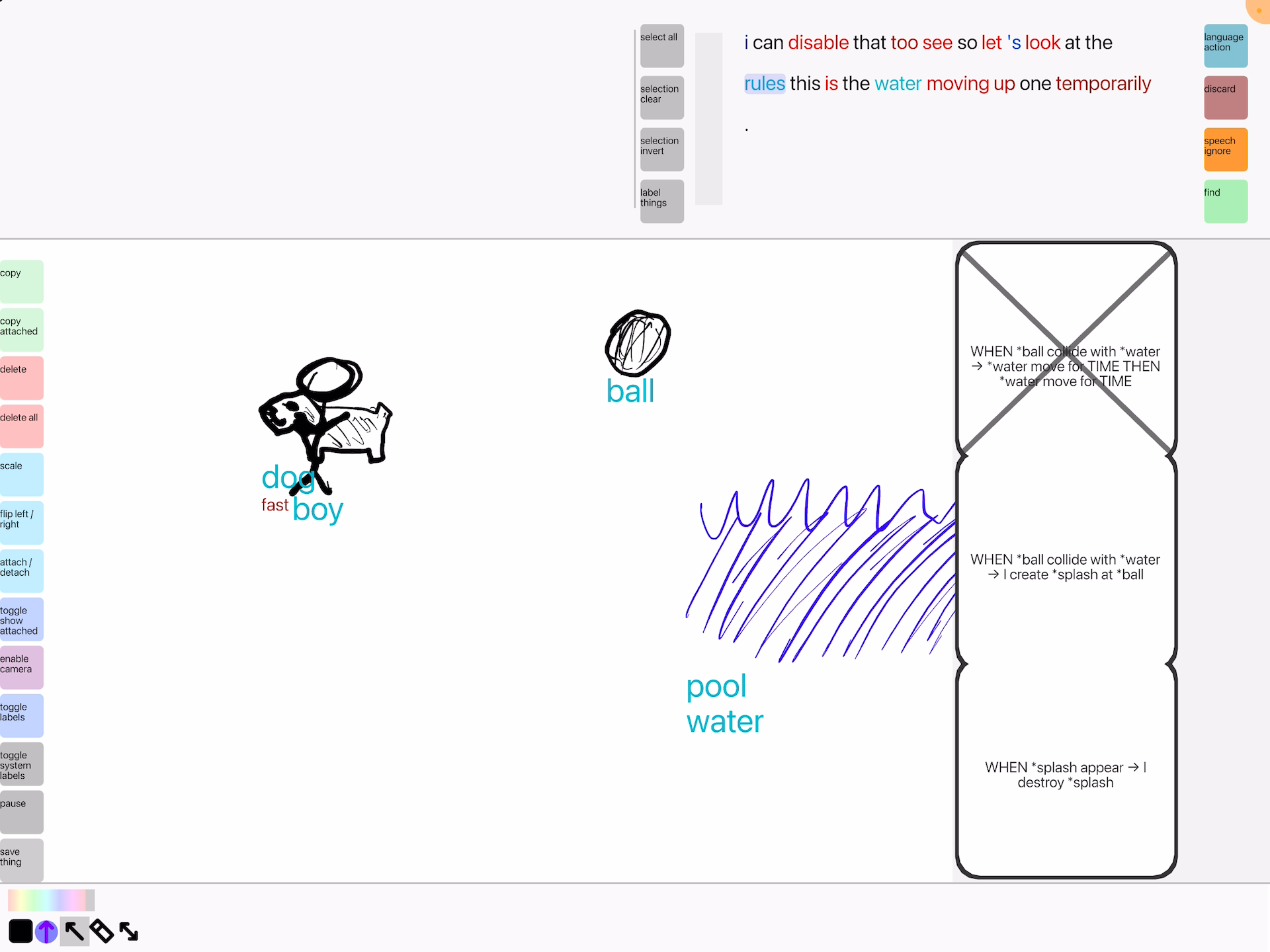}}
\caption{Rule Toggling, Deleting}
\Description[Toggling on/off states for user-defined rules using the rules search functionality in the "find panel"]{Toggling on/off states for user-defined rules using the rules search functionality in the "find panel": a scene containing a boy, dog, ball, and lake/pool/ The transcript view has the words, "I can disable that too see so let's look at the rules this is the water moving up one temporarily." The word "rule" in that sentence is highlighted, indicating that the find panel will show the rules. The panel has three entries for existing rules: 1: "WHEN *ball collide with *water -> *water move for TIME THEN *water move for TIME"" is crossed-out, meaning the rule is inactive. 2: "WHEN" *ball collide with *water -> I create *splash at *ball. 3: "WHEN *splash appear -> I destroy *splash"}
\end{subfigure}
\caption{\textit{Find Panel}: for performing search queries on sketches' noun and adjective labels. (a) Selecting the word "tree" searches for all objects with that label; tapping on an entry warps to the sketch; the eraser deletes the sketch instead; pen dragging copies entries (b) Forgetting active actions and rules: selecting the word "actions" (in the transcript) and an object in the scene will display a panel of all current actions affecting the object. Deleting an action using the pen will stop the action immediately. Selecting the word "rules" (in the transcript) will display active rules. Tapping with the pen will toggle the rule off/on.
}
\label{fig:interface:find}
\end{figure*}

Complementary to the sketching and language workflows, the find panel along with the \TextView{} enables the user to find and teleport to objects by semantic search, which helps avoid losing track of objects. (This use of speech is inspired by real-world interaction, in which we use language to talk about real, distant, invisible, imaginary, or conceptual objects.) Additionally, it enables a means to view, toggle, or delete the current commands and rules acting on the world simulation (See \autoref{fig:interface:find}).

In sum, all three of these sketching-language interface components combine to support the user's process via our design goals. Together, they provide transparency into the system's state, multiple forms of editability and error-tolerance, and multiple optional levels for user control.

\subsection{Sketching and Labeling}
\label{sec:sketching_and_labeling}
Sketches are independently-movable freehand drawings, text, or numbers created by the user. The user labels their sketches to make them controllable: nouns (names) for unique identification and adjectives and adverbs (properties) for modulating sketch behavior. For flexibility, we offer two direct ways to label (or unlabel).
\begin{enumerate}
    \item tap 1 to many sketches and speak with deixis \cite{stapleton_deixis_2017}
    (e.g. \textit{"this/that is a <noun>"}, \textit{"this/those are <noun>s"}).
    \item touch sketches + pen-tap words to link at any time with any text in the history, as opposed to only the current speech, enabling freer narration.
\end{enumerate}

Labeling works in tandem with the user interface elements (\ref{sec:user_interface_elements}). See \autoref{fig:dog_fetch} for an example workflow.

\subsection{Language Commands and Functionality}
\label{sec:language_commands_and_functionality}
DrawTalking interprets the structure of language input into commands built from primitives. 
These primitives are intended to be a small sample of possible functionality that demonstrate our working concept. (They are partly inspired by existing software and design spaces \cite{shi_understanding_2021, game:Dreams, resnick_scratch_2009}.) (See the Info Sheet supplemental material.)
We emphasize that our contribution is the interactive \textit{way} of controlling elements. Our specific application and the fidelity and comprehensiveness of the supported behaviors are implementation details meant as a minimal demonstration of the DrawTalking \textit{concept}. In other words, animation complexity and visual fidelity aren’t limitations inherent to the interface concept. These can be improved by implementers without rethinking the interaction design or contribution. We focus on DrawTalking's expressiveness as a control mechanism for any potential behaviors.

\textbf{Verbs} are actions performable by sketches and the system, either built-in or \textit{user-defined in terms of other verbs by composition of existing primitives}. Examples of implemented verb primitives include:
\begin{itemize}
    \item animations 
    (e.g. \textit{move}, \textit{follow}, \textit{rotate}, \textit{jump}, \textit{flee})
    \item state changes
    (e.g. \textit{create}, \textit{transform})
    \item events
    (e.g. \textit{collide with}, \textit{press})
    \item inequalities 
    (e.g. \textit{equal}, \textit{exceed})
\end{itemize}

\begin{figure}
\centering
 \frame{\includegraphics[width=0.45\textwidth]{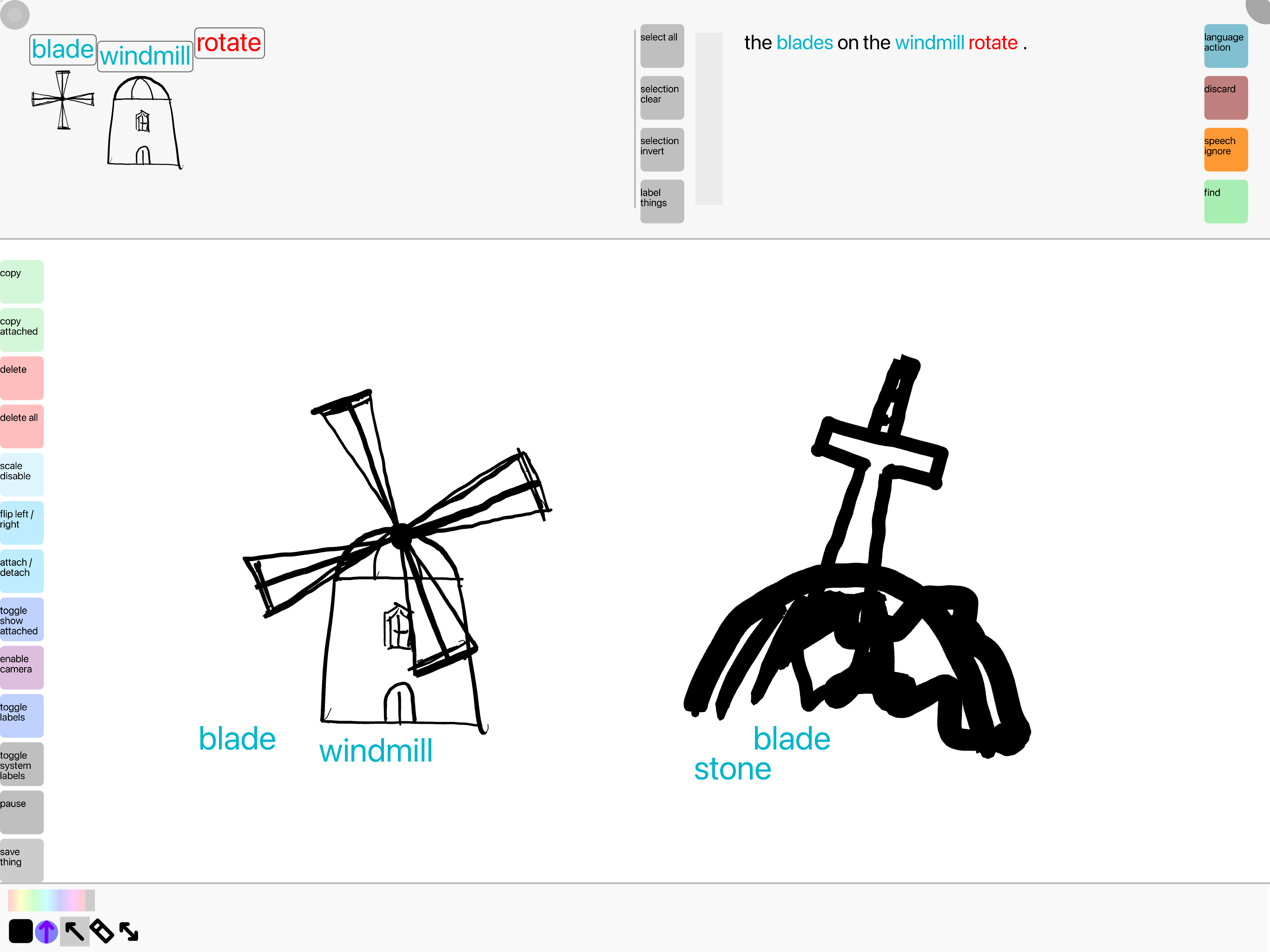}}
\caption{\textbf{Hierarchical Disambiguation}: Here, two sketches named "blade" are attached to different parent objects, but the system picks the one with the correct relationship by looking at the names in the objects' hierarchies. (The verb, "attach to" was used to attach the sketches beforehand.)}
\label{fig:hierarchy}
\Description[Disambiguation of searched objects with same names, by scene hierarchy. The DrawTalking user interface is shown.]{Disambiguation of searched objects with same names, by scene hierarchy. The DrawTalking user interface is shown. The canvas has a windmill base object, a windmill blade object (attached to the windmill base), a stone object, a blade (sword) object (attached to the stone). The user input in the transcript view (the command input) reads "the blades on the windmill rotate;" The semantics diagram (command) shows: "blade windmill rotate" Underneath the word "blade" is a miniature of the blade object that is attached to the windmill. Underneath the word "windmill" is a miniature of the windmill object. This all shows that the system correctly picked the blade object specified in the command for a blade on the windmill.}
\end{figure}

Verb behavior changes based on other parts of the sentence:

\begin{description}
\item[Conjunctions] run simultaneously
(e.g. "The dog jumps \textit{"and"} the cats jump").
\item[Sequences] run in-order
(e.g. "The dog jumps \textit{"and then"} the cats jump).
\item[Stop commands] cancel an ongoing operation
(e.g. "The square stops moving").

\item[Prepositions] (like \textit{on}, \textit{to}, \textit{under}) cause verbs to exhibit different behavior, e.g. \textit{"The dog jumps on/under the bed"} impacts the dog's final position relative to the target. All verbs can use any such prepositions as input. Prepositions also might describe the spatial relationships between objects in a command (\autoref{fig:hierarchy}).
\item[Timers] specify the duration of a verb, e.g. \textit{"the square moves up for 11.18 seconds and then jumps."}.
\item[Loops] repeat an action, either forever (e.g. \textit{"endlessly the dog jumps"}), or finitely (e.g. \textit{"10 times the dog jumps excitedly"}).
\end{description}
Special verbs include \textit{"become"}, which modifies the sketch's labels, and \textit{"transform into"}, which also instantly replaces a sketch's visual representation with another's to support state changes, e.g. \textit{"the sun transforms into a moon"} or \textit{"the frog transforms into a prince"}. "Follow" also offers a simple specialization on nouns — if a system object "view" is commanded to follow a given object, the system camera will track that object. e.g. \textit{"The view follows the hero."}

\textbf{Nouns, pronouns, and deixis} refer to object labels and are used to pick \textit{specific} objects or specify \textit{types} of objects. Using deixis while selecting an object will select the object immediately; pronouns can refer-back to objects, enabling commands that read less repetitively and more naturally: (e.g. "The dog jumped. \textit{She} jumped again.")

The following choose 1+ \textbf{specific} objects for a command by matching object labels (nouns, optional adjectives, etc.) with words:
\begin{itemize}
    \item \textit{1 or more objects:} \textit{... "the" <object label(s)> ...} 
    \item \textit{all objects:} \textit{... "all" <object label(s)> ...}
    \item \textit{a number of objects}: \textit{... <number> <object label(s)> ...}
    \item \textit{a single random object:} \textit{... "a"/"an" <object label(s)> ...}
\end{itemize}

\begin{figure*}[h!]
    \centering
    \includegraphics[width=0.897\linewidth]{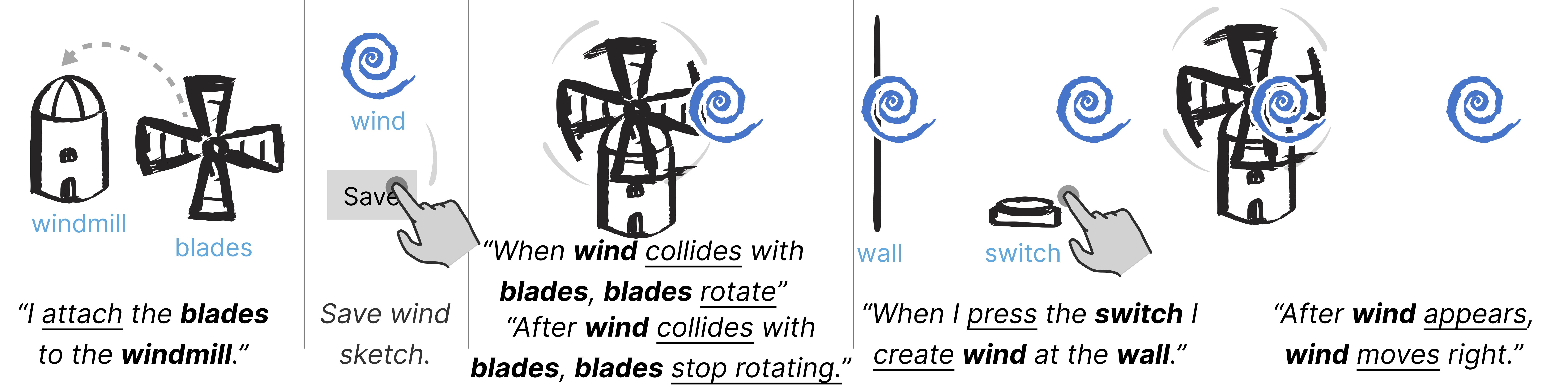}
    \caption{\textbf{Windmill Simulation}: Flexible process for constructing an interactive windmill built from user-defined rules, sketches, and triggers. The user can do this in \textit{any} order and can try results at each step. This works on \textit{any} sketch labeled "blade."
    }
    \label{fig:wind}
    \Description[A process showing the construction of a windmill whose blades rotate as wind sketches spawn-in and move past the blades, all scripted-in using speech commands for defining rules, along with hand-drawn sketches]{A process showing the construction of a windmill whose blades rotate as wind sketches spawn-in and move past the blades, all scripted-in using speech commands for defining rules, along with hand-drawn sketches; Panel 1 shows construction of the windmill with the base labeled "windmill," and the blades labeled "blades." The blades are attached to the windmill via the command (captioned below) "I attach the blades to the windmill." Panel 2 shows the user saving a hand-drawn wind spiral sketch so the system knows what to spawn when "wind" is specified for a "create" command. Panel 3 shows the definition of rules for collision between wind and blades, captioned: "When wind collides with blades, blades rotate." and "After wind collides with blades, blades stop rotating." Panel 4 shows that extra "switch" and left-side wall sketches have been drawn. Pressing the switch causes wind to spawn from the left wall and move to the right. Several clones of the previously saved wind sketch are shown moving to the right and colliding with the windmill blades, causing them to spin clockwise. Captioned are the commands to define all of this: "When I press the switch I create wind at the wall." "After wind appears, wind moves right."}
\end{figure*}

There are a few special cases. The pronoun "I" is reserved; it allows the user to take-part in the narrative of a command (e.g. "I destroy the wall"). "Thing" is also a reserved noun that can refer to any object regardless of label. Plural nouns with no modifiers (e.g. as in "blades rotate") refer to labels used to define the interactions between objects with that label. Specifically, this is used to construct rules, as described below and in \autoref{fig:wind}.

\textbf{Adjectives and adverbs} \label{sec:adjectives} are usable as labels that define the properties of objects. Verbs receive adjectives as arguments, which are evaluated continuously to modulate values and effects (e.g. magnitude, speed, size, distance, height).
For example, \textit{"fast", "slow," "excited"} impact jump height and/or movement speed. Adjectives also disambiguate like-noun labeled-objects (e.g. \textit{"first house"} vs. \textit{"second house"}). Adjectives can be removed programmatically via negation, e.g. \textit{The thing is \textbf{not} fast}.

Adverbs heighten adjective effects multiplicatively -- e.g. \textit{"very"} and \textit{"slightly"} -- and are chainable -- e.g. \textit{"very, very."} Special adjectives offer system-control: e.g. "static" fixes sketches to the screen like a GUI element, useful for buttons, d-pads, score displays, etc.; "visible"/"invisible" toggle sketch visibility by command (e.g. \textit{"The <...> becomes visible/invisible"}) or automatically when saying \textit{"This is a(n) visible/invisible <thing...>"}.

\textbf{Rules} are conditionals that run any kind of command in the future when objects with certain labels satisfy the condition. This enables the user to specify automated commands for the future "when," "as," or "after" an event has completed, without needing to know what they want or create objects in advance,

e.g., {``\textit{When} arrows collide with balloons, arrows destroy balloons.''}.
Rules also allow definition of new verbs in terms of existing primitives,
e.g. "flicker:" \textit{When lights flicker, forever lights disappear for 0.1 seconds and then lights appear for 0.1 seconds.}

\subsection{Examples}

\begin{figure*}[h!]
    \centering
    \includegraphics[width=0.897\linewidth]{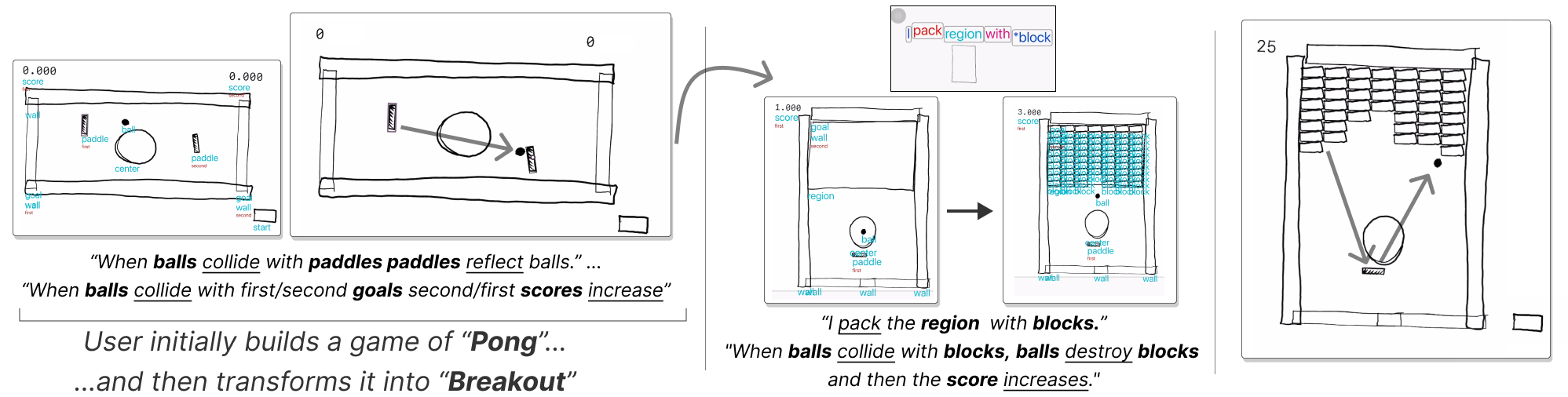}
    \caption{\textbf{Pong into Breakout}: \textit{Left:} per-player paddles and points; \textit{Middle, Right:} scene reoriented, second player's objects removed, destructible bricks logic defined and objects added in-bulk into the desired region. (Note: not all commands shown.)}
    \label{fig:pong}
    \Description[Steps for creation of a game of Pong that is then repurposed and extended using the same sketches into the similar game of Break-out]{Steps for creation of a game of Pong that is then repurposed and extended using the same sketches into the similar game of Break-out: Panel 1: Pong contains 4 wall sketches and score counters for each of two players atop the walls. There is a start game switch in the bottom right that starts the game when pressed. Two paddles within the walls are being manipulated via touch controls. A ball sketch is shown bouncing towards the right paddle, which is about to reflect it left. The current score is 0-0. A version of the same scene is shown with object labels enabled and then a version with them turned off. The speech commands are written below: "When balls collide with paddles, paddles reflect balls."... "When balls collide with first/second goals second/first scores increase." (The slashes indicate these are two commands with differences in the words in the slashed positions.) Panel 2: additional steps transforming Pong into breakout. Objects corresponding to the second player have been removed so it becomes a one-player game. The scene is rotated vertically, a rectangular region is added on the top. Next the semantics diagram displays "I pack region with *block" with the rectangle region in miniature under "region." Last the result of that command is shown where several blocks have been packed into the rectangular region and the region has been removed. (The point was to fill the game region with several breakable blocks.) The ball is shown being reflected off the paddle into the blocks, some of which have already been broken by this state. The score is 3. Panel 3: the final state with object names hidden. By now, the scene has been transformed into Breakout. The scene has been rotated vertically, and the second score counter has been deleted. The user has created several blocks filling the top part of the inner-walled area of the game area they've created. Some of the blocks have been destroyed by the user's bouncing ball. Currently, the ball has bounced off the user's paddle and is heading towards another breakable brick. The score counter reads 25 for the total number of points earned from breaking the blocks. All of these mechanics have been created from user-specified rules in speech commands.}
\end{figure*}

\begin{figure*}[h!]
    \centering
    \includegraphics[width=0.897\linewidth]{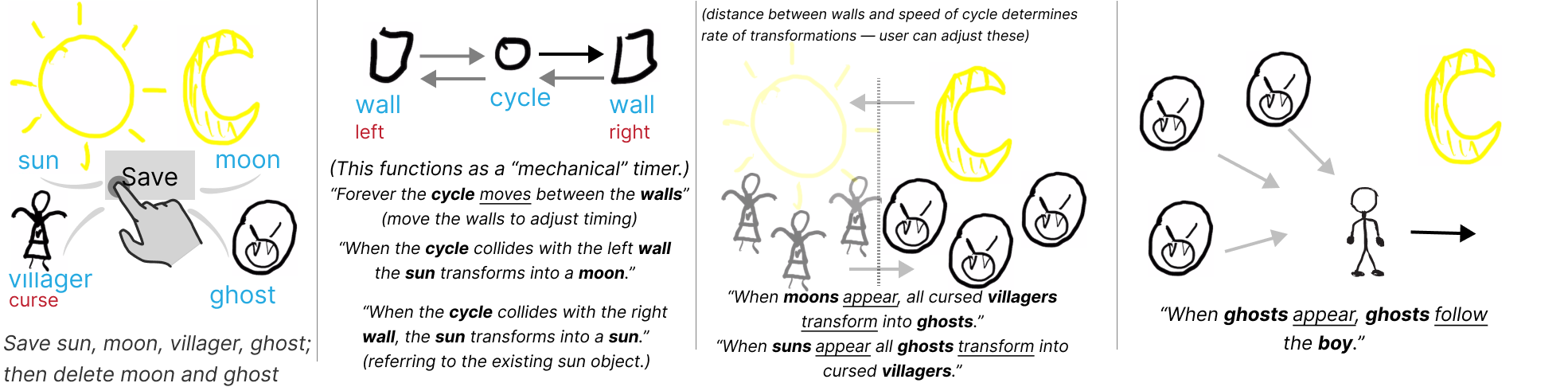}
    \caption{\textbf{Day/Night Cycle with Transforming Villagers}: A sun/moon transforms into the other in-time with collision events. When the moon appears, the villagers transform into ghosts, which chase the boy. This shows event-based behavior changes.}
    \label{fig:example:villager_ghost_transformation}
    \Description[Day/Night Cycle with Transforming Villagers - creation steps]{Day/Night Cycle with Transforming Villagers - creation steps. Panel 1: sun, moon, cursed villager (labeled with the word "curse" as a property), ghost sketches created. The user taps the save button, indicating all 4 sketches have been saved by name into the system for spawning later. Caption: "Save sun, moon, villager, ghost; then delete moon and ghost." Panel 2: There's a left wall sketch (small rectangle), a cycle (circular shape) sketch, a right wall sketch (small rectangle. Arrows point from the left wall to the cycle to the right wall, and the right wall back to the cycle back to the left wall. These indicate a back-and-forth motion of the cycle between the left and right walls. The speech command is shown: "When the cycle collides with the left wall, the sun transforms into a "moon," "When the cycle collides with the right wall; Caption: the sun transforms into a "sun." (referring to the existing sun object.); Panel 3: top caption: "distance between walls and speed of cycle determines rate of transformations --- user can adjust these). The panel is divided left/right by a vertical line separator. On left is the sun, with three cursed villagers below. On right is the moon with three ghosts below, corresponding to the cursed villagers that will transform into ghosts when the moon appears. Arrows between left and right indicate the periodic changes between these two scene states due to the aforementioned logic. Below read the two commands, "When moons appear, all cursed villagers transform in ghosts" and "When suns appear, all ghosts transform into cursed villagers." Panel 4. Three ghost sketches chase a boy sketch, the moon is shown, indicating it's the night-time state. A right arrow pointing from the boy indicates that the user might move the boy around. Arrows from the ghosts towards the boy indicate their following-the-boy behavior. The speech command for completing this set-up reads, "When ghosts appear, ghosts follow the boy"}
\end{figure*}

\begin{figure*}[h!]
    \centering
    \includegraphics[width=0.897\linewidth]{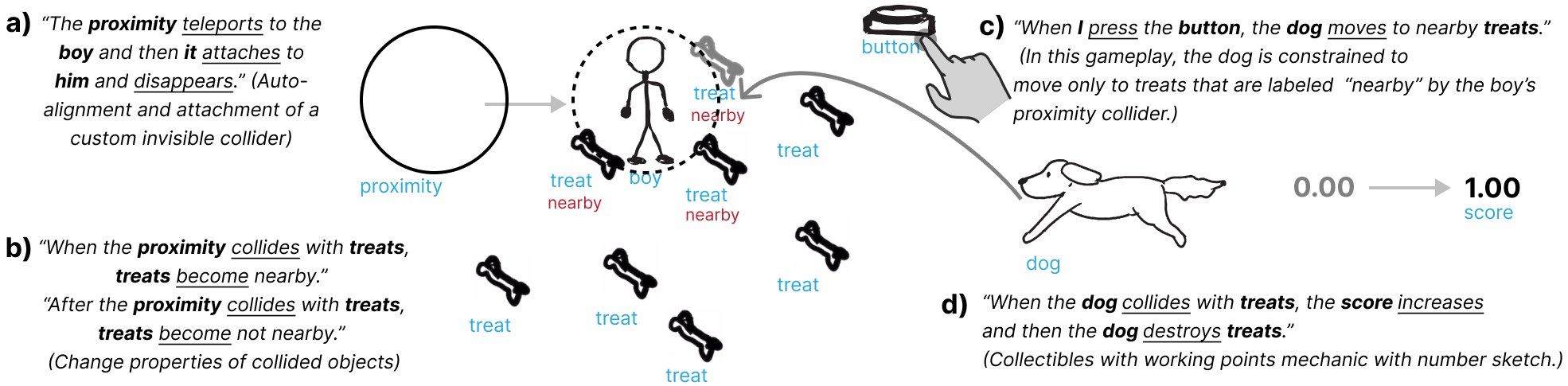}
    \caption{\textbf{"Dog Plays Fetch" as a Prototype Game Mechanic (demonstrating adjectives for dynamic logic)}: The dog interaction is re-imagined as an interactive game mechanic.}
    \label{fig:example:dog_fetch_as_game_mechanic}
    \Description[Constructing a game mechanic in which a button causes a dog to collect point-giving treat objects near a main character sketch.]{Constructing a game mechanic in which a button causes a dog to collect point-giving treat objects near a main character sketch. The scene shows a circular sketch named "proximity" moving to the center of a "boy" sketch, corresponding to the command described in description a): "The proximity teleports to the boy and then it attaches to him and disappears" (Auto-alignment and attachment of a custom invisible collider; in the scene are several objects labeled "treat", a button sketch, and a dog sketch. The treats colliding with the circular proximity collider described before receive the label "nearby" as a result of the rules defined in description b): "When the proximity collides with treats, treats become nearby.", "After the proximity collides with treats, treats become not nearby" (Change properties of collided objects). Description c) for another command for this setup reads: "When I press the button, the dog moves to nearby treats." (In this gameplay, the dog is constrained to move only to treats that are labeled "nearby" by the boy's proximity collider.) An arrow points from the dog to a treat labeled "nearby" within the boy's proximity collider radius, with the user's finger tapping the aforementioned button, indicating that the dog has been commanded by the rule to move to a treat that is nearby. There is also a score (number) sketch labeled "score," mentioned in description d) for the rule: "When the dog collides with treats, the score increases and then the dog destroys treats." (Collectibles with working points mechanic with number sketch.) The score sketch is shown increasing from 0.00 to 1.00 by the dog collecting a nearby treat.}
\end{figure*}

We demonstrate DrawTalking with illustrated examples showing multiple procedural and programming-like capabilities.

\subsubsection{Pond scene} A simple example demonstrating randomized behavior is shown in \autoref{fig:teaser}. Here a user sketches a frog, lily pads, water, and a butterfly. To cause the frog to hop randomly between any existing or future lily pad, they say: \textit{"Forever the frog hops to \textbf{a} lily,"}. The butterfly is then commanded to follow the frog, and the user adjusts the speed of the butterfly: \textit{"this butterfly is slow."}. The user can pause the simulation to edit any objects before resuming the actions.

\subsubsection{Dog and boy's infinite game of fetch} This highlights sequences, loops, and user participation. The user sketches a boy, dog, water, and ball and commands the boy and dog to interact with the ball (\autoref{fig:dog_fetch}). The user can play the role of puppeteer by interactively moving the objects (e.g. the ball) to influence the other sketches' actions dynamically while the simulation is ongoing.
For additional effect, the water could rise upon collision with the ball, with a command like: \textit{"When water collides with balls water moves up for 0.2 seconds and then water moves down for 0.2 seconds."}

\subsubsection{Windmill simulation} This example demonstrates rules, custom object saving and spawning, and buttons. (See \autoref{fig:wind} for steps.) A key point is the flexibility to iterate on rules. In fact, here the rules can be defined in any order. If the user only defines the condition for on-collision, the blades will continue spinning by design. The user, however, can quickly rectify this by defining a rule to stop the blades \textit{after} collision between wind and blades.

\subsubsection{Creating the game "Pong" and turning it into "Breakout"} This example shows how one can quickly reuse functionality to turn one idea into another, in this case, a version of the game "Pong" \cite{noauthor_50_2022} into a version of the game "Breakout" \cite{atariadmin_new_2022} (\autoref{fig:pong}).

Create a ball and walls, then paddles, goals, and points for each player. For points, say \textit{"I want the number 0"} and touch+pen the number to spawn a number object. Label with \textit{"This is the first/second score, goal."}. Make points screen-space UI with \textit{"This thing is static"} and set-up rules for collisions between the ball and goals: \textit{"When balls collide with first/second goals second/first scores increase".} Last, add collision logic: \textit{"when balls collide with walls walls reflect balls"}, \textit{"when balls collide with paddles paddles reflect balls."} Now we have a playable touch-based version of Pong.

We turn this into Breakout by rotating the canvas, deleting the second-player-related sketches, and adding a breakable blocks mechanic. To speed-up our process, we can sketch a temporary "region" and say \textit{"I pack the region with blocks"} to fill it with sketches e.g. a custom "block" sketch. To make the blocks destructible and increase the point count, create a rule \textit{"When balls collide with blocks balls destroy blocks and then the score increases."} We are done. In a few steps we have revised the scene into a different game.

\subsubsection{Day/Night Cycle: Transforming Objects and Periodic Collision Logic}
The example in \autoref{fig:example:villager_ghost_transformation} demonstrates the ability to command objects to transform into other saved objects automatically, which not only changes objects' representation, but also enables automatic behavior and state changes. Further, the user can easily adjust periodic rates by using objects as a directly-manipulable, visual timing mechanism. By moving the walls in the figure closer or farther, we can opt to adjust the rate of transformation without the need for a precise timing value.

\subsubsection{"Dog plays fetch revisited" as a playable gameplay mechanic}

The dog interaction is re-imagined as an interactive game mechanic. The example demonstrates adjectives being used for logical control (\autoref{fig:example:dog_fetch_as_game_mechanic}). The user hits a button to trigger the dog's fetch ability: collect a treat for points, but only if the treat is in the proximity of the boy's collider. This example demonstrates a mix of logical behavior-programming using adjectives, as well as the beginnings of quick game-prototyping iteration.

\input{implementation}

\section{Open-Ended User Study}
\label{sec:open_ended_user_study}

We conducted an exploratory discussion-focused study to gauge understandability of interaction, discover use cases and directions, and learn users' perceptions. We believe this form of verbal feedback is valuable because it can extend our own understanding of the tool with specific anecdotal and experiential feedback. The goal was to give the participant enough exposure to the interface to arrive at working artifacts from the process and elicit meaningful discussion.

To that end, the study was an open-ended exploration between the researcher and the participant. To mitigate the potential for biasing participants towards certain answers, the researcher did \textbf{not} reveal the design goals or indicate particular use cases for the tool. Participants were informed only that the tool was a prototype with animation features controlled by sketching and speaking.
We let participants draw their own conclusions as to the usefulness, use cases, and potential for the interactions.

We chose this approach because we were interested in qualitative, early stage feedback on the \textit{concept} and \textit{interactions} behind DrawTalking, rather than feedback on our implementation, user performance, or based on comparisons. There is no known baseline for comparison with DrawTalking's interactions, and at this stage, we felt it would not contribute towards achieving the goals of our research to compare the DrawTalking prototype with production tools in terms of usability (i.e. the "usability trap" as in Olsen's work \cite{olsen_evaluating_2007}). 
We wanted user feedback on potential use cases based on discussions rather than specializing on a specific use case.

\subsection{Participants and Procedure}

We invited 9 participants (students, professors, artists, game design) chosen by responses to an online recruitment form. We wanted candidates with relatively-high confidence in speaking and sketching, with interest in the topic, and with a fair mix of professional/academic backgrounds. For participant details, see \autoref{appendix:exploratory_story_additional_content}.

Each study lasted 1 hour and 15 minutes, and each participant was compensated by 30 USD. For each session, the researcher sat alongside the participant in front of DrawTalking running on an iPad, and taught the drawing features. Next, the participant was told to draw 5 objects of their choice, then taught both object labeling methods. Then, the session was improvised using the objects to explore all language features in rough increasing complexity. The researcher could help or suggest ideas, but participants mainly guided their own exploration using a provided language features list (the Info Sheet supplemental material), and each experience was different and ended with different interactive scenes. Participants were allowed to think-aloud and comment. After, the participant was asked to reflect on the experience. The researcher ended by showing additional pre-recorded videos and soliciting final feedback.
We analyzed the results by taking note of common key words (e.g. fluidity) and of unique anecdotes, suggestions, and comments from the session and post-discussion. We recorded each session and took screen-shots for post-viewing (For samples, see \autoref{sec:study_select_screens}). In the following section, we report the discussion-oriented results.

\section{Results and Discussion}
All participants understood and learned the mechanics, and produced thoughtful discussion. They tried their own ideas for how to make things or test their own understanding out of curiosity. \textit{All became comfortable with the core mechanics (controls, touch+deixis, commands, etc.) in 10-15 minutes. Some were excited to continue playing with the tool}. They identified use cases including educational applications, rapid video game and design prototyping, paper prototyping, user interface design, presentation, language learning, and visual-oriented programming.

We explore the qualitative experiences of participants using DrawTalking. The following come from the open-ended discussions and roughly covers topics related to the design goals.

\subsection{As programming-like capability}
DrawTalking's style of control shows promise for empowering both programming and non-programming audiences (D4).

Several participants with programming knowledge and experience teaching programming compared DrawTalking functionality to constructs like variables, loops, and conditionals, indicating the patterns were familiar (D4). Some were reminded of Scratch \cite{maloney_scratch_2010} if it did not have the explicit programming interface.

In contrast, P7 (an experienced digital artist without programming experience) appreciated DrawTalking as an accessible tool for non-programmers that could reduce frustration encountered with existing programming tools: \textit{"Me instructing a game engine -- someone like me who's not a programmer and who is intimidated by doing C\# or ... using [a] visual scripting language like blueprints -- this is a really clean interface that I think can achieve you can get 90- or 80\% the way there. It just makes the user experience cleaner than having to use all these knobs and buttons or things like that or using scripting language or having to actually write code. You [the researcher] \do not force me to work with the scripting language.
"}.
P7 here favors the \textit{simplicity} and \textit{accessibility} of language control as a complement to more powerful programming interfaces.

The takeaway of interest is that \textbf{not including an explicit coding interface was appealing for its simplicity and accessibility.} Nevertheless, it would be interesting to explore balancing a greater level of complexity (in terms of interface) to serve more expert users, without compromising the simplicity for non-programmers. P7, for example, is aware of more powerful tools, but prefers something simpler. DrawTalking, in concept, seems to serve the role as abstraction layer between language- and code-based control. We could think about different modes of information-hiding depending on the task and audience.

\subsection{Semantics-visuals correspondence is valued}
Participants valued the visual correspondence between visual elements and language embedded in DrawTalking's sketches and components as a strength in the design. (P5: \textit{"Mixing visual elements with language helps me understand it. More of that [in future explorations] could be good... drawing connections between [things] visually as I'm speaking."})

For the semantics diagram, participants understood it as a helpful means to introspect and correct the system's understanding of user input. P2: \textit{"[The diagram] is pretty important. Especially if you're getting to used to the program, you might say something wrong and it interprets it in a different way and you can correct it."} P8: \textit{"This is a really cool debug thing. I immediately understand what happens when I read it."} Additionally, the interface \textit{simplifies} the input, making it easier to read: P7: \textit{"[It's] showing what it's going to do. It's taking the verbs and cutting all the other stuff out."}
The feedback suggests that \textbf{the diagram is valuable for instilling confidence in the system}, and that the interface achieved D3.

For the transcript view, we observed users (e.g. P1, P5) habitually clearing the transcript using "discard" on their own. We asked why and some participants reported wanting to keep the interface clean. We did not foresee this, but this user behavior serves a dual-purpose: the user is aware of what they say, which means they are perhaps likelier to catch errors, and they also help by keeping the system input from growing too long. We can take this as a positive: the system helps the user, and the user helps the system. We can also take this as room for improvement: how to keep the transcript view easier to clean without user intervention?

\subsection{Physicality, Prototyping, Playfulness}
A reported strength of DrawTalking was the combination of physicality with logic and animation: P5: \textit{"I can execute different rules of things that are happening. The scene is playing out here AND physical thing is happening."} 
P2, a games student, likened DrawTalking to a form of digital, semi-automatic \textit{paper-prototyping}: \textit{"When we make games there's something called a paper prototype where we make a bunch of pieces of the players and the objects. We just move it around by hand to kind of simulate what it would be. [DrawTalking] is kind of like that but on steroids a bit. So it's very nice to be able to kind of have those tools to help you with it without needing to manually make it."} Relatedly, P1 said, \textit{"It’s like language and Legos put-together."}

This suggests that \textbf{the design decision to give users interactive control via spatial direct manipulation in tandem with automation was successful}; it enabled a form of playful exploration. This supports something akin to what professional designers and kids use to tinker with creative ideas. Note that P2 values "help with making," but doesn't ask for full automation. To them, interactivity was important for the playful prototyping feel. 

\subsection{Creative problem-solving}
P1 was able to discover creative use of rule functionality by their own curiosity. In their case, they drew a boy, house, and tree. By creating two co-dependent rules, they wanted to see if they could recreate infinite loop functionality without using "over and over" or "forever." Their solution was \textit{"When boys collide with trees, boys move to houses"} and \textit{"When boys collide with houses, boys move to trees."}. This resulted in an infinite loop once the boy sketch collided with either the tree or the house. The example demonstrates imaginative composition of language-controllable primitives (\autoref{fig:study_p1}).

A highlight in P2's session involved working-around a then-limitation of one of the verbs, "climb." The scene contained a tree and squirrel, and we wanted the squirrel only to climb the tree if it were not already on-top.  However, our simple version of "climb" was (at the time) programmed simply to move an object to the base of a target and then move it to the top, so the squirrel would inadvertently climb down then up. To work around this issue, a solution was devised in which an invisible collision box placed at the top of the tree programmed with \textit{"When squirrels collide with the collider, squirrels stop climbing."}. Then, upon \textit{"The squirrel climbs the tree"}, the discovery was made that the collider could also be used as an in-situ stopping mechanism the user could simply drag onto the squirrel at any time to stop it (\autoref{fig:study_p2}).

We believe this means DrawTalking's interactions \textbf{successfully enable creativity, discoverability, and playful problem-\\solving, independent of the limitations of the featureset.}

\subsection{Re-applicability to other toolsets}
Users saw potential for DrawTalking as generic functionality for integration with other applications in a creative pipeline, e.g. as a playful ideation phase from which to import/export assets or runnable code; attaching speech control to production tools to support faster workflows. P1: \textit{"You could sit here and have a conversation and build up just using language your interactions and then you send that out for [exporting] code."} P3: (comparing with an Adobe AfterEffects workflow) \textit{"here, it just takes one second for me to 'say it' -- [this could be a] built-in function for any tool/interactive AI"}. P3 felt the interactions were fluid. P8 emphasized language control: \textit{"just incorporate the language and control interface here. We can easily create animations with the fancy stuff they have."} 

In other words, \textbf{participants perceive DrawTalking as an interaction that is independent from a specific application.} The suggestions for outputting code or plugging-into other applications with more advanced functionality hint at an interest in DrawTalking as a general control mechanism. Participants expressed excitement over how the ideas might be re-applied.

\subsection{Observations and Feedback on Current Limitations of Language Understanding}
Participants occasionally encountered some of the limitations of the semantic parsing described in \autoref{sec:limitation_semantic_parsing}. For instance, P2 initially said, \textit{"When the squirrel collides with the collider stop."} The prototype cannot currently infer that the rule should stop the squirrel. The subject (squirrel) needs to be re-specified as the target object to stop. The correct form would be: \textit{"When the squirrel collides with the collider the squirrel stops."}. Similarly, P1 tried, \textit{"The boy moves to the house and then to the tree."} The prototype does not infer the second subject and verb, so the input must be: \textit{"The boy moves to the house and then the boy moves to the tree."} Other user errors included using imperative form without a subject (P5 \textit{"Make a star"}) (as imperatives currently require an explicit object), or expecting alternative behavior or synonyms (e.g. "touch" as a substitute for "collide with" or a variant of "follow" that moves to a target only once, not continuously). Most of the unsupported cases suggest the following categories for improvement in future prototypes: reducing the need for user-specification, even in ambiguous or less-grammatically-correct cases; inferring implicit meaning from context; making the vocabulary and tenses more flexible. These are not tied to the DrawTalking concept, but could improve the experience of a concrete implementation.

As for the study experience, participants needed to become acquainted with the supported language structures. This was expected in the natural process of learning a new interface. In spite of the limitations, participants did not noticeably express frustration; they were excited. We suspect this is due to the following main factors:

\textit{Interaction design}: invalid commands do not cause surprising or destructive side-effects (nothing happens) and redoing commands is fast. Pressing the discard button and speaking again does not incur too much of a time cost before the participant can see results. The world isn't interrupted; it continues running and is interactive even while staging a command. In short, the quick and fluid error recovery might have helped reduce wait-time.

\textit{Participants were able to focus on the potential of the interactions and design}: they were informed of the fact that they were being introduced to a prototype and they formulated responses with this in-mind. This is evidenced in the fact that each participant provided insight on potential improvements, integrations, and use cases assuming a more robust system. P1 and P5 especially, as seasoned computer scientists and educators, focused on the potential to integrate the workflow, feel, and functionality into other software.

This increases our confidence that participants focused their feedback on the concept rather than the implementation details.

\subsection{Naming by speech preferred to text linking}
Labeling by speech (deixis) was unanimously preferred to linking and was used most often or exclusively.  (See \autoref{sec:sketching_and_labeling} to recap labeling functionality.) It was considered more intuitive, direct, and similar to how people talk. On the other hand, pen+touch linking could be useful when freer speech (without explicitly naming objects) is preferred, e.g. for flexibility while giving talks or to reference previous discussion without repeating words.  Labeling by linking directly to a word instead of by speaking could allow for freer use of language, as the desired labels just need to be present in the narration. P5 (an interactive computer graphics professor) gave such as example: \textit{"I often talk about arrows in the context of vectors in my teaching"}. Here, no deixis is used. Instead, the user would perform a pen+touch operation between objects and the words "arrows" and "vectors" to do the labeling. The speaker can perform this labeling in the background without needing to use a specific grammar.

However, if the user does prefer the speech approach to the linking approach, this means it might be possible to reduce reliance on the transcript view, which could open-up opportunities to hide the transcript for contexts when it might be inconvenient. e.g. for eyes-free interfaces or XR spatial systems.

\subsection{User control of physical elements and robots}
Participant P9 was a roboticist who could give specialized feedback. When asked about possibly using DrawTalking for controlling robots, P9 said that for real-world environments, they might want to have even \textbf{more control and self-specification of commands} due to safety concerns, in spite of the loss of automation:

P9: For potential tangible \textit{interfaces, physical things like robots might need more specification (for safety? accuracy?). [I] would be more tolerant of extra specification to make sure it's correct. Definitely don't want surprises. People won't want it to be fully magical. [They want] more control.}

\textbf{This suggests that our emphasis on user control is likely important}  for reapplication of DrawTalking concepts in  use cases involving physical objects. Additional safety considerations emerge and there are trade-offs between automation and control.

\section{Future Work}
\label{sec:future_work}
In sum, users described DrawTalking as fluid and flexible; a natural language way of achieving programming-like functionality; a rapid prototyping environment; an independent general interaction technique; capable of integration with other applications; physical, tangible, spatial; accessible to kids; a new approach to working-out visual problems.

We believe DrawTalking works, then, because it successfully captures some of the playful and creative attributes of programming, spatial manipulation and sketching and language, owing to our initial goals and to our designs.

Based on the primary contribution of our work (the interaction mechanism enabling sketching+speaking control) concrete implementations could use the concept to drive deeper levels of functionality and programmability than what is capable with our application prototype. 
The overarching idea, above all, is about exploring how to provide controllable feedback and facilitate a kind of \textit{vernacular} style of programming, to extend our creative capabilities.

There are many  directions:
improving the system design and vocabulary; further-exploring visual-linguistic mappings; integrating with other applications; longer-term studies applying our approach to domain-specific creative or educational workflows; multi-user collaborations; application to spatial experiences for interaction with the real-world. We describe several future possibilities:

\textit{Integration with other applications and extensibility.} A DrawTalking-like sketching application could be a part of a prototyping/open-ended phase within a creative process. Future work can explore how to take the interaction with  named rough sketches and convert them for use or replacement in later-stage production phases, in which goals have become more specific. The names and even the history of users' direct manipulation and narration could be used to inform such a conversion process by providing context. We can also explore how to convert back and forth between different representations of objects and behaviors to support moving seamlessly between ideation and production phases. For example, we suspect that our strategy of naming objects independent of their representation could be useful as a way of generating scenes from the sketch representation.

Additionally, as P9 (\autoref{sec:open_ended_user_study}) suggests, DrawTalking could either provide animations to other tools, or else control external tools' capabilities (not just limited animations) via plugin or remote API. This way, we treat DrawTalking as a controller rather than as a standalone application. (This also helps avoid vendor lock-in.) Vocabulary and functionality are potentially unlimited.

\textit{Deeper Exploration of relationships and mappings between semantics and visuals.} There are several unexplored concepts relevant to semantics-visual mapping. For example, the use of sketches to represent abstract concepts could be evaluated as additional means for controlling content, in addition to the concrete sketches and language we've used. The semantics diagram, further, could be extended to support greater levels of editability. Another future area might involve context-sensitivity: What if the user's culture and assumptions should produce different behaviors and visual-semantics mappings? How might DrawTalking evolve when considering not only English, but also other languages, modes of communication, and cultural contexts?

\textit{Supporting language input that is more natural.} Future work could work towards achieving increasingly natural and free-flowing speech input for interactive interfaces by implementing a more robust and feature-complete language processing and multi-modal input back-ends. We suspect that a translation layer from natural, more fragmented speech could convert input into simpler, structured, and deterministic instructions similar to what our interface uses. This direction could allow existing interfaces with deterministic instructions like ours to accept more expressive, less restricted language input, independent of their functionality. This could lead to reduced user-specification and a better approximation fully "natural" language interfaces.

\textit{Making trade-offs: reasons why future interface designers might not want to use or rely on generative-artificial-intelligence models (e.g. LLMs), and why we did not use them.}
To achieve fully natural input, a suggestion might be to use generative AI models such as large language models (LLMs) to perform the aforementioned proposed translation of language input. These models could also generate visual content alongside user-drawn sketches. (A number of study participants made this suggestion.)

However, as per our design goals for this interface (\autoref{sec:designGoals}), the system \textit{must} be interactive and instant, and the user must be in-control of the content's representation. We considered LLMs, but they contradicted our design goals: reducing fluidity, user control over content, and system transparency and understandability. Interface designs with similar criteria might also encounter the same challenges and opt to make trade-offs.

LLMs are currently too slow to provide instant feedback, even if we use LLMs only to transform natural input into structured commands as we suggest. (This also raises additional research questions: How well can LLMs rewrite users' speech and reflect intent?) In contrast, the NLP technology we chose responds nearly instantly.

Also, LLMs often exhibit unpredictable or unreproducible behavior, whereas the chosen NLP returns the same results for the same language input. We wanted determinism so command outcomes would be easy to reason about, especially to keep the scope of possibility manageable for the user study. Not using LLMs allowed for feasibility and stable development. 

Still, generative-AI is promising given the right context and requirements. Future work could see how LLMs might be complementary to other methods (traditional NLP, procedural generation, and others). We believe a good direction for research would be to find the balance between the many approaches and solutions we have developed over time.

\textit{Spontaneous authoring of new primitives without programming.} Exploring interactions for richer behavior-authoring interactions compatible with our approach would be another direction.
We can imagine a number of possibilities:
exploring extensions to the existing semantics diagram interface for lightweight editing;
using DrawTalking's controls to drive other applications with their own unique primitives; combining procedural methods with the capabilities of generative models by finding the right balance;
crowd-sourcing programmed-functionality through shared libraries of domain-specific functionality;
introducing multiple layers of programming capability in the same interface for different audiences (e.g. artists, programmers), trading-off simplicity for granular control as is common in many applications. 

\textit{Possibilities for generalizability and scalability.} We envision that combinations of DrawTalking-like functionality and different implementations and domain-specific applications could lead to a more-scalable ecosystem of interfaces. We also believe that by decoupling natural language processing implementations from the interface (rather than building directly around specific NLP implementations or probabilistic results) it might be easier to support reproducible research within interactive interface design. This is in-part why we chose to compile NLP information into intermediary, generic command structures. These could feasibly be used in future research as a compilation target by any other NLP or alternative approach without relying on a specific implementation. For example, this is where a gen-AI model could be explored without changing the rest of the system.

Lastly, although we've focused on rough sketching, conceptually there is no limitation preventing sketches from being replaced with other forms of controllable digital or tangible objects. Sketches can represent collections of variables and properties mapped to some objects, whatever forms they might take.

We are excited to see ideas taken in new directions for playful and creative human-machine interfaces beyond our initial scope. We think that there is an opportunity to seek-out new human-computer interactions that draw from our natural behavior for inspiration.

\section{Conclusion}
We have introduced DrawTalking, an approach to building and controlling interactive worlds by sketching and speaking while telling stories, and have shown its potential. Our interface was inspired by our natural interplay between sketching and language, and our ability to communicate via make-believe. There are many possible directions, and we are excited to see future research build on our approach and uncover other human-centered approaches to extending natural human abilities. We consider this project just one possible step and we hope that it will foster fruitful discussion and research.

\begin{acks}

We would like to thank the anonymous reviewers (of this version and an earlier CHI LBW version of this work \cite{DrawTalkingShortRosenberg}) for their constructive comments;
Professor Lakshminarayanan Subramanian for his feedback; Devamardeep Hayatpur for additional feedback on writing and visuals for the CHI LBW; Daniel Zhang for brainstorming and discussions;
the user study participants for their time and insights;
and the many others who participated in conversations and gave insights during the journey of this work.

\end{acks}

\bibliographystyle{ACM-Reference-Format}
\bibliography{main.bib}

\appendix

\section{Formative Steps Additional Content}
\label{appendix:formative_steps_additional}

\begin{figure}[h!]
\centering
\includegraphics[width=\linewidth]{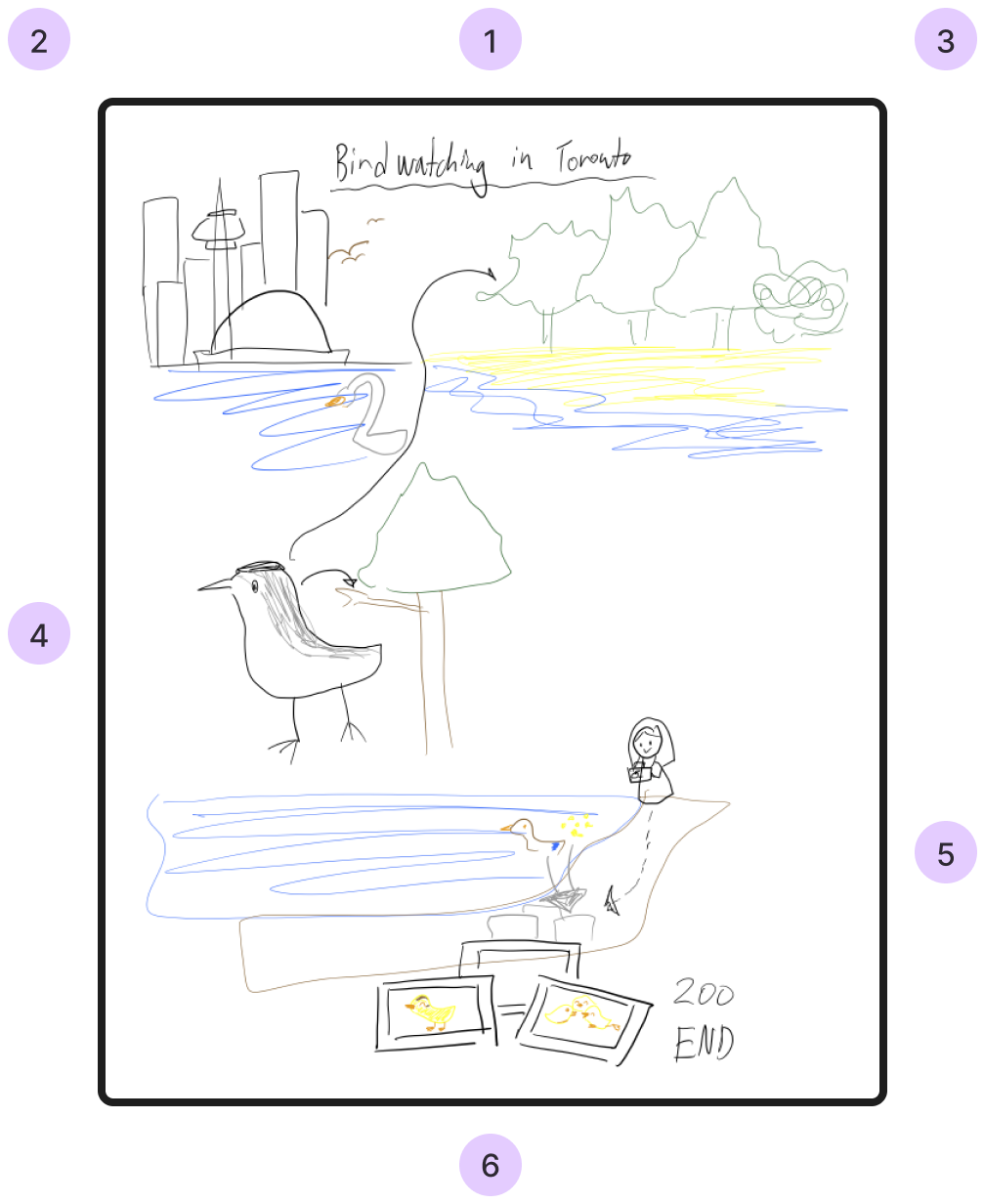}
\caption[formativeSketchingSessionP1]{\textbf{Formative Sketching Session, P1 final image}: P1 narrated their own bird-watching story while sketching. Their iterative workflow and visual style of moving between "islands of locations" helped inspire some features: e.g. discrete objects, object hierarchies, and spatially-flexible object relationships. Drawn roughly in-order: 1) Inline text titling the "story"; 2) drawing of the city Toronto to set the location; 3) P1 visited the forest in Toronto; 4) P1 saw a bird, drew arrows indicating object relation (on a tree, in the forest); 5) P1 (the person sketch) found a pond and watched ducklings; 6) P1 took 200 photos.
\Description[Formative Sketching Session, P1's final image result]{Formative Sketching Session, P1's final image result: P1 narrated their own bird-watching story while sketching. Their iterative workflow and visual style of moving between "islands of locations" helped inspire some features: e.g. discrete objects, object hierarchies, and spatially-flexible object relationships. Drawn roughly in-order: 1) Inline text titling the "story"; 2) drawing of the city Toronto to set the location; 3) P1 visited the forest in Toronto; 4) P1 saw a bird, drew arrows indicating object relation (on a tree, in the forest); 5) P1 (the person sketch) found a pond and watched ducklings; 6) P1 took 200 photos.}
}
\label{fig:formative-style-p1}
\end{figure}

\autoref{fig:appendix:formative-content} shows a sample of motivating content. \autoref{fig:formative-style-p1} shows P1's result from the formative exercises.

\aptLtoX[graphic=no,type=html]{
\begin{figure*}[h!]
\centering
\includegraphics[width=\textwidth]{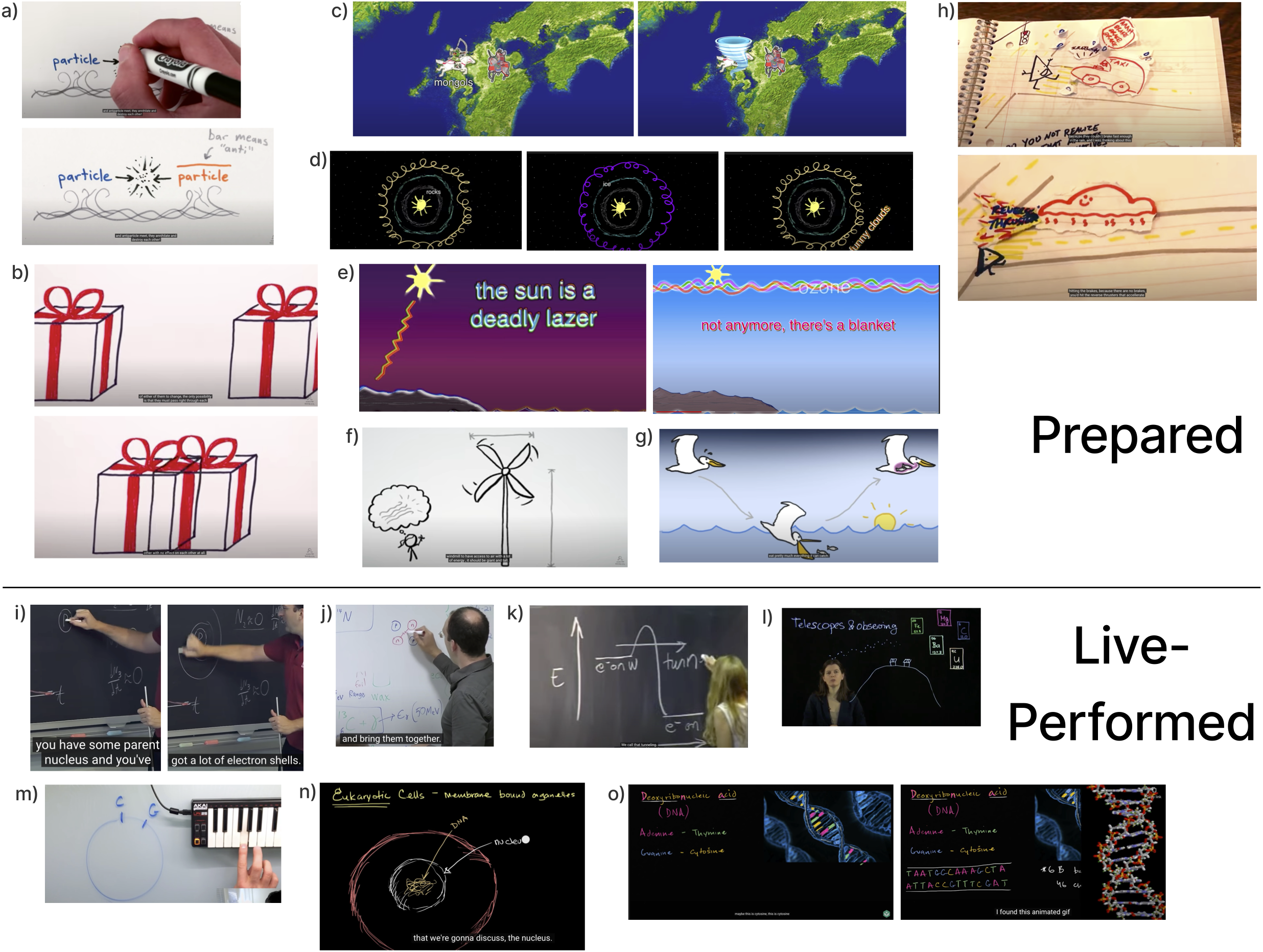}
    \caption[formativeContentExamples]{\textbf{ Formative Content Exploration:} (top) subset of prepared and (bottom) live-performed sketch-based content} \textbf{Attribution and Disclaimer:} \textit{All rights belong to the original creators of the referenced video content. The screen-shots are for the sole purpose of providing examples of content used as partial inspiration for our research.}
    \begin{description}
    \item[a] \textcopyright Neptune Studios — minutephysics — Antimatter Explained[39]
  \item[b] \textcopyright Neptune Studios — minutephysics — Immovable Object vs. Unstoppable Force - Which Wins?38
    \item[c] \textcopyright Bill Wurtz — history of japan[5]
    \item[d, e] \textcopyright Bill Wurtz — history of the entire world, i guess[6]
    \item[f] \textcopyright Neptune Studios — minutephysics — The Physics of Windmill Design[40]
    \item[g] \textcopyright Neptune Studios — MinuteEarth — Why These Bears “Waste” Food[37]
    \item[h] \textcopyright Vi Hart — The Case for Hovercars[67]
    \item[i] \textcopyright MIT OpenCourseWare — MIT 22.01 Introduction to Nuclear Engineering and Ionizing Radiation, Fall 2016 — 11. Radioactivity and Series Radioactive Decays[43]
    \item[j] \textcopyright MIT OpenCourseWare — MIT 22.01 Introduction to Nuclear Engineering and Ionizing Radiation, Fall 2016 — 1. Radiation History to the Present — Understanding the Discovery of the Neutron[43]
    \item[k] \textcopyright MIT OpenCourseWare — Atomic Theory of Matter — Lec 1 | MIT 5.111 Principles of Chemical Science, Fall 2005[41]
    \item[l] \textcopyright MIT OpenCourseWare — MIT RES.8-007 Cosmic Origin of the Chemical Elements, Fall 2019 — Ep. 6: Element Production (Fusion) -- Part 1[44]
    \item[m] \textcopyright Michael New — The Circle of Fifths - How to Actually Use It[35]
    \item[n] \textcopyright Khan Academy — Organelles in eukaryotic cells | The cellular basis of life | High school biology | Khan Academy[21]
    \item[o] \textcopyright Khan Academy — DNA | Biomolecules | MCAT | Khan Academy[20]
    \end{description}
    \label{fig:appendix:formative-content}
    \Description[Prepared and Live-Performed Example Motivating Content]{Prepared and Live-Performed Example Motivating Content: These examples are screen-shots from videos (attribution in the main figure text). Two panels of example content. Panel "Prepared": a) live-sketched animation demonstrating particle-antiparticle collision and annihilation; b) animation of sketches of gift boxes with red ribbons colliding into each other; c) world map of japan with cartoon images of the mongols (represented as a horseback soldier and labeled with the text "mongols") advances towards a figure of a Japanese solider, only to be swept-up in a tornado cartoon image; d)  sequence of top-down  sketch art of the solar system with the labels "rocks," "ice," and "funny clouds." beside the objects; e) 2D art, sideview, showing the early Earth with the sun shooting a laser at the Earth to represent how dangerous it was before the atmosphere developed. The sentence "the sun is a deadly lazer" is written above. Another frame from the same video shows the ozone layer now developed (labeled with the text "ozone") between the sun and the Earth. Text is shown: "not anymore, there's a blanket." (referring to the ozone layer); f) hand-drawn sketch art of a person beside a spinning windmill; g) a sequence of gulls (side-view) diving into the ocean to catch fish. All at once the steps are shown in an arrow flow diagram: gull before diving, arrow from to, gull in water getting fish, arrow from to, gull above water with fish in gull's stomach; h) a triangular character drawn with markers on physical paper (resting above a physical notebook) encounters a car and (in a second panel) a hovercar with a smiling face - all physical paper; i) two panels of a lecturer narrating over drawing an atomic nucleus with electron shells, "you have some parent nucleus and you've" (the nucleus is drawn on the blackboard) "got a lot of electron shells" (the electron shells have been drawn around the nucleus); j) lecturer showing a formation of particles labeled n and p (two each) being brought together while speaking, "and bring them together"; k) lecturer black-board drawing with chalk while speaking - the drawing is of a function on a graph; l) a lecturer presenting while facing the audience of the video through a transparent whiteboard with text for "Telescopes & observing" and sketches of periodic table  elements (rectangular entries) and a hill; m) music-theory-related drawing of circle of fifths on a whiteboard, with the presenter playing the notes on a physical MIDI piano keyboard on-screen; n) digital blackboard lesson,  sketch of a cell. text above: "Eukaryotic Cells - membrane bound organelles. The parts of the cell have text labels with arrows pointing to them. (e.g. DNA, nucleus). The narrator says he's about to discuss the nucleus; o) two panels of a digital blackboard lesson on DNA structures. Screen 1 shows a sketch of 3 DNA helices. The middle one has the bases (adenine thymine, guanine cytosine) highlighted. On the left is the text: Deoxyribonucleic Acid (DNA). Below is a color key for the base pairs. Adenine (written in red) - Thymine (written in green), below, Guanine (written in blue) - Cytosine (written in yellow). The second panel shows a pre-rendered animated GIF image of a DNA structure. (The narrator says "I found this animated gif" to introduce it. On the left, an example of DNA mappings has been drawn in the form of 2 rows of A T G C all contained in a rectangle. The columns of the rows form T,A, A,T, A,T, T,A, G,C, G,C C,G, A,T, A,T, A,T, G,C, C,G, T,A, A,T.}
\end{figure*}}{
\begin{figure*}[h!]
\centering
\includegraphics[width=\textwidth]{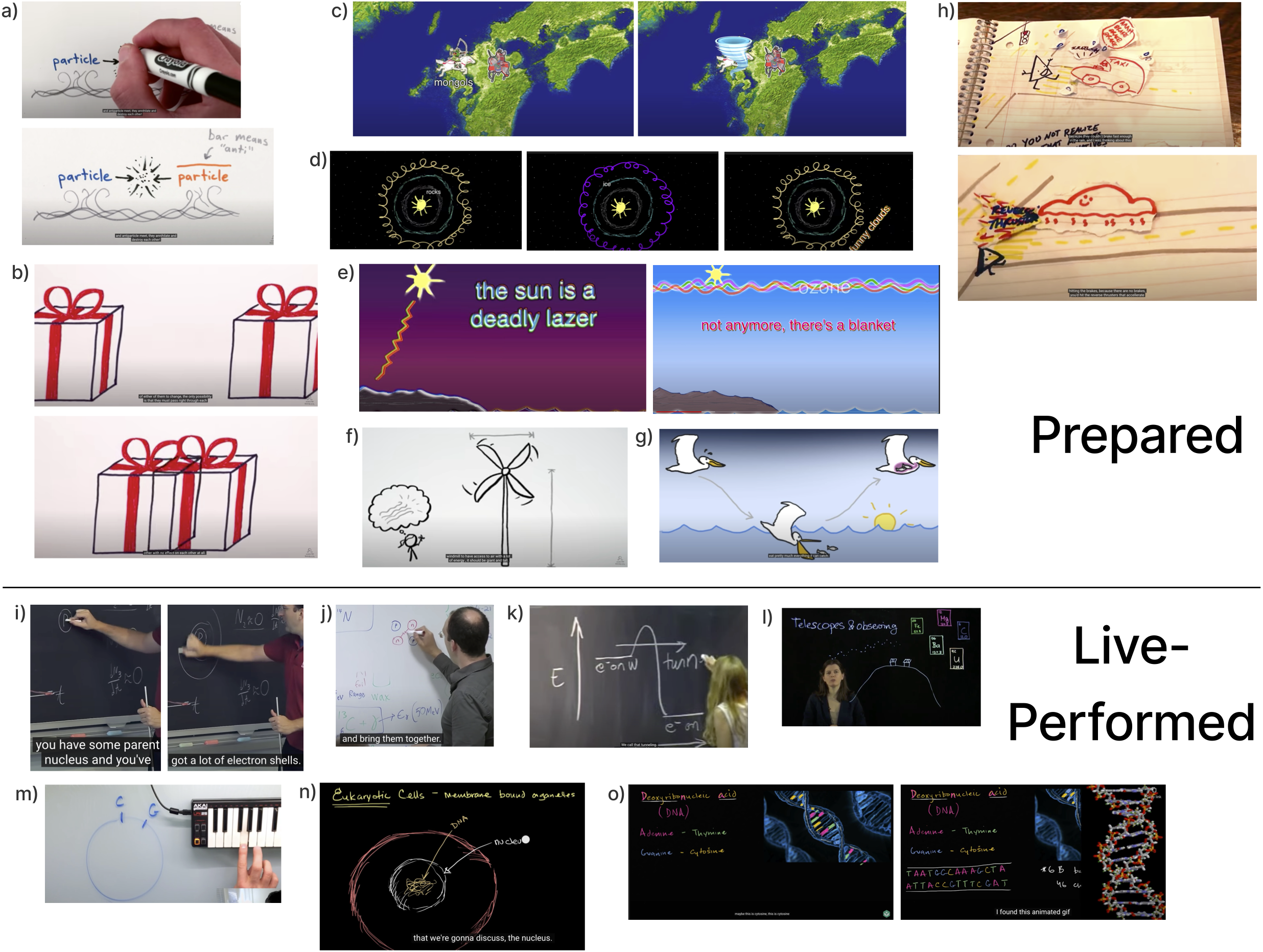}
    \caption[formativeContentExamples]{\textbf{ Formative Content Exploration:} (top) subset of prepared and (bottom) live-performed sketch-based content} \textbf{Attribution and Disclaimer:} \textit{All rights belong to the original creators of the referenced video content. The screen-shots are for the sole purpose of providing examples of content used as partial inspiration for our research.}
    \begin{description}
    \item[a] \textcopyright Neptune Studios — minutephysics — Antimatter Explained\cite{minutephysics_antimatter_2014}
  \item[b] \textcopyright Neptune Studios — minutephysics — Immovable Object vs. Unstoppable Force - Which Wins?\cite{minutephysics_immovable_2013}
    \item[c] \textcopyright Bill Wurtz — history of japan\cite{bill_wurtz_history_2016}
    \item[d, e] \textcopyright Bill Wurtz — history of the entire world, i guess\cite{bill_wurtz_history_2017}
    \item[f] \textcopyright Neptune Studios — minutephysics — The Physics of Windmill Design\cite{minutephysics_physics_2021}
    \item[g] \textcopyright Neptune Studios — MinuteEarth — Why These Bears “Waste” Food\cite{minuteearth_why_2021}
    \item[h] \textcopyright Vi Hart — The Case for Hovercars\cite{vihart_case_hovercars_2016}
    \item[i] \textcopyright MIT OpenCourseWare — MIT 22.01 Introduction to Nuclear Engineering and Ionizing Radiation, Fall 2016 — 11. Radioactivity and Series Radioactive Decays\cite{mit_opencourseware_11_2019_radioactivity_and_series}
    \item[j] \textcopyright MIT OpenCourseWare — MIT 22.01 Introduction to Nuclear Engineering and Ionizing Radiation, Fall 2016 — 1. Radiation History to the Present — Understanding the Discovery of the Neutron\cite{mit_opencourseware_1_2019_radiation_history}
    \item[k] \textcopyright MIT OpenCourseWare — Atomic Theory of Matter — Lec 1 | MIT 5.111 Principles of Chemical Science, Fall 2005\cite{mit_opencourseware_lec_2008_principles_of_chemical}
    \item[l] \textcopyright MIT OpenCourseWare — MIT RES.8-007 Cosmic Origin of the Chemical Elements, Fall 2019 — Ep. 6: Element Production (Fusion) -- Part 1\cite{mit_opencourseware_ep_2019_element_production_fusion}
    \item[m] \textcopyright Michael New — The Circle of Fifths - How to Actually Use It\cite{michael_new_circle_2015}
    \item[n] \textcopyright Khan Academy — Organelles in eukaryotic cells | The cellular basis of life | High school biology | Khan Academy\cite{khan_academy_organelles_2017}
    \item[o] \textcopyright Khan Academy — DNA | Biomolecules | MCAT | Khan Academy\cite{khan_academy_dna_2014}
    \end{description}
    \label{fig:appendix:formative-content}
    \Description[Prepared and Live-Performed Example Motivating Content]{Prepared and Live-Performed Example Motivating Content: These examples are screen-shots from videos (attribution in the main figure text). Two panels of example content. Panel "Prepared": a) live-sketched animation demonstrating particle-antiparticle collision and annihilation; b) animation of sketches of gift boxes with red ribbons colliding into each other; c) world map of japan with cartoon images of the mongols (represented as a horseback soldier and labeled with the text "mongols") advances towards a figure of a Japanese solider, only to be swept-up in a tornado cartoon image; d)  sequence of top-down  sketch art of the solar system with the labels "rocks," "ice," and "funny clouds." beside the objects; e) 2D art, sideview, showing the early Earth with the sun shooting a laser at the Earth to represent how dangerous it was before the atmosphere developed. The sentence "the sun is a deadly lazer" is written above. Another frame from the same video shows the ozone layer now developed (labeled with the text "ozone") between the sun and the Earth. Text is shown: "not anymore, there's a blanket." (referring to the ozone layer); f) hand-drawn sketch art of a person beside a spinning windmill; g) a sequence of gulls (side-view) diving into the ocean to catch fish. All at once the steps are shown in an arrow flow diagram: gull before diving, arrow from to, gull in water getting fish, arrow from to, gull above water with fish in gull's stomach; h) a triangular character drawn with markers on physical paper (resting above a physical notebook) encounters a car and (in a second panel) a hovercar with a smiling face - all physical paper; i) two panels of a lecturer narrating over drawing an atomic nucleus with electron shells, "you have some parent nucleus and you've" (the nucleus is drawn on the blackboard) "got a lot of electron shells" (the electron shells have been drawn around the nucleus); j) lecturer showing a formation of particles labeled n and p (two each) being brought together while speaking, "and bring them together"; k) lecturer black-board drawing with chalk while speaking - the drawing is of a function on a graph; l) a lecturer presenting while facing the audience of the video through a transparent whiteboard with text for "Telescopes & observing" and sketches of periodic table  elements (rectangular entries) and a hill; m) music-theory-related drawing of circle of fifths on a whiteboard, with the presenter playing the notes on a physical MIDI piano keyboard on-screen; n) digital blackboard lesson,  sketch of a cell. text above: "Eukaryotic Cells - membrane bound organelles. The parts of the cell have text labels with arrows pointing to them. (e.g. DNA, nucleus). The narrator says he's about to discuss the nucleus; o) two panels of a digital blackboard lesson on DNA structures. Screen 1 shows a sketch of 3 DNA helices. The middle one has the bases (adenine thymine, guanine cytosine) highlighted. On the left is the text: Deoxyribonucleic Acid (DNA). Below is a color key for the base pairs. Adenine (written in red) - Thymine (written in green), below, Guanine (written in blue) - Cytosine (written in yellow). The second panel shows a pre-rendered animated GIF image of a DNA structure. (The narrator says "I found this animated gif" to introduce it. On the left, an example of DNA mappings has been drawn in the form of 2 rows of A T G C all contained in a rectangle. The columns of the rows form T,A, A,T, A,T, T,A, G,C, G,C C,G, A,T, A,T, A,T, G,C, C,G, T,A, A,T.}
\end{figure*}}

\section{Exploratory Study with DrawTalking — Additional Content}
\label{appendix:exploratory_story_additional_content}

\subsection{Participants' Backgrounds}

The following summarizes the participants' backgrounds:
\begin{itemize}
\item[] $\textbf{P1}_{expl}$ \demographics{male}{46}: An experienced professor (at a higher-level institution) in computer science, interactive graphics, and new visual media  (e.g. AR, VR storytelling and games) --- focuses on teaching students new to programming. Uses many visual aids in classes.
\item[]$\textbf{P2}_{expl}$ \demographics{female}{21}: Digital fine arts university student (design and computer science), drawing is a hobby. Uses sketching for game design; planning gameplay features and figuring out how they should work. e.g. diagramming and framework design. Experienced with drawing on tablets.
\item[] $\textbf{P3}_{expl}$ \demographics{male}{23}: Experience as a digital designer, economics and digital art university student,
\textit{not}\\professionally-trained in art, uses a digital drawing application (Procreate) to record visual ideas.
\item[] $\textbf{P4}_{expl}$ \demographics{female}{20}: Tech / Design university student. Does design for web, desktop, apps in-general. Has done digital illustration and created drawings since childhood.
\item[] $\textbf{P5}_{expl}$ \demographics{male}{49}: An experienced professor in computer programming for interactive graphics. Does live-coding and streaming for internet-based education as well. Uses and authors open-source coding libraries for interactive graphics. Teaches programming for creating visuals rather than designing visuals by hand.
\item[] $\textbf{P6}_{expl}$ \demographics{female}{18}: Student in interactive media. Draws and paints using physical media (e.g. paintbrush and canvas), with lifelong interest and experience. Experienced with digital graphical design work. Low experience with programming.
\item[] $\textbf{P7}_{expl}$ \demographics{male}{36}: Industry creative expert. 15+ years experience as a set designer and in the theater industry. Digital illustrator, spatial experience designer, AR/VR immersive projects, projects in entertainment and fine-arts. Creative directing, collaborations with companies and academia. Non-programmer, some experience with visual blocks-based interfaces.
\item[] $\textbf{P8}_{expl}$ \demographics{female}{26}: Robotics and machine-learning researcher, works with developing ML models, often needs to visualize ML inputs/outputs and illustrates rough sketches to think-through policies (for robots). No prior experience using a tablet.
\item[] $\textbf{P9}_{expl}$ \demographics{female}{24}: Robotics and machine-learning PhD student, used tablet-based sketching interfaces.
\end{itemize}

\subsection{Select Screenshots}
\label{sec:study_select_screens}
\begin{figure}[h!]
    \centering

\begin{subfigure}[s]{0.45\textwidth}
\includegraphics[width=\textwidth]{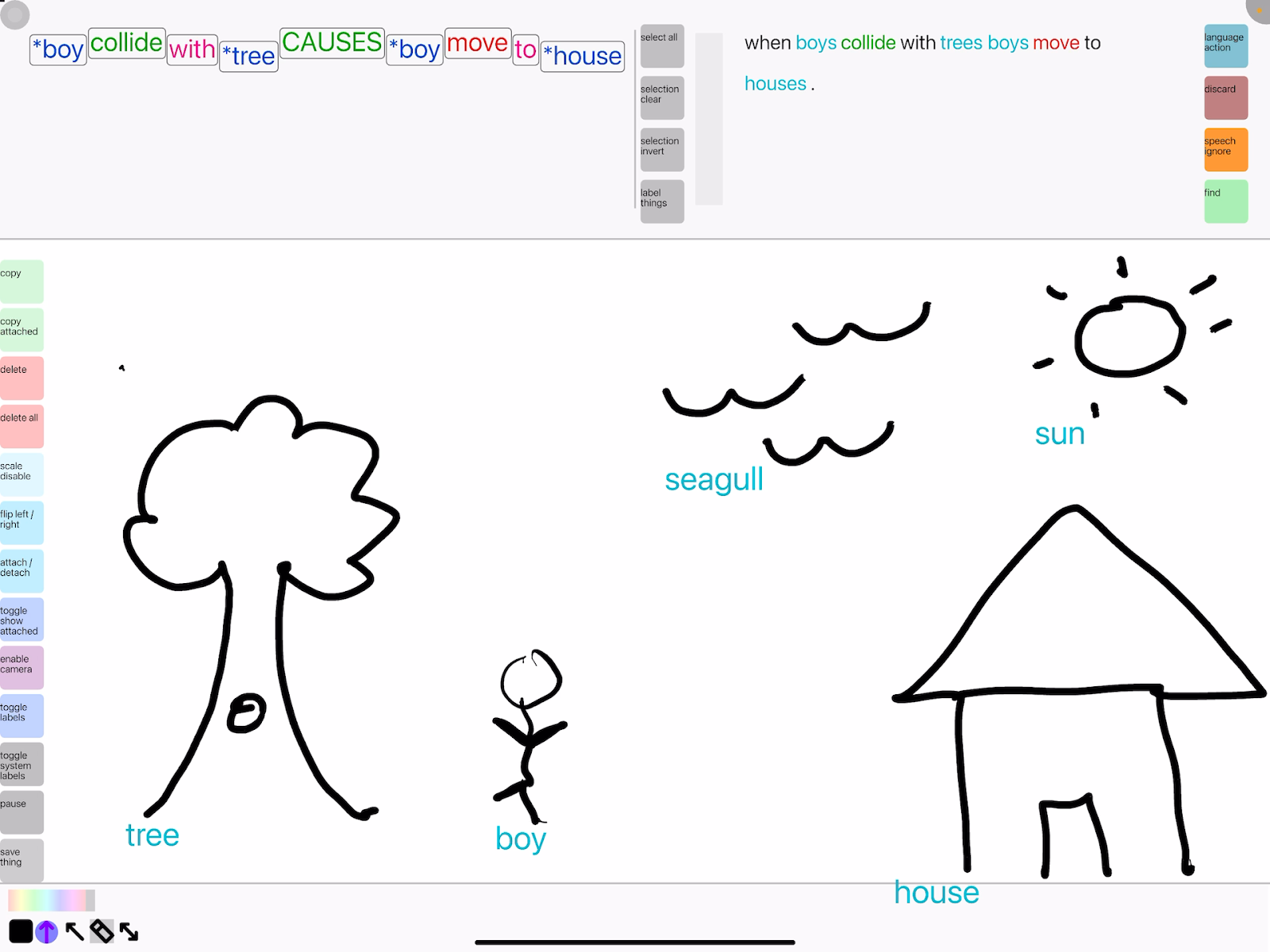}
\caption[p1ScreenA]{Define: upon collision with trees, boys move to houses.}
\Description[p1ExploreRule1]{Rule 1 is created from the user input (shown in the transcript view): "When boys collide with trees, boys move to houses."}
\end{subfigure}

\hfill

\begin{subfigure}[s]{0.45\textwidth}
\includegraphics[width=\textwidth]{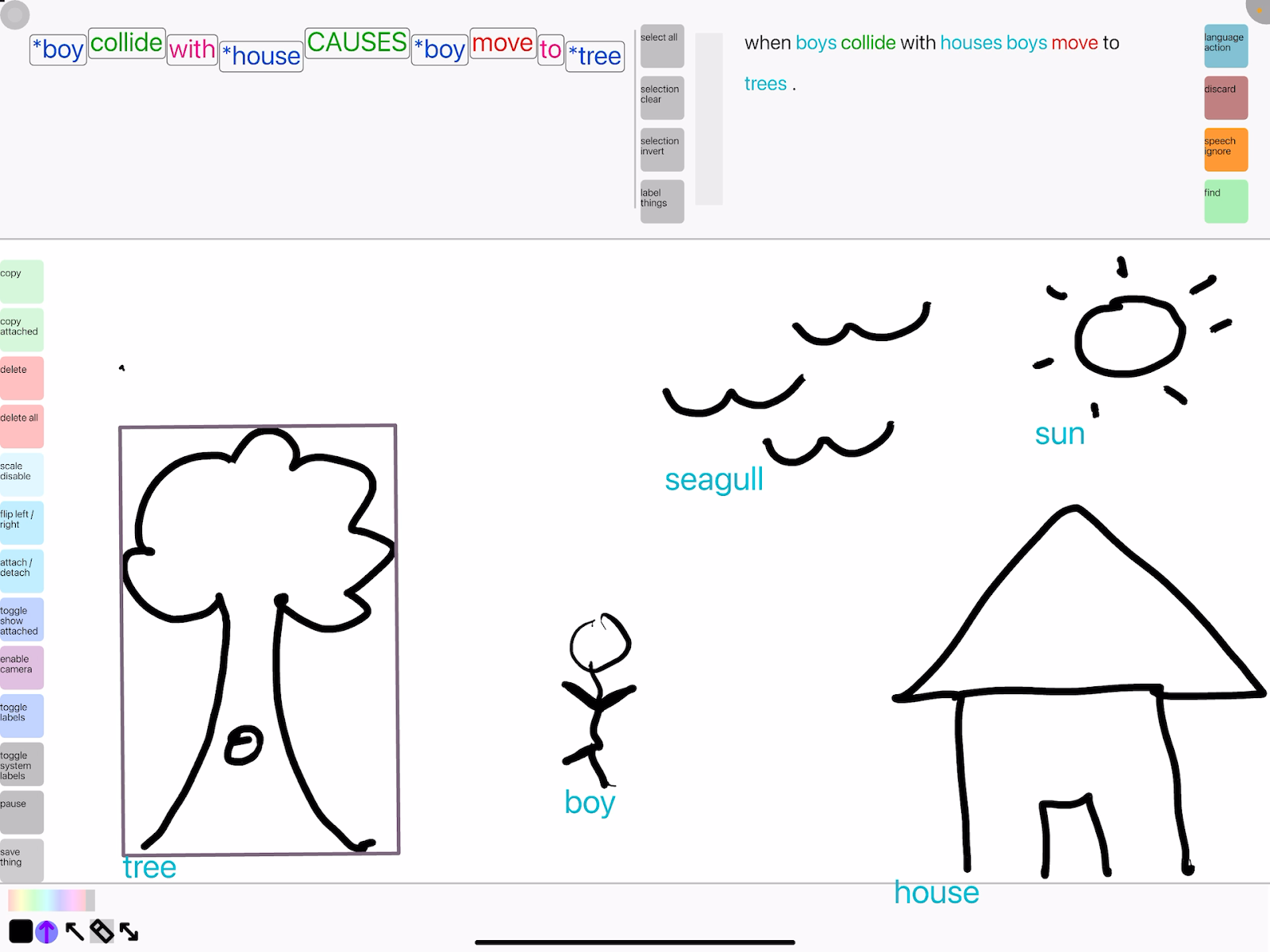}
\caption[p1ScreenB]{Define: upon collision with houses, boys move to trees.}
\Description[p1ExploreRule2]{Rule 2 is created from the user input (shown in the transcript view): "When boys collide with houses, boys move to trees."}
\end{subfigure}
   
    \caption[p1ScreenOverview]{\textit{\textbf{P1: Creatively defining a custom loop using rules}.} P1 discovered a way to make a kind of infinite loop using 2 rules to cause an infinite sequence of collision/response movements back and forth between the house and the tree.}
    \label{fig:study_p1}
    \Description[Participant P1 defines infinite loop-like behavior using rules.]{Participant P1 defines infinite loop-like behavior using rules. The scene contains a tree, seagulls, sun, boy, and house sketch.}
\end{figure}

\begin{figure}[h!]
    \centering
    \includegraphics[width=1.0\linewidth]{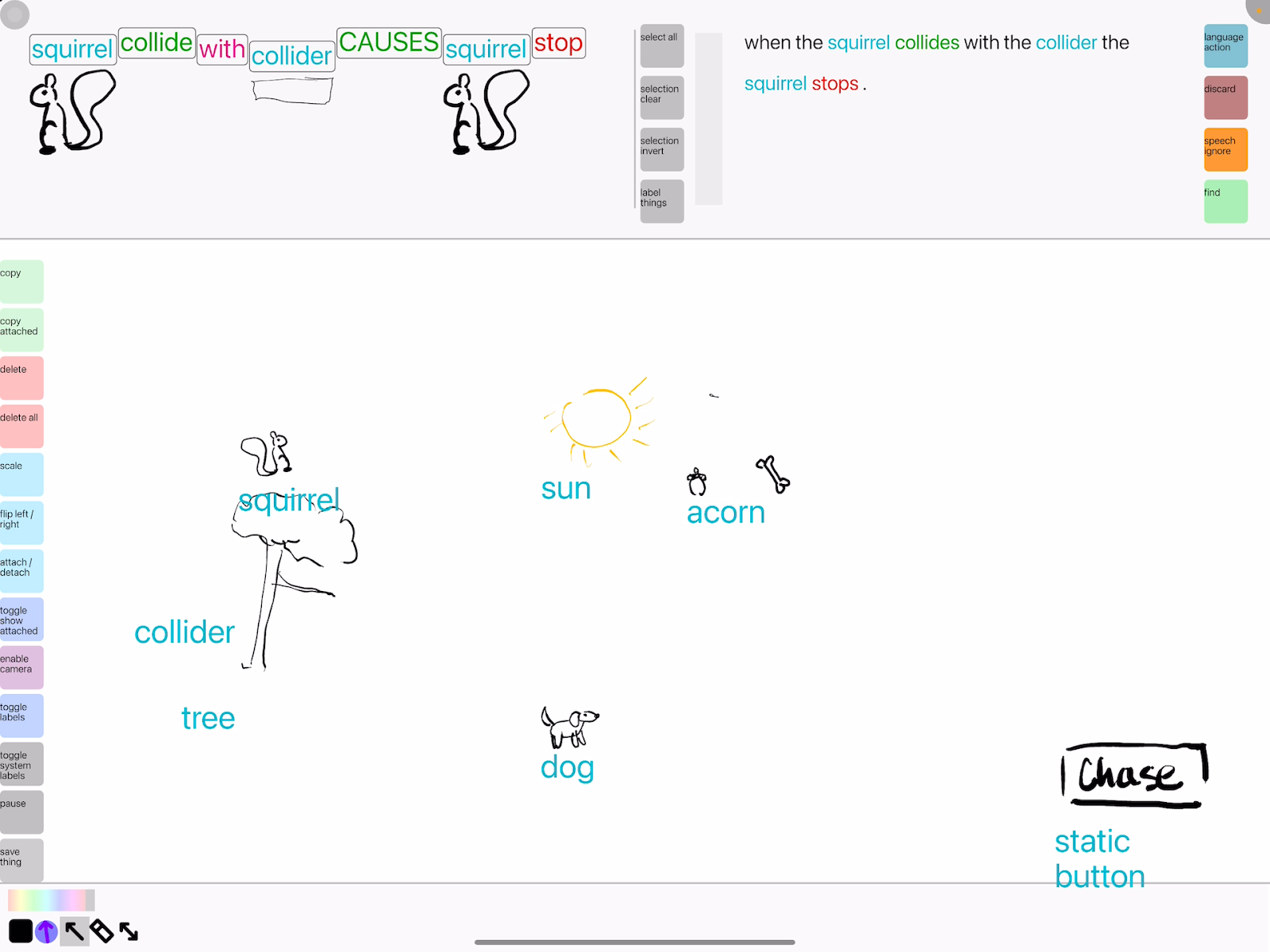}
   
    \caption[p2Screen]{\textit{\textbf{P2: Squirrel Climbing Tree Stopped by Collider}.} A rule is assigned to the collider to make it a prop the user can move around themselves to stop the squirrel from climbing upon collision with the collider. (The collider is shown in the semantics diagram as a rectangle, but invisible in the canvas on-purpose — where the label "collider" is.) P2 also created a button that causes the dog to move.
    }
    \label{fig:study_p2}
    \Description[Squirrel climbing scene]{Squirrel climbing scene: canvas: sketches for a squirrel, sun, acorn, a button with the word "chase" on it, a dog, a tree, and an invisible collider. Transcript view user input: "when the squirrel collides with the collider the squirrel stops."}
\end{figure}

\begin{figure}[h!]
    \centering
    \includegraphics[width=1.0\linewidth]{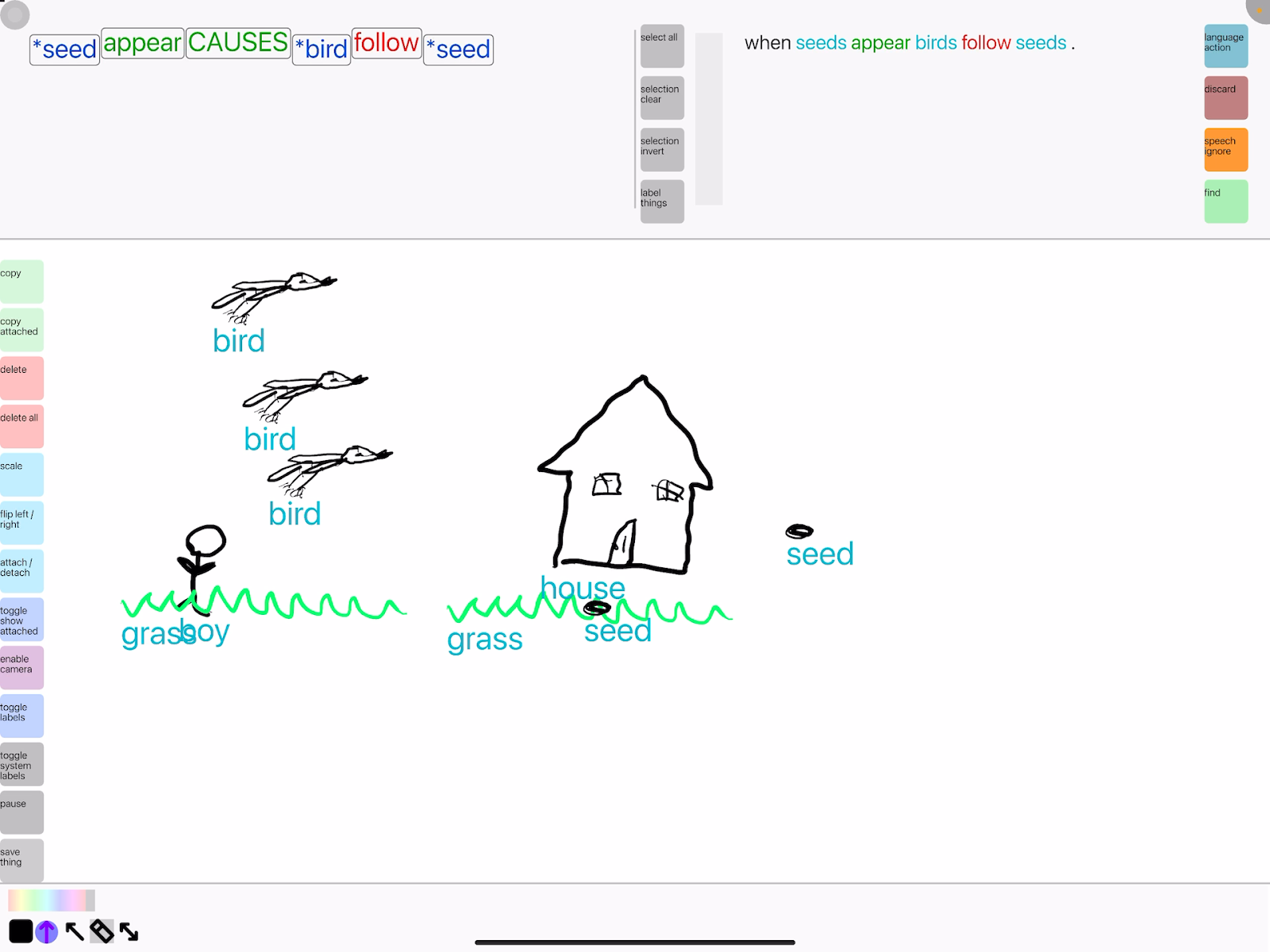}
   
    \caption[p3Screen]{\textit{\textbf{P3: "When seeds appear birds follow seeds."}}}
    \label{fig:study_p3}
    \Description[A scene in which birds follow spawning seeds]{A scene in which birds follow spawning seeds: canvas: sketch of a boy, two grass lawns, two seeds (one on the rightmost grass, a house above the rightmost grass, three birds, vertically stacked above boy on left, boy is on the left grass. Transcript view input: "when seeds appear birds follow seeds."}
\end{figure}

\begin{figure}[h!]
    \centering
    \includegraphics[width=1.0\linewidth]{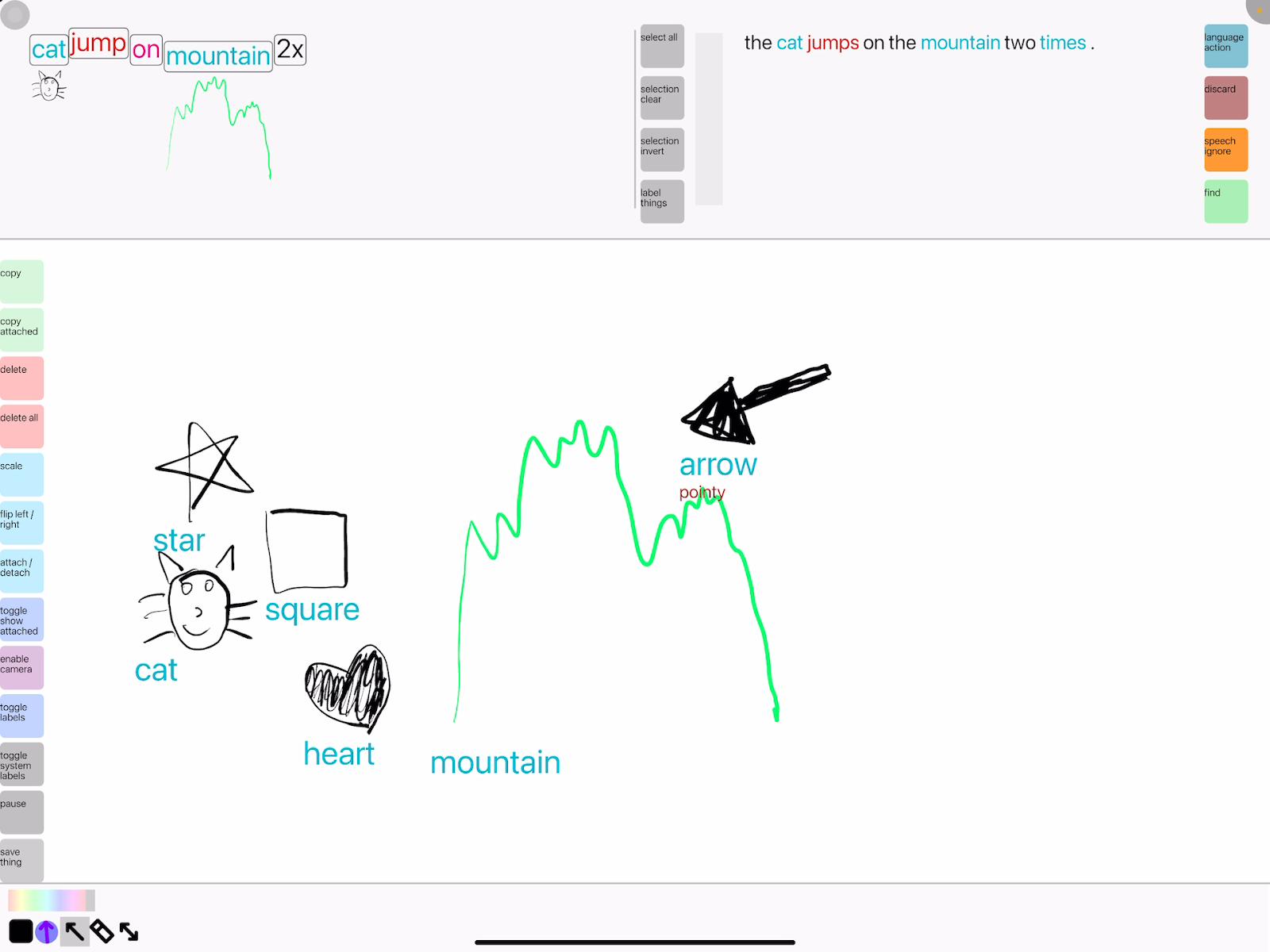}
   
    \caption[p5Screen]{\textit{\textbf{P5: "The cat jumps on the mountain two times."}}}
    \label{fig:study_p5}
    \Description[This scene shows a cat jumping on a mountain.]{This scene shows a cat jumping on a mountain. The canvas consists of sketches for a cat, star, square, heart, mountain, arrow (labeled with adjective "pointy") (roughly left to right). Transcript view input: "The cat jumps on the mountain two times."}
\end{figure}

\begin{figure}[h!]
    \centering
    \includegraphics[width=1.0\linewidth]{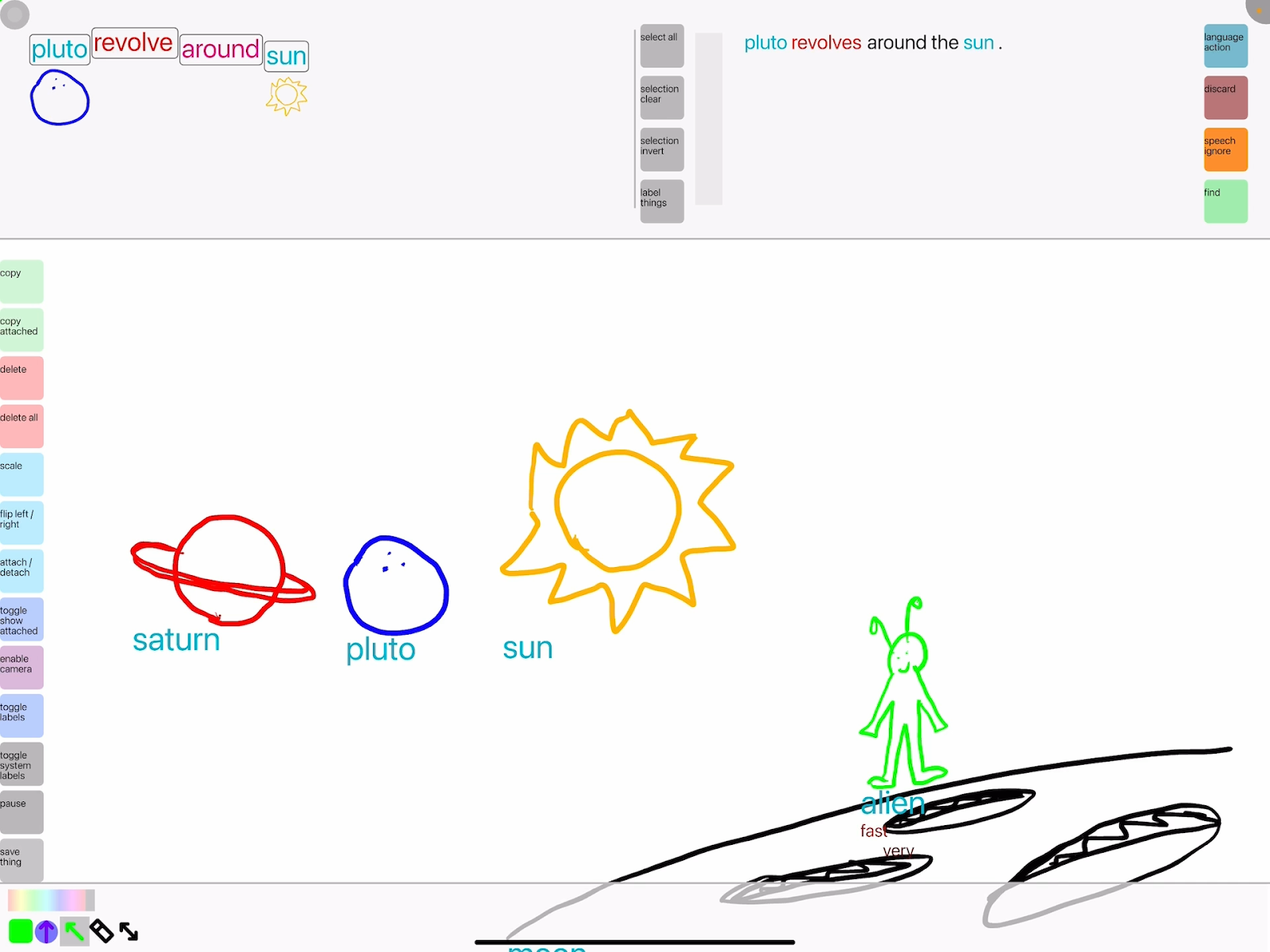}
   
    \caption[p6Screen]{\textit{\textbf{P6: "Pluto revolves around the sun."}}}
    \label{fig:study_p6}
    \Description[A revolving planet]{A revolving planet: sketches: Saturn, Pluto, Sun, alien (very fast), moon - alien stands on the moon on right. Transcript view input: "pluto revolves around the sun."}
\end{figure}

\begin{figure}[h!]
    \centering
    \includegraphics[width=1.0\linewidth]{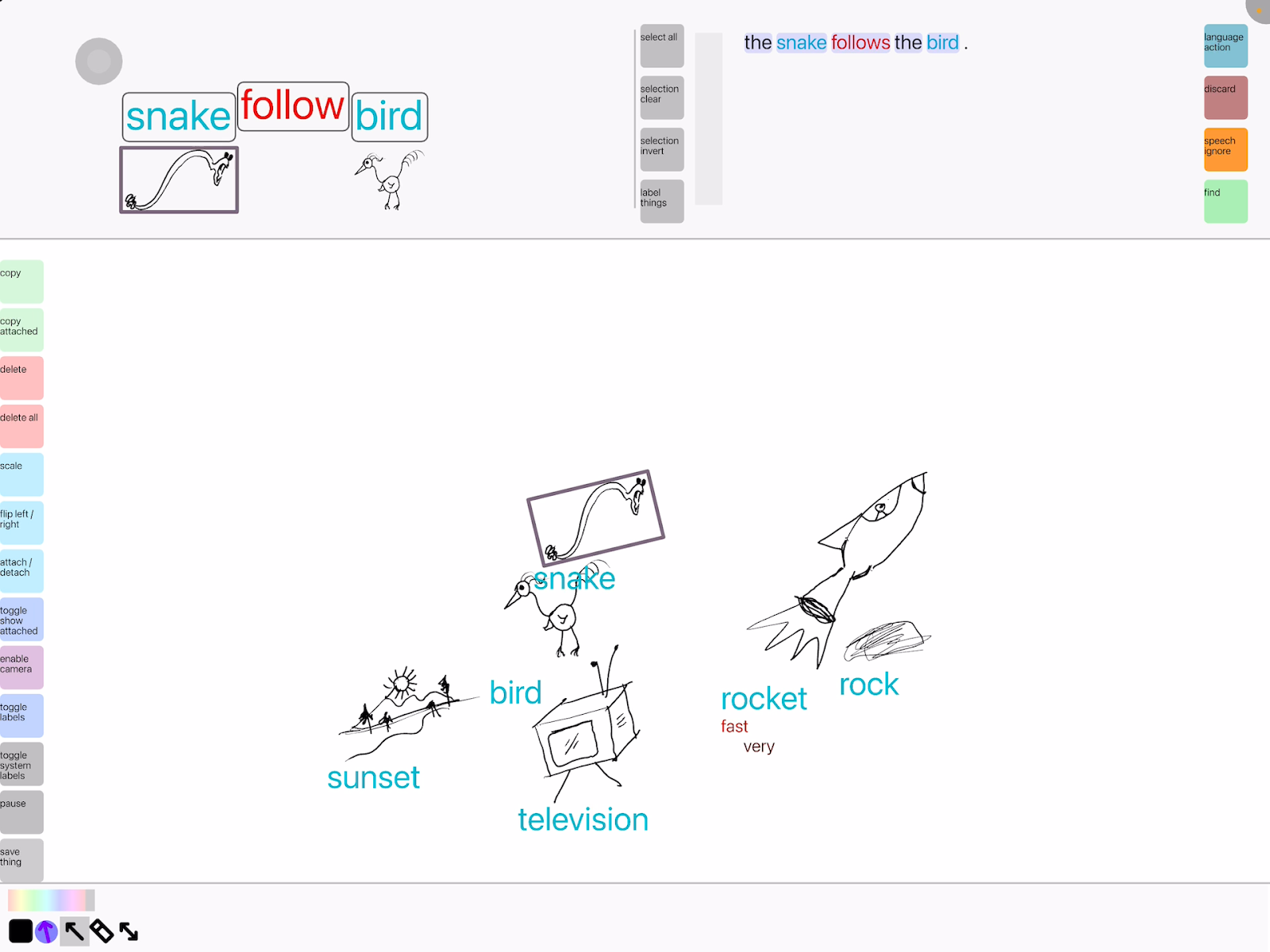}
   
    \caption[p7Screen]{\textit{\textbf{P7: "The snake follows the bird."}}}
    \label{fig:study_p7}
    \Description[A snake follows a bird]{A snake follows a bird. Cartoon characters and props: sketches: a sunset backdrop, on left of vertically-stacked television, bird, snake, rocket (very fast) on right, a rock on right. Transcript view (input): "the snake follows the bird." The snake is selected by touch.}
\end{figure}

\begin{figure}[h!]
    \centering
    \includegraphics[width=1.0\linewidth]{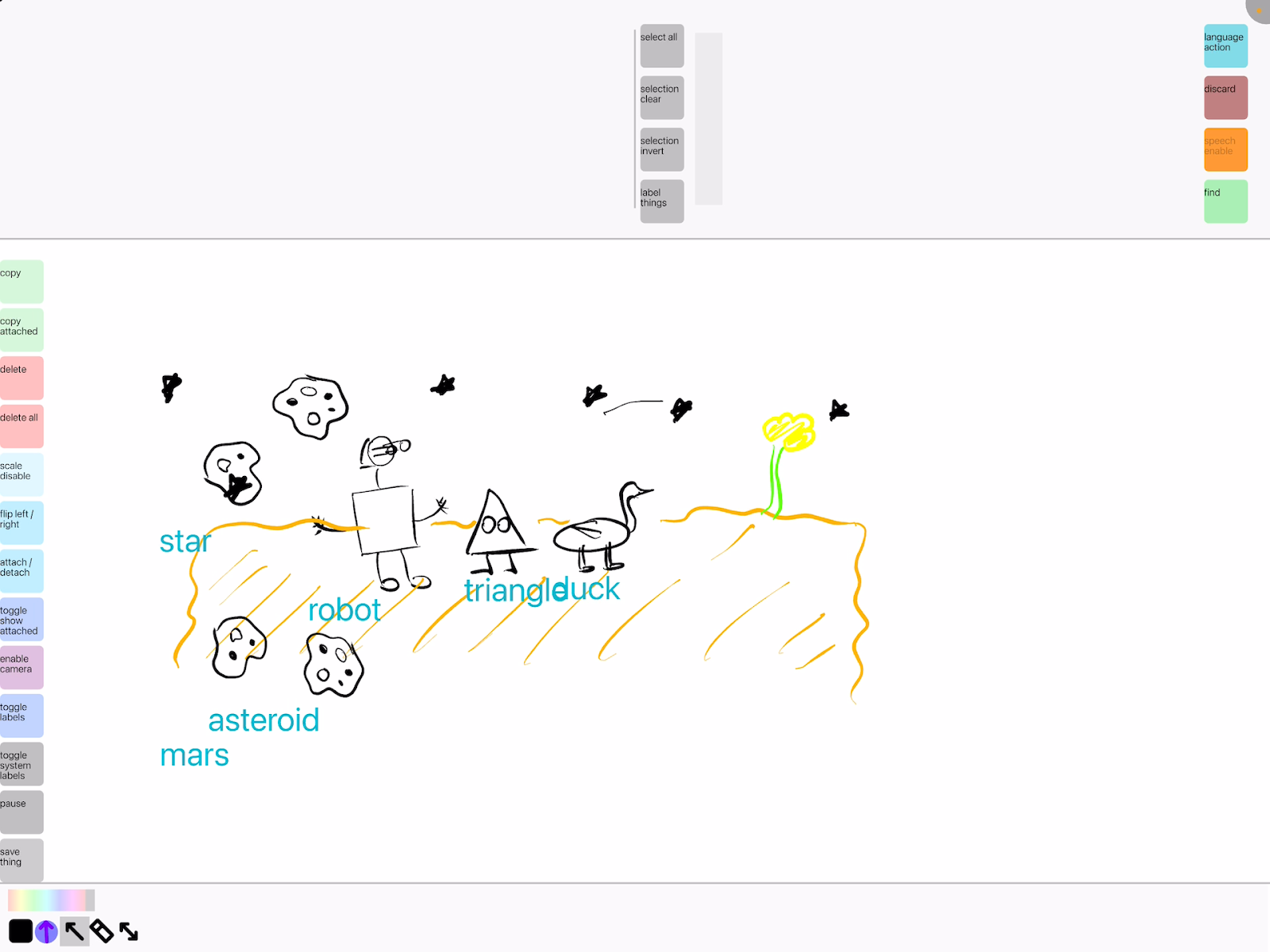}
   
    \caption[p8Screen]{\textit{\textbf{P8: Space scene with robots, ducks, asteroids, and stars near Mars}.}}
    \label{fig:study_p8}
    \Description[Outer-Space scene on the planet Mars]{Outer-Space scene on the planet Mars: sketches for the surface of Mars, stars (several stars in a single sketch), asteroids, a robot, a triangle character, a duck, and an unnamed flower.}
\end{figure}

\begin{figure}[h!]
    \centering
    \includegraphics[width=1.0\linewidth]{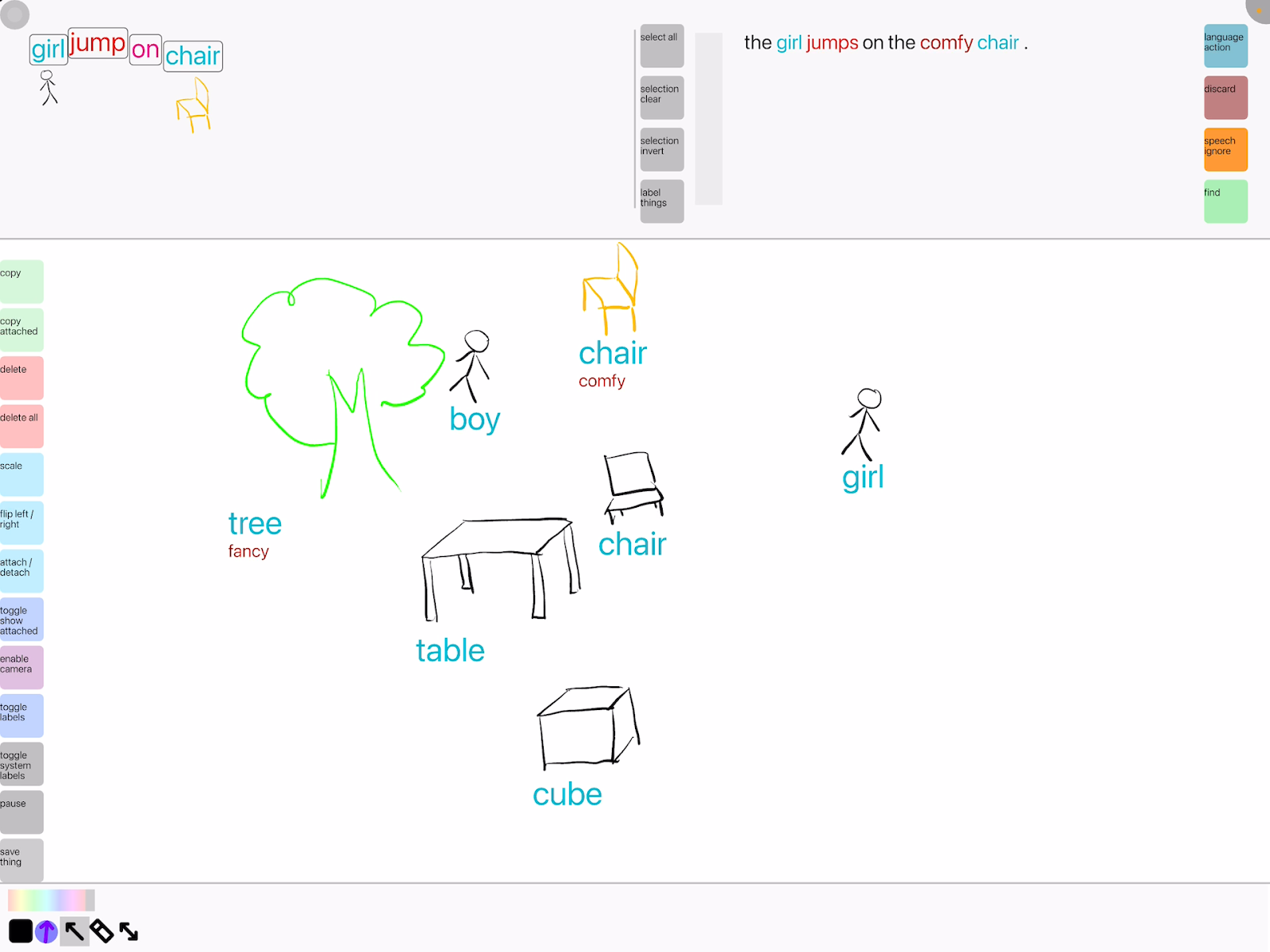}
   
    \caption[p9Screen]{\textit{\textbf{P9: "The girl jumps on the comfy chair."}}}
    \label{fig:study_p9}
    \Description[A scene with characters and props at-home/out-doors]{A scene with characters and props at-home/out-doors: tree, boy, chair, table, cube, chair (comfy), girl; Transcript view (input): "The girl jumps on the comfy chair." The semantics diagram shows the system correctly selecting the comfy chair sketch instead of the chair that wasn't called comfy.}
\end{figure}

A selection of examples from the user sessions are shown:
P1 \autoref{fig:study_p1},
P2 \autoref{fig:study_p2},
P3 \autoref{fig:study_p3},
P4 \autoref{fig:interface-overview},
P5 \autoref{fig:study_p5},
P6 \autoref{fig:study_p6},
P7 \autoref{fig:study_p7},
P8 \autoref{fig:study_p8},
P9 \autoref{fig:study_p9}.

\section{Exploratory Features and Examples for the Record}
During the development of this project, we explored several ideas mixing interactive visuals and text. Some of these did not contribute to the core ideas and this leg of the research was complete without them, but the examples might be useful to include for readers interested in combining more direct-manipulation techniques.

\begin{figure}
    \centering
    \includegraphics[width=1.0\linewidth]{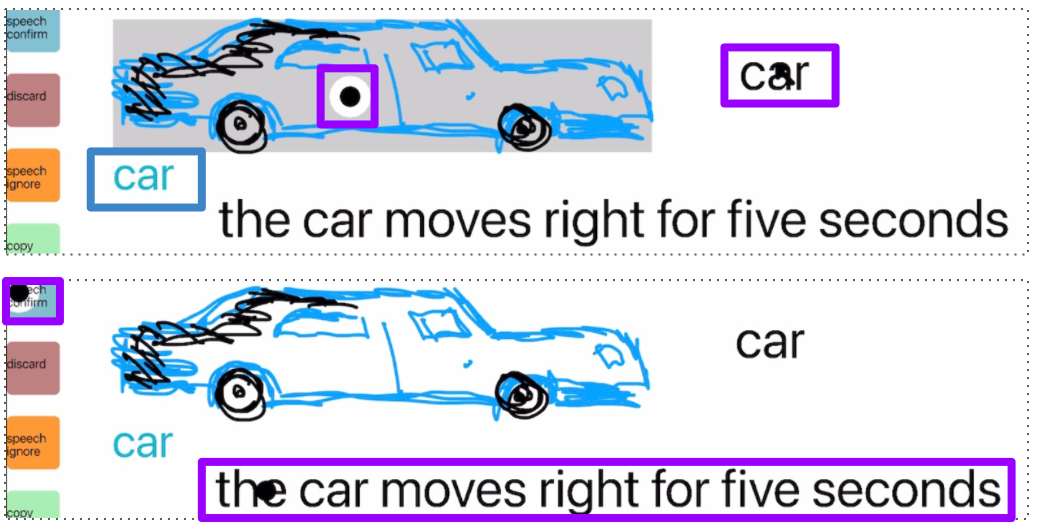}
   
    \caption[textObjectsAsMacros]{\textbf{Text objects as macros}:
    We can reduce repetitive speech by storing text as inline objects that work exactly as text in the semantics diagram (\Cref{fig:semantics-diagram}). They're created by linking the text transcript's current content and the canvas. Here, inline text objects (in purple) have been created from 1) the word "car" and 2) the sentence, "the car moves right for five seconds." The user can link a new object with 1) to label it "car" and link 2) and the language action button to treat the text as a reusable command. (Running "the car moves right..." again.) There is potential to explore this further. However, we decided to focus on the more general capability of triggering commands with rules, e.g. upon pressing a button.}
    \label{fig:UX_I_text_objects}
    \Description[This shows how a text sketch can be used to repeat the same command repeatedly on a car sketch by user interaction between the text sketch and the car sketch]{This shows how a text sketch can be used to repeat the same command repeatedly on a car sketch by user interaction between the text sketch and the car sketch; a car sketch named "car", a first text sketch with the text "car," a second text sketch with the text "the car moves right for five seconds." 2 panels. panel 1: Pen+touch between the unnamed car sketch and text sketch 1 gives the unnamed car sketch the label, "car." panel 2: Pen+touch between text sketch 2 and the language action button in the DrawTalking interface causes DrawTalking to treat the text sketch's raw "the car moves right for for five seconds" text as input for a command.}
\end{figure}

\begin{figure}
    \centering
    \includegraphics[width=1.0\linewidth]{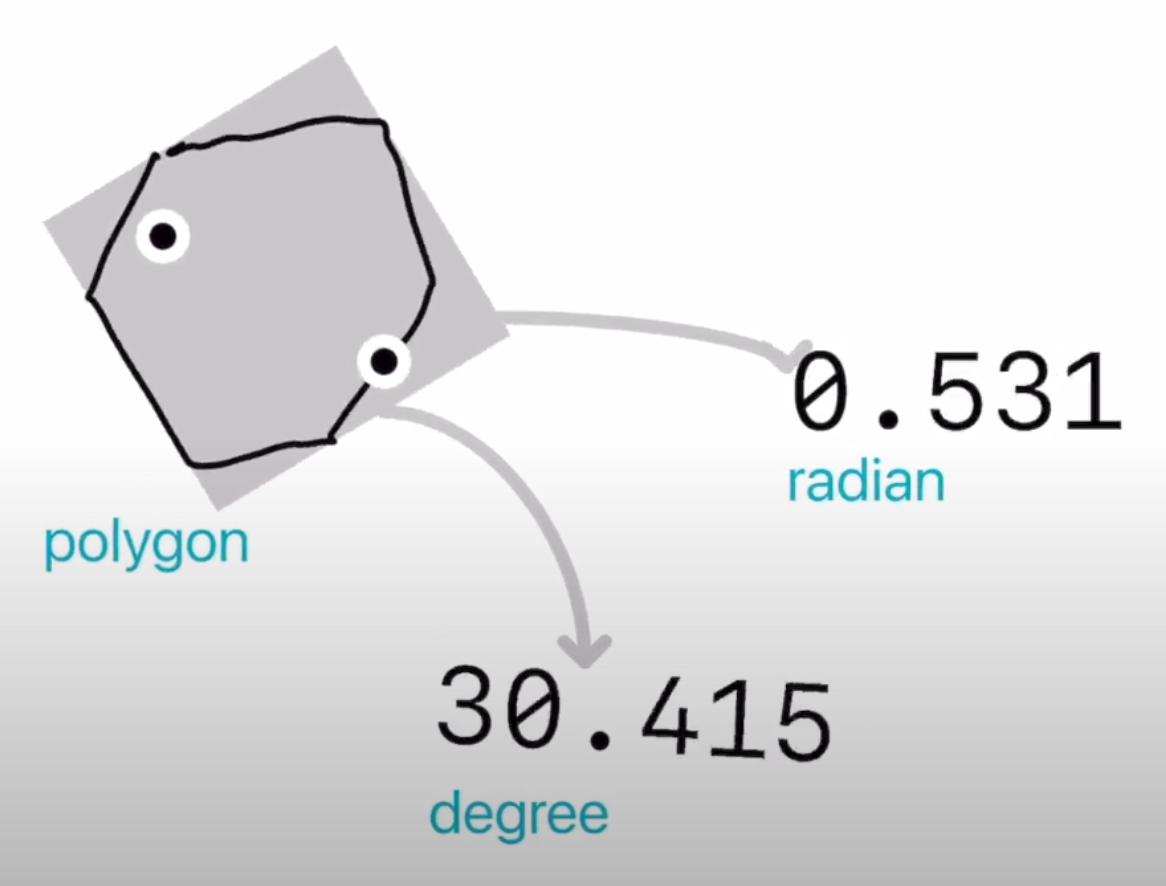}
    \caption[contextSensitiveRepresentationsExploration]{\textbf{Context-sensitive representations based on labels}: We briefly explored using the labels to control the visual representation of "number sketches." In this case, arrows connected from a freehand sketch output the sketch's angle to the numbers, and the labels determine whether to display the angle in degrees or radians. "Dynamic representations" was left to future work.}
    \label{fig:math_rep}
    \Description[A polygon sketch connected to 2 number sketches labeled for radians and degrees respectively is shown being rotated by the user, causing the number sketches to display the shape's angle. ]{A polygon sketch connected to 2 number sketches labeled for radians and degrees respectively is shown being rotated by the user, causing the number sketches to display the shape's angle. Arrows from the polygon to two number sketches labeled "degree" and "radian" indicate data flow of the numerical rotation angle of the polygon sketch to the number targets. The degree sketch shows a value of 30.415, whereas the radian sketch shows the equivalent value in radians 0.531. This shows that the number sketches interpret the same input value differently according to their names.}
\end{figure}

For example: reusable text commands in \autoref{fig:UX_I_text_objects} and context-sensitive representations of numerical outputs based on labels in \autoref{fig:math_rep}.

\begin{figure}[h!]
    \centering
    \includegraphics[width=0.9\linewidth]{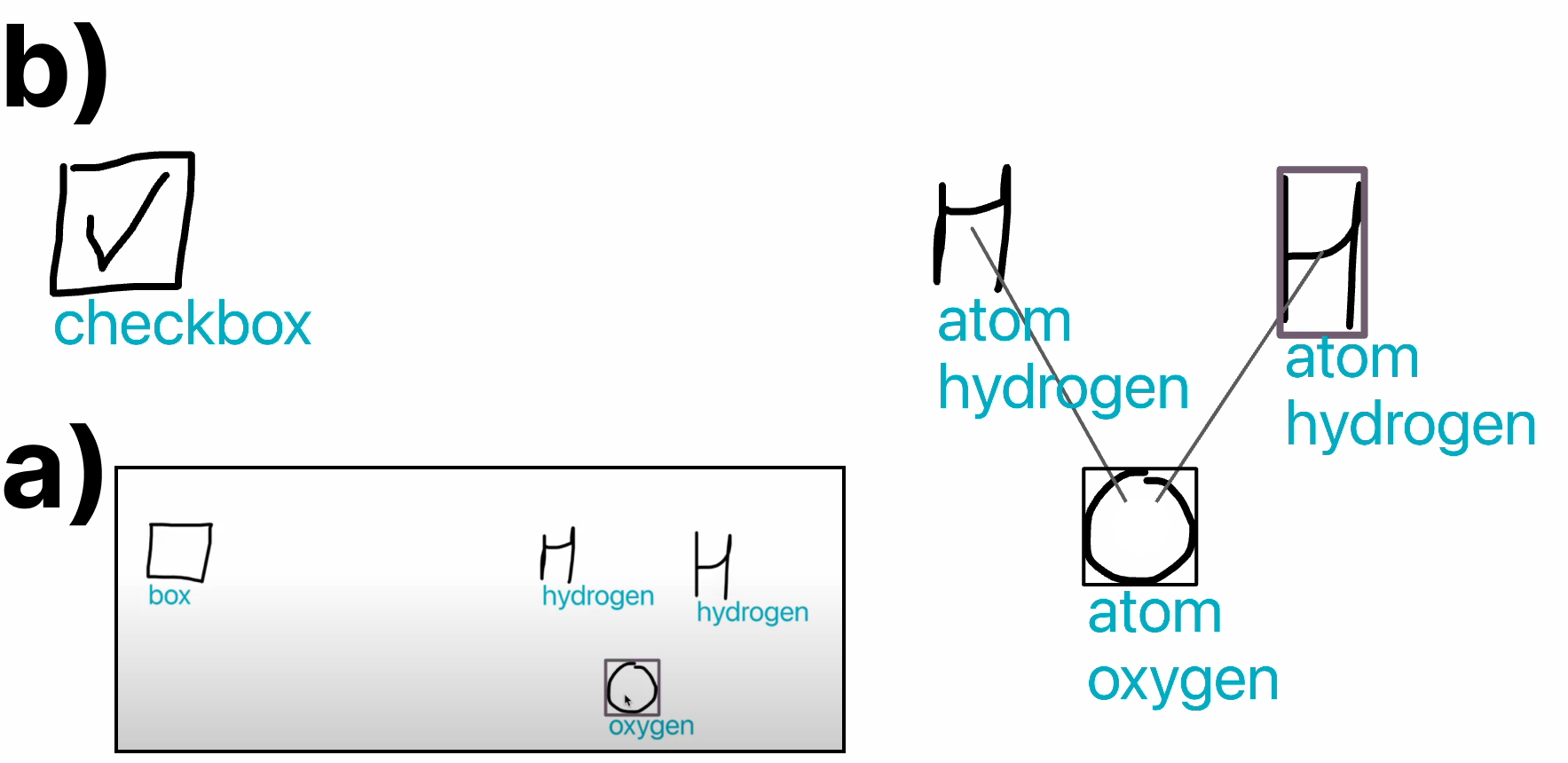}
    \caption[moleculeMatching]{\textbf{Molecule Matching Lesson prototype}: This example shows a checkbox (initially empty: a) that has automatically been filled (b) after the user has recreated the structure of a sketched water ($H_2O$) molecule formed from sketches labeled atoms, hydrogen, and oxygen.}
    \label{fig:screen:molecule_matching}
    \Description[A molecule matching game is shown consisting of atoms and a checkbox that is marked when the user correctly recreates the user-drawn target molecule (H-2-O)]{A molecule matching game is shown consisting of atoms and a checkbox that is marked when the user correctly recreates the user-drawn target molecule (H-2-O); panel a) - initial state, a box is empty (the empty checkbox) beside three sketches named hydrogen, hydrogen, oxygen. panel b) - final state, the user has named the aforementioned sketches as "atom" and connected them into one structure. The box has transformed into a checkbox because the sketch structure matches that of water (H-2-O).}
\end{figure}

We also prototyped an additional working example specializing in "digital interactive assignments" in which the user builds a molecule-matching game. See \autoref{fig:screen:molecule_matching}. An empty "box" sketch, a "checkbox" sketch, and a hierarchical 
"water" sketch have been saved prior that looks like the structure shown, with "water" labeled. (These were deleted prior.) Next, the rule was created: \textit{"When atoms \textbf{form} water, the box transforms into a checkbox."}. This creates a script that checks whether the user has created any sketch structure matching the names and hierarchy of the target. We believe the matching functionality of "form" would be useful to have available in an interactive lesson notebook or textbook.

\begin{figure}
    \centering
    \includegraphics[width=0.5\linewidth]{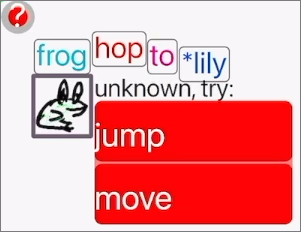}
    \caption[substitutionSuggestions]{\textbf{Substitution suggestions for unknown verbs}: This extension to the semantics diagram is an early exploration in error-tolerance that could be developed. When unknown verbs are used in a command, the semantics diagram displays a list of potentially-equivalent words from the existing set, for each unknown verb. The user can select one or opt to cancel. Once the verb is substituted, the mapping persists so as not to interrupt the user again. (Here, the command is \textit{"The frog hops to a lily"}, but "hop" is undefined.)}
    \label{fig:verb-substitution}
    \Description[This screen-shot shows an example of verb substitution suggestions that appear on the semantics diagram when the verb is unknown by the system.]{This screen-shot shows an example of verb substitution suggestions that appear on the semantics diagram when the verb is unknown by the system. The semantics diagram has been created for "frog," "hop," "to," "*lily" (meaning "The frog hops to a lily." A frog sketch is displayed in miniature under the word "frog" to indicate that the object has been selected for the command. A question mark is displayed atop the diagram on the top left to indicate that there is an error in the diagram. Beneath "hop" is written, "unknown, try:" (meaning the word "hop" is currently unknown to the system). Beneath is a tappable list of alternative verbs from which to choose, displayed as vertical red blocks. The verb suggestions are "jump" and "move."}
\end{figure}

Towards additional exploration of error-tolerance, we implemented an early version of unknown verb substitution for an extension of the semantics diagram~\autoref{fig:verb-substitution}. Unknown verbs must be substituted in the semantics diagram for the command to proceed. Once the verb is substituted, the mapping persists so as not to interrupt the user again. The idea is to interrupt the user only when necessary (i.e. there is ambiguity.) (The suggestions are found simply using WordNet \cite{miller_wordnet_1995} within NLTK\cite{loper_nltk_2002}.) Context-sensitive error recovery components like this could be explored more deeply.

\section{Abstract Semantic Structure}
\label{section:abstract_semantic_structure}

The following defines the basic structure of S2 (\autoref{lang:struct2}), which is a simplified semantic structure graph representing an interpretable command. The concrete implementation of DrawTalking traverses this structure to generate final execution commands S3 (\autoref{lang:struct3}).

\clearpage

\begin{lstlisting}[caption=S2 Data Structure]
S2_Element : $\textbf{Type\_Definition}$ {
    type := 
        CMD_LIST |
        ACTION | 
        AGENT | DIRECT_OBJECT | OBJECT | INDIRECT_OBJECT |
        PREPOSITION |
        TRIGGER_RESPONSE | TRIGGER | RESPONSE |
        SEQUENCE_SIMULTANEOUS | SEQUENCE_THEN |
        PROPERTY | PLURAL | COUNT | SPECIFIC_OR_UNSPECIFIC | TIME |
        COREFERENCE

    value :=
        Number | // e.g. any of Float64, Float32, Uint64, etc.
        Thing_ID |
        Thing_Type |
        String |
        Boolean |
        // e.g. pointer or int ID to dynamic-allocated object
        Reference |
        List[Value_Type]

    // S2_Elements should be allocated with stable pointers, or use stable IDs
    parent : Reference(S2_Element)
    // Similar to JSON, but elements are always lists
    // (can have multiple children for the same key
    // (although the layout is not a hard requirement)
    key_to_value := Map[String : List[S2_Element]]
    // property can refer to another property e.g. for coreference
    refers_to : Reference(S2_Element)
    // usually refers to some user feedback UI element
    user_feedback_ref : Reference(Anything)
    // optional: stable pointer or ID to the raw token in 
    // the language input used to create this element
    token : Reference(Token) 
}
\end{lstlisting}

\begin{lstlisting}[caption=Possible structures for S2]
// Note that each right-hand-side can be a list.

CMD_LIST -> any of the rest

ACTION -> any combination of 
            SOURCE,
            DIRECT_OBJECT,
            INDIRECT_OBJECT,
            OBJECT,
            PREPOSITION,
            // contains a "trait" with a string value for the property name
            // e.g. how adverbs or adjectives are used.
            PROPERTY, 
            SEQUENCE_SIMULTANEOUS,
            SEQUENCE_THEN,
            // usually how long the action should last
            TIME
            COREFERENCE,

TRIGGER_RESPONSE -> TRIGGER + RESPONSE

TRIGGER, RESPONSE -> ACTION
// equivalent to ACTION.
// (but handled differently to generate rules.
// TRIGGER should be used to generate rules.
// RESPONSE should be used to generate commands 
// invoked in the future with arguments generated from rule evaluation).
            
PREPOSITION -> OBJECT

SOURCE, DIRECT_OBJECT, INDIRECT_OBJECT, OBJECT -> any combination of
PLURAL, // whether plural or not
COUNT, // number of elements
SPECIFIC_OR_UNSPECIFIC // referring to a specific object or not
PROPERTY 

PLURAL, COUNT, SPECIFIC_OR_UNSPECIFIC, TIME are terminal

SEQUENCE_THEN -> ACTION
SEQUENCE_SIMULTANEOUS -> ACTION

COREFERENCE // contains a pointer to another node in the value, usually noun-like.
\end{lstlisting}

\begin{lstlisting}[
    basicstyle=\tiny, %or \small or \footnotesize etc.
    caption={Output from "Forever the person throws the ball into the pond and then the dog gives the ball to her." prior to object selection},captionpos=t
]
{
    label=[], tag=[], type=[], kind=[], key=[], idx=[0] id=[1931]
    [CMD_LIST] = [
    {
        label=[], tag=[], type=[CMD], kind=[], key=[CMD_LIST], idx=[0] id=[1932]
        [ACTION] = [
        {
            label=[throw], tag=[VERB], type=[ACTION], kind=[ACTION], key=[ACTION], idx=[0] id=[1933]
            [PREPOSITION] = [
            {
                label=[into], tag=[], type=[into], kind=[], key=[PREPOSITION], idx=[0] id=[1934]
                [OBJECT] = [
                {
                    label=[pond], tag=[NOUN], type=[], kind=[THING_INSTANCE], key=[OBJECT], idx=[0] id=[1935]
                    value={
                        THING_INSTANCE=[609]                            
                    }
                    [SPECIFIC_OR_UNSPECIFIC] = [
                    {
                        label=[the], tag=[DET], type=[VALUE], kind=[SPECIFIC], key=[SPECIFIC_OR_UNSPECIFIC], idx=[0] id=[1936]
                        value={
                            FLAG=[true]                                
                        }
                    }
                    ,
                    ]
                    [COUNT] = [
                    {
                        label=[], tag=[], type=[VALUE], kind=[], key=[COUNT], idx=[0] id=[1937]
                        value={
                            NUMERIC=[1.000000]                                
                        }
                    }
                    ,
                    ]
                    [PLURAL] = [
                    {
                        label=[], tag=[], type=[VALUE], kind=[], key=[PLURAL], idx=[0] id=[1938]
                        value={
                            FLAG=[false]                                
                        }
                    }
                    ,
                    ]
                }
                ,
                ]
            }
            ,
            ]
            [SEQUENCE_THEN] = [
            {
                label=[give], tag=[VERB], type=[ACTION], kind=[ACTION], key=[SEQUENCE_THEN], idx=[0], @=[MUST_FILL_IN_AGENT] id=[1939]
                [PREPOSITION] = [
                {
                    label=[to], tag=[], type=[to], kind=[], key=[PREPOSITION], idx=[0] id=[1940]
                    [OBJECT] = [
                    {
                        label=[person], tag=[NOUN], type=[], kind=[THING_INSTANCE], key=[AGENT], idx=[0] id=[1968]
                        <coreference substitution> id=[1960]
                        value={
                            THING_INSTANCE=[606]                                
                        }
                        [SPECIFIC_OR_UNSPECIFIC] = [
                        {
                            label=[the], tag=[DET], type=[VALUE], kind=[SPECIFIC], key=[SPECIFIC_OR_UNSPECIFIC], idx=[0] id=[1969]
                            value={
                                FLAG=[true]                                    
                            }
                        }
                        ,
                        ]
                        [COUNT] = [
                        {
                            label=[], tag=[], type=[VALUE], kind=[], key=[COUNT], idx=[0] id=[1970]
                            value={
                                NUMERIC=[1.000000]                                    
                            }
                        }
                        ,
                        ]
                        [PLURAL] = [
                        {
                            label=[], tag=[], type=[VALUE], kind=[], key=[PLURAL], idx=[0] id=[1971]
                            value={
                                FLAG=[false]                                    
                            }
                        }
                        ,
                        ]
                    }
                    ,
                    ]
                }
                ,
                ]
                [DIRECT_OBJECT] = [
                {
                    label=[ball], tag=[NOUN], type=[], kind=[THING_INSTANCE], key=[DIRECT_OBJECT], idx=[0] id=[1964]
                    <coreference substitution> id=[1955]
                    value={
                        THING_INSTANCE=[607]                            
                    }
                    [SPECIFIC_OR_UNSPECIFIC] = [
                    {
                        label=[the], tag=[DET], type=[VALUE], kind=[SPECIFIC], key=[SPECIFIC_OR_UNSPECIFIC], idx=[0] id=[1965]
                        value={
                            FLAG=[true]                                
                        }
                    }
                    ,
                    ]
                    [COUNT] = [
                    {
                        label=[], tag=[], type=[VALUE], kind=[], key=[COUNT], idx=[0] id=[1966]
                        value={
                            NUMERIC=[1.000000]                                
                        }
                    }
                    ,
                    ]
                    [PLURAL] = [
                    {
                        label=[], tag=[], type=[VALUE], kind=[], key=[PLURAL], idx=[0] id=[1967]
                        value={
                            FLAG=[false]                                
                        }
                    }
                    ,
                    ]
                }
                ,
                ]
                [AGENT] = [
                {
                    label=[dog], tag=[NOUN], type=[], kind=[THING_INSTANCE], key=[AGENT], idx=[0] id=[1951]
                    value={
                        THING_INSTANCE=[608]                            
                    }
                    [SPECIFIC_OR_UNSPECIFIC] = [
                    {
                        label=[the], tag=[DET], type=[VALUE], kind=[SPECIFIC], key=[SPECIFIC_OR_UNSPECIFIC], idx=[0] id=[1952]
                        value={
                            FLAG=[true]                                
                        }
                    }
                    ,
                    ]
                    [COUNT] = [
                    {
                        label=[], tag=[], type=[VALUE], kind=[], key=[COUNT], idx=[0] id=[1953]
                        value={
                            NUMERIC=[1.000000]                                
                        }
                    }
                    ,
                    ]
                    [PLURAL] = [
                    {
                        label=[], tag=[], type=[VALUE], kind=[], key=[PLURAL], idx=[0] id=[1954]
                        value={
                            FLAG=[false]                                
                        }
                    }
                    ,
                    ]
                }
                ,
                ]
            }
            ,
            ]
            [DIRECT_OBJECT] = [
            {
                label=[ball], tag=[NOUN], type=[], kind=[THING_INSTANCE], key=[DIRECT_OBJECT], idx=[0] id=[1955]
                value={
                    THING_INSTANCE=[607]                        
                }
                [SPECIFIC_OR_UNSPECIFIC] = [
                {
                    label=[the], tag=[DET], type=[VALUE], kind=[SPECIFIC], key=[SPECIFIC_OR_UNSPECIFIC], idx=[0] id=[1956]
                    value={
                        FLAG=[true]                            
                    }
                }
                ,
                ]
                [COUNT] = [
                {
                    label=[], tag=[], type=[VALUE], kind=[], key=[COUNT], idx=[0] id=[1957]
                    value={
                        NUMERIC=[1.000000]                            
                    }
                }
                ,
                ]
                [PLURAL] = [
                {
                    label=[], tag=[], type=[VALUE], kind=[], key=[PLURAL], idx=[0] id=[1958]
                    value={
                        FLAG=[false]                            
                    }
                }
                ,
                ]
            }
            ,
            ]
            [PROPERTY] = [
            {
                label=[modifier], tag=[ADV], type=[PROPERTY], kind=[], key=[PROPERTY], idx=[0] id=[1959]
                value={
                    TEXT=[forever]                        
                }
            }
            ,
            ]
            [AGENT] = [
            {
                label=[person], tag=[NOUN], type=[], kind=[THING_INSTANCE], key=[AGENT], idx=[0] id=[1960]
                value={
                    THING_INSTANCE=[606]                        
                }
                [SPECIFIC_OR_UNSPECIFIC] = [
                {
                    label=[the], tag=[DET], type=[VALUE], kind=[SPECIFIC], key=[SPECIFIC_OR_UNSPECIFIC], idx=[0] id=[1961]
                    value={
                        FLAG=[true]                            
                    }
                }
                ,
                ]
                [COUNT] = [
                {
                    label=[], tag=[], type=[VALUE], kind=[], key=[COUNT], idx=[0] id=[1962]
                    value={
                        NUMERIC=[1.000000]                            
                    }
                }
                ,
                ]
                [PLURAL] = [
                {
                    label=[], tag=[], type=[VALUE], kind=[], key=[PLURAL], idx=[0] id=[1963]
                    value={
                        FLAG=[false]                            
                    }
                }
                ,
                ]
            }
            ,
            ]
        }
        ,
        ]
    }
    ,
    ]
}
\end{lstlisting}

\begin{lstlisting}[
    basicstyle=\tiny, %or \small or \footnotesize etc.
    caption={Output from "Every few seconds the frog hops to a lily." prior to object selection},captionpos=t
]
{
    label=[], tag=[], type=[], kind=[], key=[], idx=[0] id=[1018]
    [CMD_LIST] = [
    {
        label=[], tag=[], type=[CMD], kind=[], key=[CMD_LIST], idx=[0] id=[1019]
        [ACTION] = [
        {
            label=[hop], tag=[VERB], type=[ACTION], kind=[ACTION], key=[ACTION], idx=[0] id=[1005]
            [PREPOSITION] = [
            {
                label=[to], tag=[], type=[to], kind=[], key=[PREPOSITION], idx=[0] id=[1013]
                [OBJECT] = [
                {
                    label=[lily], tag=[NOUN], type=[], kind=[THING_INSTANCE], key=[OBJECT], idx=[0] id=[1014]
                    value={
                        THING_INSTANCE=[0]                        
                    }
                    [SPECIFIC_OR_UNSPECIFIC] = [
                    {
                        label=[a], tag=[DET], type=[VALUE], kind=[UNSPECIFIC], key=[SPECIFIC_OR_UNSPECIFIC], idx=[0] id=[1016]
                        value={
                            FLAG=[false]                            
                        }
                    }
                    ,
                    ]
                    [COUNT] = [
                    {
                        label=[], tag=[], type=[], kind=[], key=[COUNT], idx=[0] id=[1017]
                        value={
                            NUMERIC=[1.000000]                            
                        }
                    }
                    ,
                    ]
                    [PLURAL] = [
                    {
                        label=[], tag=[], type=[VALUE], kind=[], key=[PLURAL], idx=[0] id=[1015]
                        value={
                            FLAG=[false]                            
                        }
                    }
                    ,
                    ]
                }
                ,
                ]
            }
            ,
            ]
            [TIME] = [
            {
                label=[second], tag=[TIME], type=[INTERVAL], kind=[], key=[TIME], idx=[0] id=[1006]
                [PROPERTY] = [
                {
                    label=[trait], tag=[ADJ], type=[PROPERTY], kind=[], key=[PROPERTY], idx=[0] id=[1008]
                    value={
                        TEXT=[few]                        
                    }
                }
                ,
                ]
            }
            ,
            ]
            [AGENT] = [
            {
                label=[frog], tag=[NOUN], type=[], kind=[THING_INSTANCE], key=[AGENT], idx=[0] id=[1009]
                value={
                    THING_INSTANCE=[0]                    
                }
                [SPECIFIC_OR_UNSPECIFIC] = [
                {
                    label=[the], tag=[DET], type=[VALUE], kind=[SPECIFIC], key=[SPECIFIC_OR_UNSPECIFIC], idx=[0] id=[1011]
                    value={
                        FLAG=[true]                        
                    }
                }
                ,
                ]
                [COUNT] = [
                {
                    label=[], tag=[], type=[VALUE], kind=[], key=[COUNT], idx=[0] id=[1012]
                    value={
                        NUMERIC=[1.000000]                        
                    }
                }
                ,
                ]
                [PLURAL] = [
                {
                    label=[], tag=[], type=[VALUE], kind=[], key=[PLURAL], idx=[0] id=[1010]
                    value={
                        FLAG=[false]                        
                    }
                }
                ,
                ]
            }
            ,
            ]
        }
        ,
        ]
    }
    ,
    ]
}
\end{lstlisting}

\end{document}

%% file: implementation.tex
\section{Implementation}

\subsection{Hardware and Software}

\begin{figure}
    \centering    
    \includegraphics[width=\linewidth]{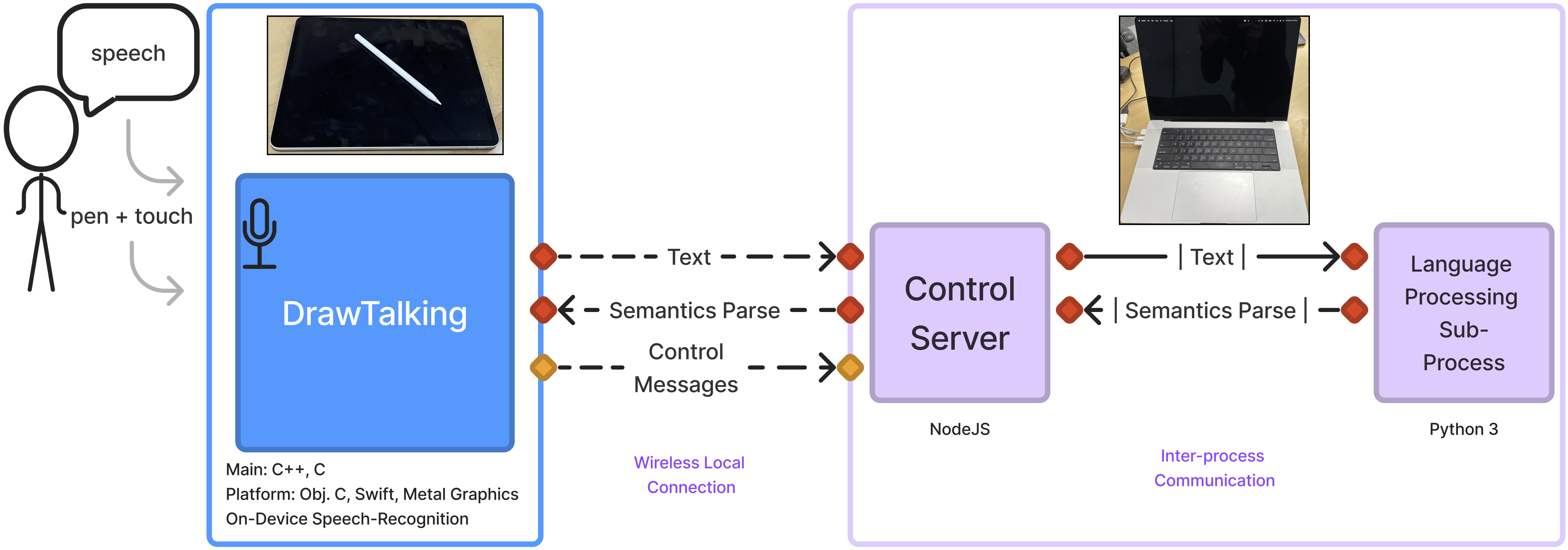}
    \caption{\textbf{Concrete Client/Server Details}: Text data are sent to a server for natural language processing over a data channel and forwarded to an NLP-focused sub-process on the same machine. The semantic parse data return to the client. A command channel handles user events independently (e.g. discarding commands interrupts in-flight stale NLP).}
    \label{fig:arch_implementation_network}
    \Description[A system diagram for the client-server architecture and data flow]{A system diagram for the client-server architecture and data flow. The user is shown as a stick-figure with a quote bubble reading "speech" and the text "pen+touch" both connected to the main DrawTalking client node in the system diagram (indicating possible sources of user input.) The client node contains a real-life photo of the researcher's iPad Pro used for the prototype, atop a blue square with the name "DrawTalking" on it, as well as a small cartoon microphone icon. Below the square reads: "Main C++, C; Platform: Obj C, Swift, Metal Graphics; On-Device Speech Recognition". 3 arrows connect the client node to a server node. The arrows are referred to as "Wireless Local Connection" (to the server. Arrow 1 (outgoing) is for text data. Arrow 2(incoming) is for the semantics parse. Arrow 3 (outgoing) is for control messages to the server; In the server node is a real-life photo of the Macbook Pro used for the server, above two nodes: left: the control server labeled NodeJS; right the Language Processing Sub-Process labeled Python 3. 2 arrows between server nodes (both surrounded by an enclosing rectangle): arrow 1: (outgoing) is for text. arrow 2: (incoming) is for the resulting semantics parse, which is sent back to the control server and back to the DrawTalking client along the other arrow incoming to the DrawTalking client.}
\end{figure}

We implemented DrawTalking natively on the iPad Pro M2, with the main code written in C++ and C, bridged with Objective C++ and Swift to access platform-specific APIs. Speech recognition is on-device. A local NodeJS+Python server-side for NLP runs on a MacBook Pro. Text is continuously sent to the server where 1) a dependency-tree is created, 2) sent to the client and compiled into a key-value format for semantic roles (e.g. \textit{AGENT}, \textit{OBJECT}), 3) either interpreted in real-time by our custom engine to drive interactive simulations, or used to find objects by deixis.
In final stages, the system looks-up and maps objects to roles in the command structure and executes it to produce animations, effects, and state changes. (See \autoref{fig:arch_implementation_network} for a diagram of the the concrete client-server implementation and data-flow.)

\subsection{Processing Steps}
\begin{figure}
    \centering    
    \includegraphics[width=\linewidth]{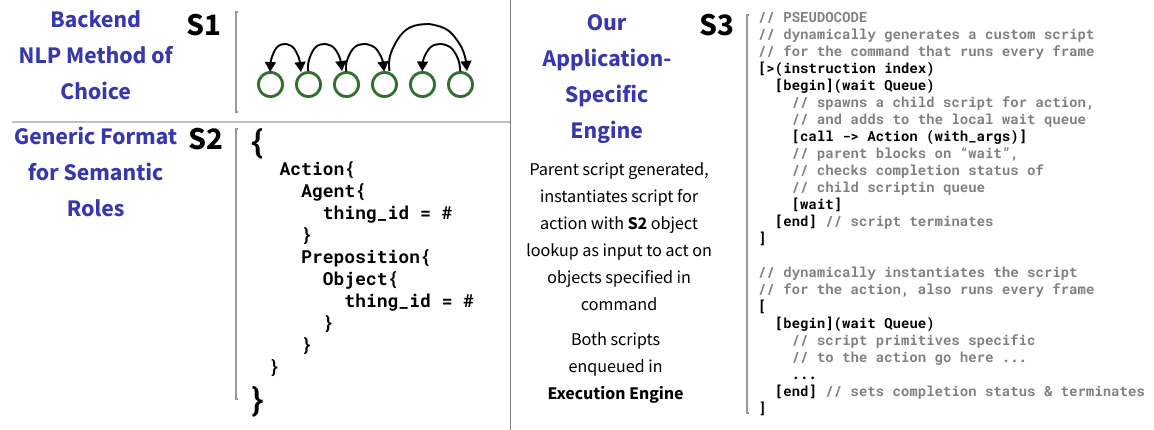}
    \caption[System Language Structures]{\textbf{Language processing into commands} (S1 to S3): S1 is any implementation-specific NLP, e.g. a dependency tree; S2 stores generic semantic role labels created from S1; S3 is an app-specific engine that receives and interprets S2. }
    \label{fig:language_structures}
    \Description[Language processing data structures pipeline]{Language processing data structures pipeline. 3 parts, S1, S2, S3: S1 "Backend NLP Method of Choice" - accompanied by an abstract NLP dependency tree diagram. S2: Generic Format for Semantic Roles, accompanied by the following object:
    "
    {
      Action{
        Agent{
          thing_id = \#
        }
        Preposition{
          Object{
            thing_id = \#
          }
        }
      }
    }
    ";
    S3: Our Application-Specific Engine, accompanied by a text description and pseudocode for the engine code. The description reads: "Parser script generated, instantiates script for action with \textbf{S2} object lookup as input to act on objects specified in command
    Both scripts enqueued in \textbf{Execution Engine}. The pseudocode reads:
    "
    // PSEUDOCODE
// dynamically generates a custom script
// for the command that runs every frame
[>(instruction index)
  [begin](wait Queue)
    // spawns a child script for action,
    // and adds to the local wait queue
    [call -> Action (with_args)]
    // parent blocks on “wait”,
    // checks completion status of 
    // child script in queue
    [wait]
  [end] // script terminates
]

// dynamically instantiates the script 
// for the action, also runs every frame
[
  [begin](wait Queue)
    // script primitives specific
    // to the action go here ...
    ...
  [end] // sets completion status \& terminates
]
    "
    }
\end{figure}

The language processing component comprises roughly 4 steps that translate a natural language data structure into a command for the application (\autoref{fig:language_structures}). These user-initiated commands are scripts constructed and executed at run-time. As they're executed, the scripts handle data states, sequences, and dynamic instantiations of scripts corresponding to verbs in our built-in library. Verb scripts are modules comprising arbitrary code and accepted parameters, as opposed to generated code. The system draws from ideas in visual programming and virtual machine-based engines \cite{Max:Cycling74, resnick_scratch_2009,AnotherWorld:CodeReview, game:LittleBigPlanet2}. Note that implementers can add functionality without modifying the rest of the system. For supplementary examples, see \autoref{section:abstract_semantic_structure}.

\paragraph{Natural language raw text $\rightarrow$ S1}
\label{lang:struct1} The system receives language input and outputs a directed graph data structure encoding the dependencies between words and the words' semantic roles. 

Our implementation uses the Spacy (v 3.1.4)\cite{montani_explosionspacy_2023} library to output a dependency tree per-sentence, along with a library called "coreferee" to fill-in coreference information. Accuracy of this phase is tied to the chosen NLP method, not our interface. Spacy tended to reproduce the same results for the same sentence structures, which made visual output predictable.

\paragraph{S1 $\rightarrow$ S2}
\label{lang:struct2} The system traverses S1 and generates a new generic graph structure S2. Each node entry in the graph structure represents a nested hierarchy of semantic units containing information such as labels (the word actually used) and part of speech (such as noun, verb, etc.).

\paragraph{Incomplete S2 $\rightarrow$ S2 with system-context feedback}
\label{lang:struct2_system_feedback}
Upon creating the structure for S2, there might be unspecified placeholders for objects, which we call "incomplete". The system will now look at application-context such as user-input and objects, and fill the structure with concrete entity IDs as-needed.

A query sub-system is needed to register and lookup objects based on their labels. For each noun-like entry in the traversal, the system queries for objects with the given noun and adjective labels, and returns the IDs of all objects in the world that match those labels.

\paragraph{Incomplete S2 $\rightarrow$ complete S2 with user feedback}
\label{lang:struct2_user_feedback}
After this process, the user should be able to do last-second editing and correction of the intermediate structure. The system can output user feedback (\autoref{fig:semantics-diagram}).

\paragraph{S2 $\rightarrow$ S3}
\label{lang:struct3} Lastly, the application traverses the structure and generates a final application-defined structure S3 that it can evaluate -- in our case, a mix between scriptable virtual machine and animation engine akin to \cite{perlin_improv_1996, resnick_scratch_2009} or a runnable node-based program \cite{Max:Cycling74}. S3 can call these scripts and retain S2 for reference. When traversing nodes in S2 labeled ACTION (the verbs) we lookup the appropriate previously-defined script for that action and insert a reference to it in S3. The arguments for that action, i.e. semantic role keys mapped to object ids or types, are inserted into a lookup table specific to the ACTION script instance so when the action executes, it knows what objects to modify. Loops and timed waits are also inserted into S3. The application evaluates the final structure according to its own interpretation. This results in any potential application-specific side-effect -- in our case, animation, simulation, rule creation, application state-changes, and so on.

The application can also retain the generic S2 structure to generate commands later.

\subsection{Current Limitations of Semantic Parsing}
\label{sec:limitation_semantic_parsing}
We focused on supporting narrative third-person form for this prototype due to our emphasis on scenarios involving explanation and storytelling. Passive tense, split infinitives, dangling prepositions, and idioms among others are unsupported. Other tenses and forms \textit{are} supported as input, but behave the same as present tense. For example, past tense (\textit{"the dog jumped"}), future tense (\textit{"the dog will jump"}), present-progressive (\textit{"the dog is jumping"}), and imperative (\textit{"dog, jump"}) are equivalent to \textit{"the dog jumps."} This allows for variety in the storytelling, but a future implementation might wish to distinguish between tenses for greater expressiveness.

Rules must sometimes be specified verbosely to avoid ambiguity in more complex cases. For example:  \textit{"When dogs collide with cats dogs jump and cats jump."} This could be expressed less verbosely, but the prototype at-present cannot disambiguate alternatives such as \textit{"When dogs collide with cats they jump."} Here, "they" is ambiguous because it could refer to "dogs" only, "cats" only, or as intended in this example, both. The first form is more verbose, but unambiguous. We support the second, but just as in English, the result might be unclear and the system will likely pick either "dogs" or "cats."

Note that the semantics diagram still appears when a command is created from unsupported grammar. However, the diagram might show incorrect or missing object mappings and semantic roles. In such cases, the user can discard the malformed command or link missing objects to the diagram as in \autoref{fig:semantics-diagram}. The user must learn and restrict themselves to the specific grammar to get the most consistent results. We cover how we might support more natural speech in \autoref{sec:future_work}.

Although commands are user-initiated by design, a potential limitation is that commands must be created and confirmed one-by-one. Continuous speech input has unclear sentence boundaries, so we do not currently integrate a component to predict where a command should end because this could be unreliable and would take away control. To avoid rushing the user, we did not rely on a timeout either. We chose to give the user control over the sentence boundaries via the speech command button and interaction with the text transcript view (\autoref{fig:text_selection_in_view}). Different implementations could try automating commands without compromising user-control.

Additionally, the procedure to transform S1 into S2 (\autoref{lang:struct2}) is coupled to the NLP library of choice by design. S2 and onwards are generic. An implementer would need to be aware of the need to build their own version of this translation step if their NLP back-ends use different dependency labels.